\documentclass[twoside,12pt]{article}
\usepackage{epsfig}

\newcommand{\be}{\begin{equation}}
\newcommand{\ee}{\end{equation}}
\newcommand{\bea}{\begin{eqnarray}}
\newcommand{\eea}{\end{eqnarray}}

\usepackage{amsmath}
\usepackage{amsfonts}
\usepackage{amssymb}
\usepackage{amsxtra}
\usepackage{supertabular}
\begin{document}
\title{ \vspace{1cm}
How do Clifford algebras show the way to the second quantized fermions with
unified spins, charges and families, and to the corresponding
second quantized vector and scalar gauge fields } 
%
\author{N.S.\ Manko\v c Bor\v stnik,$^{1}$ \\
$^1$Department of Physics, University of Ljubljana\\
SI-1000 Ljubljana, Slovenia}
\maketitle

\begin{abstract}
This contribution presents properties of the second quantized not only fermion fields
but also boson fields, if the second quantization of both kinds of fields origins in 
the description  of the internal space of fields with the ''basis vectors'' which are the superposition of odd (when describing fermions)  or even (when describing bosons) 
products of the Clifford algebra operators $\gamma^a$'s.  The tensor products of
the ''basis vectors'' with the basis in ordinary space forming the creation operators 
manifest the anticommutativty (of fermions) or commutativity  (of bosons) of 
the ''basis vectors'', explaining the second quantization postulates of both kinds of fields. 
Creation operators of boson fields have all the properties of the gauge fields of the corresponding fermion fields, offering a new understanding of the fermion and 
boson fields.
\end{abstract}


%
\section{Introduction}
\label{introduction}

%
%
%
%


In a long series of works~\cite{norma92,%
norma93,pikanorma,IARD2016,n2014matterantimatter,JMP2013,normaJMP2015,n2012scalars}
I have found, together with the collaborators~(\cite{nh02,nh03,nd2017,nh2018,%
n2019PIPII,nh2021RPPNP} and the references therein), with H.B. Nielsen and in long 
discussions with participants during the annual workshops ''What comes beyond the 
standard models'', the phenomenological 
success with the model named the  {\it spin-charge-family}  theory:  
The internal space of fermions are in this model described with the Clifford 
algebra objects of all linear superposition of odd products of $\gamma^a$'s in 
$d=(13+1)$. Fermions  
 interact with only gravity --- with the vielbeins and the two kinds of the spin 
connection fields (the gauge fields of $S^{ab}=\frac{i}{4}(\gamma^a \gamma^b- 
\gamma^b \gamma^a)$ and $\tilde{S}^{ab}=\frac{1}{4} (\tilde{\gamma}^a 
\tilde{\gamma}^b - \tilde{\gamma}^b  \tilde{\gamma}^a)$~%
\footnote{
If there are no fermion present the two kinds of the spin connection fields are 
uniquely described by the vielbeins~\cite{prd2018}.}). Spins from higher 
dimensions, $d>(3+1)$, described by $\gamma^a$'s,  manifest in $d=(3+1)$ as 
charges of the {\it standard model} quarks and leptons and antiquarks and 
antileptons, appearing in (two times four) families, the quantum numbers of which
are determined by the second kind of the Clifford  algebra object $\tilde{S}^{ab}$'s.
Gravity in higher dimensions manifests as the {\it standard model} vector gauge 
fields as well as the scalar Higgs  and Yukawa couplings~\cite{nd2017,IARD2016,%
n2014matterantimatter,JMP2013,normaJMP2015,n2012scalars,nh02,nh03,nd2017,%
nh2018,nh2021RPPNP,normaBled2020,n2019PIPII,2020PartIPartII}, 
predicting new scalar fields, which offer the explanation besides for higgs scalar 
and Yukawa couplings also for the asymmetry between matter and antimatter 
in our universe and for the dark matter (represented by the stable of the upper 
group of four families), predicting a new family --- the fourth family to the 
observed three.

In this contribution I shortly repeat the description  of the internal space of the 
second quantized fermion fields with the odd products of the Clifford operators 
$\gamma^a$'s,  what leads to the creation operators for fermions without 
postulating  the second quantization requirements of Dirac~\cite{Dirac,BetheJackiw,Weinberg}.
The creation operators for fermions, which are superposition of tensor products 
of the ordinary basis and  the ''basis vectors'' describing the internal space of 
fermions, anticommute, explaining correspondingly the postulates of Dirac, 
offering also a new understanding of fermion fields~(\cite{nh2021RPPNP} and
references therein).  The creation operators of fermions  appear in families, 
carrying either left or right handedness, their Hermitian conjugated partners 
belong  to another set of Clifford odd ''basis vectors'' carrying the opposite 
handedness.

The main part of this contribution discusses properties of the second quantized
boson fields, which are the gauge fields of the corresponding second quantized 
fermion fields. The internal space of bosons is described by the superposition
of even products of $\gamma^a$'s. The boson fields correspondingly commute.
The corresponding creation operators and their Hermitian conjugated partners
belong to the same set of ''basis vectors'', carrying all the quantum numbers 
in adjoint representations. They interact among themselves and with the 
corresponding fermion fields.

In Sect.~\ref{GrassmannClifford} the anticommuting Grassmann and Clifford 
algebras are presenting and the relations among them discussed. The ''basis 
vectors'' are defined as the egenvectors of the Cartan subalgebra of the 
Lorentz algebra for the Grassmann and the two Clifford algebras for odd
and for even products of algebras members, and their anticommutation or 
commutation relations presented, Subsect.~\ref{basisvectors}.  The reduction 
of the two kinds of the Clifford algebras to only one makes the Clifford odd
''basis vectors'' anticommuting,  giving to different irreducible representations
of the Lorentz algebra the family quantum numbers, Subsect.~\ref{reduction}.
To make understanding of the properties of the Clifford odd and Clifford even
''basis vectors'' easier in Subsect.~\ref{cliffordoddevenbasis5+1} the case of 
$d=(5+1)$-dimensional space is chosen and the ''basis vectors'' of odd, 
\ref{odd5+1}, and even, \ref{even5+1}, Clifford character are presented 
in details and then generalized
to any even $d$, Subsect.~\ref{generalbasisinternal}.

In Sect.~\ref{fermionsbosons} the creation operators of the second quantized 
fermion and boson fields are discussed, as well as their Hermitian conjugated 
partners. In Subsect.~\ref{fermionbosonaction} the simple action for fermion 
interacting with bosons and for corresponding bosons, as assumed in the
{\it spin-charge-family} theory is presented.
 
Sect.~\ref{conclusions} reviews shortly what one can learn in this contribution.

Both algebras, Grassmann and Clifford, offer "basis vectors" for 
the description of the internal space of fermions~\cite{norma92,norma93,%
nh2021RPPNP} and the corresponding bosons with which fermions interact. 
The oddness or evenness of ''basis vectors'', transfered to 
the creation operators, which are tensor products of the finite number of 
''basis vectors'' and the (continuously) infinite number of momentum 
(or coordinate) basis, and  to their Hermitian conjugated partners annihilation 
operators, offers the second quantization of fermions and bosons without 
postulating the 
second quantized conditions~\cite{Dirac,BetheJackiw,Weinberg} for either the 
half integer spin fermions or integer spin bosons, enabling the explanation 
of the Dirac's postulates. 
Further investigations are needed in both case, for the boson case in particular,
although promising, the time for this study was too short. 


%
%

%
%
%

%
\section{Grassmann and Clifford algebras}
\label{GrassmannClifford}

To describe the internal space of fermions and bosons one can use either the 
Grassmann or the Clifford algebras.
 
In Grassmann $d$-dimensional space there are $d$ anticommuting operators 
$\theta^{a}$, $\{\theta^{a}, \theta^{b}\}_{+}=0$, $a=(0,1,2,3,5,..,d)$, and 
$d$ anticommuting derivatives with respect to $\theta^{a}$, 
$\frac{\partial}{\partial \theta_{a}}$, $\{\frac{\partial}{\partial \theta_{a}}, 
\frac{\partial}{\partial \theta_{b}}\}_{+} =0$, offering
together $2\cdot2^d$ operators, the half of which are superposition of products of  
$\theta^{a}$ and another half corresponding superposition of 
$\frac{\partial}{\partial \theta_{a}}$.
\begin{eqnarray}
\label{thetaderanti0}
\{\theta^{a}, \theta^{b}\}_{+}=0\,, \, && \,
\{\frac{\partial}{\partial \theta_{a}}, \frac{\partial}{\partial \theta_{b}}\}_{+} =0\,,
\nonumber\\
\{\theta_{a},\frac{\partial}{\partial \theta_{b}}\}_{+} &=&\delta_{ab}\,, (a,b)=(0,1,2,3,5,\cdots,d)\,.
\end{eqnarray}
Defining~\cite{nh2018} 
\begin{eqnarray}
(\theta^{a})^{\dagger} &=& \eta^{a a} \frac{\partial}{\partial \theta_{a}}\,,\quad
{\rm leads  \, to} \quad
(\frac{\partial}{\partial \theta_{a}})^{\dagger}= \eta^{a a} \theta^{a}\,.
\label{thetaderher0}
\end{eqnarray}
$ \theta^{a}$ and $ \frac{\partial}{\partial \theta_{a}}$ are, up to the sign, Hermitian conjugated to each other. The identity is the self adjoint member of the algebra. We 
make a choice for the complex properties of $\theta^a$, and correspondingly 
of $\frac{\partial}{\partial \theta_{a}}$, as follows
\begin{eqnarray}
\label{complextheta}
\{\theta^a\}^* &=&  (\theta^0, \theta^1, - \theta^2, \theta^3, - \theta^5,
\theta^6,...,- \theta^{d-1}, \theta^d)\,, \nonumber\\
\{\frac{\partial}{\partial \theta_{a}}\}^* &=& (\frac{\partial}{\partial \theta_{0}},
\frac{\partial}{\partial \theta_{1}}, - \frac{\partial}{\partial \theta_{2}},
\frac{\partial}{\partial \theta_{3}}, - \frac{\partial}{\partial \theta_{5}}, 
\frac{\partial}{\partial \theta_{6}},..., - \frac{\partial}{\partial \theta_{d-1}}, 
\frac{\partial}{\partial \theta_{d}})\,. 
\end{eqnarray}
In $d$-dimensional space of anticommuting Grassmann coordinates and of their Hermitian conjugated partners derivatives, Eqs.~(\ref{thetaderanti0}, \ref{thetaderher0}), there 
exist two kinds of the Clifford coordinates (operators) --- $\gamma^{a}$ and 
$\tilde{\gamma}^{a}$ --- both expressible in terms of $\theta^{a}$ and their conjugate momenta $p^{\theta a}= i \,\frac{\partial}{\partial \theta_{a}}$ ~\cite{norma93}.
\begin{eqnarray}
\label{clifftheta1}
\gamma^{a} &=& (\theta^{a} + \frac{\partial}{\partial \theta_{a}})\,, \quad 
\tilde{\gamma}^{a} =i \,(\theta^{a} - \frac{\partial}{\partial \theta_{a}})\,,\nonumber\\
\theta^{a} &=&\frac{1}{2} \,(\gamma^{a} - i \tilde{\gamma}^{a})\,, \quad 
\frac{\partial}{\partial \theta_{a}}= \frac{1}{2} \,(\gamma^{a} + i \tilde{\gamma}^{a})\,,
\nonumber\\
\end{eqnarray}
offering together  $2\cdot 2^d$  operators: $2^d$ of those which are products of 
$\gamma^{a}$  and  $2^d$ of those which are products of $\tilde{\gamma}^{a}$.
%
Taking into account Eqs.~(\ref{thetaderher0}, \ref{clifftheta1}) it is easy to prove 
that they form two independent anticommuting Clifford algebras, Refs.~(\cite{nh2021RPPNP} and references therein)
\begin{eqnarray}
\label{gammatildeantiher}
\{\gamma^{a}, \gamma^{b}\}_{+}&=&2 \eta^{a b}= \{\tilde{\gamma}^{a}, 
\tilde{\gamma}^{b}\}_{+}\,, \nonumber\\
\{\gamma^{a}, \tilde{\gamma}^{b}\}_{+}&=&0\,,\quad
 (a,b)=(0,1,2,3,5,\cdots,d)\,, \nonumber\\
(\gamma^{a})^{\dagger} &=& \eta^{aa}\, \gamma^{a}\, , \quad 
(\tilde{\gamma}^{a})^{\dagger} =  \eta^{a a}\, \tilde{\gamma}^{a}\,,
\end{eqnarray}
with $\eta^{a b}=diag\{1,-1,-1,\cdots,-1\}$.

While the Grassmann algebra can be used to describe the ''anticommuting integer spin
second quantized fields'' and ''commuting integer spin second quantized fields''~\cite{%
nh2021RPPNP,n2021MDPIsymmetry}, the Clifford algebras describe the second quantized 
fermion fields, if the superposition of odd products of $\gamma^a$'s or 
$\tilde{\gamma}^a$'s are used. The superposition of even products of either 
$\gamma^a$'s or $\tilde{\gamma}^a$'s describe the commuting second quantized 
boson fields.

The reduction, Eq.~(\ref{tildegammareduced}) of Subsect.~(\ref{reduction}), of 
the two Clifford algebras --- $\gamma^a$'s and 
$\tilde{\gamma}^a$'s --- to only one  is needed --- $\gamma^a$'s are chosen --- for 
the correct description of the internal space of fermions. After the decision  that only  
$\gamma^a$'s are used to describe the internal space of fermions, the remaining ones, 
$\tilde{\gamma}^a$'s, are used to equip the irreducible representations of the Lorentz 
group (with the infinitesimal generators $S^{ab}=\frac{i}{4} \{\gamma^a, 
\gamma^b\}_{-}$) with the family quantum numbers in the case that the odd  Clifford algebra describes the internal space of the second quantized fermions. 

It then follows that the even Clifford algebra objects, the superposition of the even
products of $\gamma^a$'s, offer the description of the second quantized boson fields, 
which are the gauge fields of the second quantized fermion fields,  the internal space of which are described by the odd Clifford algebra objects.
This will be demonstrated in this contribution. 
%
\subsection{''Basis vectors'' determined by superposition of odd and even products of Clifford objects.}
\label{basisvectors}
%
 
 There are $\frac{d}{2}$ members of the Cartan subalgebra of the Lorentz 
 algebra in even dimensional spaces.
 One can choose
\begin{eqnarray}
{\cal {\bf S}}^{03}, {\cal {\bf S}}^{12}, {\cal {\bf S}}^{56}, \cdots, 
{\cal {\bf S}}^{d-1 \;d}\,, \nonumber\\
S^{03}, S^{12}, S^{56}, \cdots, S^{d-1 \;d}\,, \nonumber\\
\tilde{S}^{03}, \tilde{S}^{12}, \tilde{S}^{56}, \cdots,  \tilde{S}^{d-1\; d}\,, \nonumber\\
{\cal {\bf S}}^{ab} = S^{ab} +\tilde{S}^{ab}\,.
\label{cartangrasscliff}
\end{eqnarray}
 
 Let us look for the ''eigenstates'' of each of the Cartan subalgebra members, 
Eq.~(\ref{cartangrasscliff}), for each of the two kinds of the Clifford algebras 
separately,   
\begin{eqnarray}
S^{ab} \frac{1}{2} (\gamma^a + \frac{\eta^{aa}}{ik} \gamma^b) &=& \frac{k}{2}  \,
\frac{1}{2} (\gamma^a + \frac{\eta^{aa}}{ik} \gamma^b)\,,\quad
S^{ab} \frac{1}{2} (1 +  \frac{i}{k}  \gamma^a \gamma^b) = \frac{k}{2}  \,
 \frac{1}{2} (1 +  \frac{i}{k}  \gamma^a \gamma^b)\,,\nonumber\\
\tilde{S}^{ab} \frac{1}{2} (\tilde{\gamma}^a + \frac{\eta^{aa}}{ik} \tilde{\gamma}^b) &=& 
\frac{k}{2}  \,\frac{1}{2} (\tilde{\gamma}^a + \frac{\eta^{aa}}{ik} \tilde{\gamma}^b)\,,
\quad
\tilde{S}^{ab} \frac{1}{2} (1 +  \frac{i}{k}  \tilde{\gamma}^a \tilde{\gamma}^b) = 
 \frac{k}{2}  \, \frac{1}{2} (1 +  \frac{i}{k} \tilde{\gamma}^a \tilde{\gamma}^b)\,,
\label{eigencliffcartan}
\end{eqnarray}
 $k^2=\eta^{aa} \eta^{bb}$.
 The proof of Eq.~(\ref{eigencliffcartan}) is presented in Ref.~\cite{nh2021RPPNP}, 
 App.~(I).
 
 Let us introduce for nilpotents $\frac{1}{2} (\gamma^a + \frac{\eta^{aa}}{ik} \gamma^b), (\frac{1}{2} (\gamma^a + \frac{\eta^{aa}}{ik} \gamma^b))^2=0$ and projectors $ \frac{1}{2} (1 +  \frac{i}{k}  \gamma^a \gamma^b),
( \frac{1}{2} (1 +  \frac{i}{k}  \gamma^a \gamma^b))^2 =
 \frac{1}{2} (1 +  \frac{i}{k}  \gamma^a \gamma^b)$ 
of both algebras the notation
\begin{eqnarray}
\stackrel{ab}{(k)}:&=& 
\frac{1}{2}(\gamma^a + \frac{\eta^{aa}}{ik} \gamma^b)\,,\quad 
\stackrel{ab}{(k)}^{\dagger} = \eta^{aa}\stackrel{ab}{(-k)}\,,\quad 
(\stackrel{ab}{(k)})^2 =0\,, \quad \stackrel{ab}{(k)}\stackrel{ab}{(-k)}
=\eta^{aa}\stackrel{ab}{[k]}\nonumber\\
\stackrel{ab}{[k]}:&=&
\frac{1}{2}(1+ \frac{i}{k} \gamma^a \gamma^b)\,,\quad \;\,
\stackrel{ab}{[k]}^{\dagger} = \,\stackrel{ab}{[k]}\,, \quad \quad \quad \quad
(\stackrel{ab}{[k]})^2 = \stackrel{ab}{[k]}\,, 
\quad \stackrel{ab}{[k]}\stackrel{ab}{[-k]}=0\,,
\nonumber\\
\stackrel{ab}{(k)}\stackrel{ab}{[k]}& =& 0\,,\qquad \qquad \qquad 
\stackrel{ab}{[k]}\stackrel{ab}{(k)}=  \stackrel{ab}{(k)}\,, \quad \quad \quad
  \stackrel{ab}{(k)}\stackrel{ab}{[-k]} =  \stackrel{ab}{(k)}\,,
\quad \, \stackrel{ab}{[k]}\stackrel{ab}{(-k)} =0\,,
\nonumber\\
\stackrel{ab}{\tilde{(k)}}:&=& 
\frac{1}{2}(\tilde{\gamma}^a + \frac{\eta^{aa}}{ik} \tilde{\gamma}^b)\,,\quad 
\stackrel{ab}{\tilde{(k)}}^{\dagger} = \eta^{aa}\stackrel{ab}{\tilde{(-k)}}\,,\quad
(\stackrel{ab}{\tilde{(k)}})^2=0\,,\nonumber\\
\stackrel{ab}{\tilde{[k]}}:&=&
\frac{1}{2}(1+ \frac{i}{k} \tilde{\gamma}^a \tilde{\gamma}^b)\,,\quad \;\,
\stackrel{ab}{\tilde{[k]}}^{\dagger} = \,\stackrel{ab}{\tilde{[k]}}\,,
\quad \quad \quad \quad
(\stackrel{ab}{\tilde{[k]}})^2=\stackrel{ab}{\tilde{[k]}}\,,\nonumber\\
\stackrel{ab}{\tilde{(k)}}\stackrel{ab}{\tilde{[k]}}& =& 0\,,\qquad \qquad \qquad 
\stackrel{ab}{\tilde{[k]}}\stackrel{ab}{\tilde{(k)}}=  \stackrel{ab}{\tilde{(k)}}\,, 
\quad \quad \quad
  \stackrel{ab}{\tilde{(k)}}\stackrel{ab}{\tilde{[-k]}} =  \stackrel{ab}{\tilde{(k)}}\,,
\quad \, \stackrel{ab}{\tilde{[k]}}\stackrel{ab}{\tilde{(-k)}} =0\,,
\label{graficcliff}
\end{eqnarray}

{\bf Statement 1.} {\it One can define ''basis vectors'' to be eigenvectors of all 
the members of the Cartan subalgebras as even or odd products of nilpotents 
and projectors in any even dimensional space.}\\

Due to the anticommuting properties of the Clifford algebra objects there are
anticommuting and commuting ''basis vectors''. The anticommuting ''basis vectors''
contain an odd products of nilpotents, at least one nilpotent, the rest are then 
projectors. Let us denote the  Clifford 
odd  ''basis vectors'' of the Clifford $\gamma^a$ kind as 
$\hat{b}^{m \dagger}_{f}$, where $m$ and $f$ determine the $m^{th}$ 
member of the $f^{th} $ irreducible representation. We shall denote by 
 $\hat{b}^{m}_{f}$ $=(\hat{b}^{m \dagger}_{f})^{\dagger}$ the 
 Hermitian conjugated partner of the ''basis vector''  $\hat{b}^{m \dagger}_{f}$.
 The ''basis vectors'' of the Clifford $\tilde{\gamma}^a$ kind 
 would correspondingly  be denoted by $\hat{\tilde{b}}^{m \dagger}_{f}$ and 
 $\hat{\tilde{b}}^{m}_{f}$. 
 
 It is not difficult to prove the anticommutation relations of the Clifford odd
 ''basis vectors''  and their Hermitian conjugated partners for both algebras~%
(\cite{norma92,nh2021RPPNP} and references therein). Let us  here present
only the one of the Clifford algebras --- $\gamma^a$'s.
\begin{eqnarray}
\label{almostDirac}
\hat{b}^{m}_{f} {}_{*_{A}}|\psi_{oc}>&=& 0\, |\psi_{oc}>\,,\nonumber\\
\hat{b}^{m \dagger}_{f}{}_{*_{A}}|\psi_{oc}>&=&  |\psi^m_{f}>\,,\nonumber\\
\{\hat{b}^{m}_{f}, \hat{b}^{m'}_{f'}\}_{*_{A}+}|\psi_{oc}>&=&
 0\,|\psi_{oc}>\,, \nonumber\\
\{\hat{b}^{m \dagger}_{f}, \hat{b}^{m \dagger}_{f}\}_{*_{A}+}|\psi_{oc}>
&=&|\psi_{oc}>\,,
\end{eqnarray}
where  ${*_{A}}$ represents the algebraic multiplication of 
$\hat{b}^{m \dagger}_{f}$  and $ \hat{b}^{m'}_{f'} $  among themselves and  
with the vacuum state  $|\psi_{oc}>$ of Eq.(\ref{vaccliff}), which takes into account 
Eq.~(\ref{gammatildeantiher0}), 
\begin{eqnarray}
\label{vaccliff}
|\psi_{oc}>= \sum_{f=1}^{2^{\frac{d}{2}-1}}\,\hat{b}^{m}_{f}{}_{*_A}
\hat{b}^{m \dagger}_{f} \,|\,1\,>\,,
\end{eqnarray}
for one of the members $m$, anyone, of the odd irreducible representation $f$,
with $|\,1\,>$, which is the vacuum without any structure --- the identity.
It follows that $\hat{b}^{m}_{f}{} ||\psi_{oc}>=0$.
The relations are valid for both kinds of the odd Clifford algebras, we only have 
to replace $\hat{b}^{m \dagger}_{f}$ by $\hat{\tilde{b}}^{m \dagger}_{f}$
and equivalently for the Hermitian conjugated partners.

The Clifford odd ''basis vectors'' {\it almost} fulfil the second quantization postulates 
for fermions. There is, namely, the  property, which the  second quantized fermions 
must fulfil in addition to the relations of Eq.~(\ref{almostDirac}). If the 
anticommutation relations of  ''basis vectors'' and their Hermitian conjugated 
partners would fulfil the relation:
\begin{eqnarray} 
\label{should}
\{\hat{b}^{m}_{f}, \hat{b}^{m'\dagger}_{f'}\}_{*_{A}+}|\psi_{oc}>&=&
\delta^{m m'} \delta_{f f'} |\psi_{oc}>\,,
\end{eqnarray}
for either $\gamma^a$ or $\tilde{\gamma^a}$, then the corresponding 
creation and annihilation operators would fulfil the anticommutation relations 
for the second quantized fermions, explaining the postulates of Dirac for the
second quantized fermion fields.
For  any $\hat{b}^{m}_{f}$ and any $\hat{b}^{m'\dagger}_{f'}$ this is not 
the case. It turns out that besides $\hat{b}^{m=1}_{f=1} = \stackrel{d-1 \, d}{(-)}
\cdots  \stackrel{56}{(-)} \stackrel{12}{(-)}\stackrel{03}{(-i)}$, for example, also 
$\hat{b}^{m'}_{f'} = \, \stackrel{d-1 \, d}{(-)} \cdots  \stackrel{56}{(-)}
 \stackrel{12}{[+]}\stackrel{03}{[+i]} $ 
%
and several others give, when applied on $\hat{b}^{m=1\dagger}_{f=1}$,
nonzero contributions.
 There are namely $2^{\frac{d}{2}-1}-1$ too many annihilation operators for 
each creation operator which give, applied on the creation operator, nonzero 
contribution. 

The problem is solvable by the reduction of the two Clifford odd algebras to only one~\cite{prd2018,n2019PIPII,2020PartIPartII,nh2021RPPNP} as it is presented 
in subsection~\ref{reduction}: If 
$\gamma^a$'s are chosen to determine internal space of fermions, the 
remaining ones, $\tilde{\gamma}^a$'s, determine then quantum numbers of 
each family (described by the eigenvalues of $\tilde{S}^{ab}$ of the Cartan 
subalgebra members).  Correspondingly the creation and annihilation operators, 
expressible as tensor products, $*_{T}$, of the ''basis vectors'' and the basis 
in ordinary (momentum or coordinate) space, fulfil the anticommutation relation 
for the second quantized fermions.

Let me point out that the Hermitian conjugated partners of the ''basis vectors''
belong to different irreducible representations of the corresponding Lorentz group
than the ''basis vectors''. This can be understood, since the Clifford odd ''basis 
vectors'' have always odd numbers of nilpotents, so that an odd number of 
$\stackrel{ab}{(k)}$'s transforms under Hermitian conjugation into 
$\stackrel{ab}{(-k)}$'s, which can not be the member of the  ''basis vectors'',
since the even generators of the Lorentz transformations
transform always even number of nilpotents, keeping the number of nilpotents 
always odd. 
It is different in the case of  the Clifford even ''basis vectors'', since an even 
number of $\stackrel{ab}{(k)}$'s, transformed with the Hermitian conjugation
into  en even number of $\stackrel{ab}{(-k)}$'s  belongs to the same group 
of the ''basis vectors''.\\

{\bf Statement 2.} {\it The Clifford odd $2^{\frac{d}{2}-1}$ members of each 
of the  $2^{\frac{d}{2}-1}$ irreducible representations of ''basis vectors'' have 
their Hermitian conjugated partners in another set of $2^{\frac{d}{2}-1}$ 
$\cdot 2^{\frac{d}{2}-1}$ ''basis vectors''.  Each of the two sets of  the
$ 2^{\frac{d}{2}-1}\times 2^{\frac{d}{2}-1}$  Clifford even ''basis vectors'' 
has their Hermitian conjugated partners within the same set.}

The Clifford even  ''basis vectors''  commute.
Let us denote the Clifford even ''basis vectors'',
described by $\gamma^a$'s, by $\hat{\cal{A}}^{m \dagger}_{f}$. There is no need
to  denote their Hermitian conjugated partners by $\hat{\cal{A}}^{m}_{f}$, since in
the even Clifford sector the ''basis vectors'' and their Hermitian conjugated partners
appear within the same group.  We shall manifest this in the toy model of $d=(5+1)$.
In the Clifford even sector $m$ and $f$ are just two indexes: $f$ denotes the subgroups
within which the ''basis vectors'' do not have the Hermitian conjugated partners (Subsect.~\ref{even5+1}, Eq.~(\ref{evenproperties})).

We shall need also the equivalent
''basis vectors'' in the Clifford even part of the kind $\tilde{\gamma}^a$'s. Let these
''basis vectors'' be denoted  by $\hat{\cal{\tilde{A}}}^{m \dagger}_{f}$. 

%
%
%
%
%

These commuting even Clifford algebra objects have interesting properties.
I shall discuss the properties of even and odd ''basis vectors'' in  Sects.~\ref{even5+1}, 
\ref{generalbasisinternal}, first in $d=(5+1)$-dimensional space,  
 then in the general case.



%
%

%
\subsection{Reduction of the Clifford space}
\label{reduction}
%


The creation and annihilation operators of an odd Clifford  algebra of both kinds, of 
either $\gamma^a$'s or  $\tilde{\gamma}^{a}$'s,   turn out to obey the 
anticommutation relations for the second quantized fermions, postulated by Dirac~%
\cite{nh2021RPPNP}, provided that each of the irreducible representations of the
corresponding Lorentz group, describing the internal space of fermions, would carry a different quantum number.

But we know that a particular member $m$ has for all the irreducible representations 
the same quantum numbers, that is the same "eigenvalues" of the 
Cartan subalgebra (for the vector space of either $\gamma^a$'s or  
$\tilde{\gamma}^{a}$'s), Eq.~(\ref{graficcliff}).  

{\bf Statement 3.}
{\it The only possibility to "dress" each irreducible representation of one kind of the
two independent vector spaces with a new, let us say   "family"  quantum number, 
is that we "sacrifice" one of the two vector spaces.}\\
Let us ''sacrifice''  $\tilde{\gamma}^{a}$'s,  using $\tilde{\gamma}^{a}$'s 
to define the "family" quantum numbers for each irreducible representation of the 
vector space of ''basis vectors of an odd products of $\gamma^a$'s, while keeping the relations of Eq.~(\ref{gammatildeantiher0}) unchanged:  $\{\gamma^{a}, 
\gamma^{b}\}_{+}=2 \eta^{a b}= \{\tilde{\gamma}^{a}, 
\tilde{\gamma}^{b}\}_{+}$, $\{\gamma^{a}, \tilde{\gamma}^{b}\}_{+}=0$,
  $ (\gamma^{a})^{\dagger} = \eta^{aa}\, \gamma^{a}$, 
$(\tilde{\gamma}^{a})^{\dagger} =  \eta^{a a}\, \tilde{\gamma}^{a}$, 
$(a,b)=(0,1,2,3,5,\cdots,d)$.\\

We therefore {\it postulate}:\\
  Let  $\tilde{\gamma}^{a}$'s operate on $\gamma^a$'s as follows~%
\cite{nh03,norma93,JMP2013,normaJMP2015,nh2018}
\begin{eqnarray}
\{\tilde{\gamma}^a B &=&(-)^B\, i \, B \gamma^a\}\, |\psi_{oc}>\,,
\label{tildegammareduced}
\end{eqnarray}
with $(-)^B = -1$, if $B$ is (a function of) an odd products of $\gamma^a$'s,
 otherwise $(-)^B = 1$~\cite{nh03}, $|\psi_{oc}>$ is defined in 
Eq.~(\ref{vaccliff}).\\ 

{\bf Statement 4.} {\it After the postulate of Eq.~(\ref{tildegammareduced}) ''basis 
vectors'' which are  superposition of an odd products of $\gamma^a$'s obey all the 
fermions second quantized postulates of Dirac, presented in 
Eqs.~(\ref{should}, \ref{almostDirac}). }

 We shall see in Sect.~\ref{even5+1} that the Clifford even ''basis vectors'' obey
the bosons second quantized postulates.\\

After this  postulate  the vector  space of $\tilde{\gamma}^{a}$'s is 
"frozen out". No vector space of $\tilde{\gamma}^{a}$'s needs to be taken into account  any longer for the description of the internal space of fermions or bosons, in agreement with the observed properties of fermions.  $\tilde{\gamma}^{a}$'s obtain the role of operators
determining properties of fermion and boson ''basis vectors''. \\


Let me add that we shall still use $\tilde{S}^{ab}$ for the description of  the internal space of fermion and boson fields, Subsects.~\ref{cliffordoddevenbasis5+1},~\ref{odd5+1},~\ref{even5+1}. 
$\tilde{S}^{ab}$'s remain as operators.



One finds, using  Eq.~(\ref{tildegammareduced}),
\begin{eqnarray}
\stackrel{ab}{\tilde{(k)}} \, \stackrel{ab}{(k)}& =& 0\,, 
\qquad 
\stackrel{ab}{\tilde{(-k)}} \, \stackrel{ab}{(k)} = -i \,\eta^{aa}\,  
\stackrel{ab}{[k]}\,,\quad  
\stackrel{ab}{\tilde{(k)}} \, \stackrel{ab}{[k]} = i\, \stackrel{ab}{(k)}\,,\quad
\stackrel{ab}{\tilde{(k)}}\, \stackrel{ab}{[-k]} = 0\,, \nonumber\\
%
%
\stackrel{ab}{\tilde{[k]}} \, \stackrel{ab}{(k)}& =& \, \stackrel{ab}{(k)}\,, 
\quad 
\stackrel{ab}{\tilde{[-k]}} \, \stackrel{ab}{(k)} = \, 0 \,,  \qquad  \quad 
\quad \;\;\;\,
\stackrel{ab}{\tilde{[k]}} \, \stackrel{ab}{[k]} =  0\,,\qquad \;\;\;
\stackrel{ab}{\tilde{[- k]}} \, \stackrel{ab}{[k]} =  \, \stackrel{ab}{[k]}\,.
\label{graphbinomsfamilies}
\end{eqnarray}
Taking into account anticommuting properties of both Clifford algebras,
$\gamma^a$'s and $\tilde{\gamma}^a$'s, it is not difficult to prove  the relations in Eq.~(\ref{graphbinomsfamilies}).

\subsection{Properties of Clifford odd and even ''basis vectors''  in $d=(5+1)$}
\label{cliffordoddevenbasis5+1}

To make discussions  easier let us first look for the properties of  ''basis vectors'' in 
$d=(5+1)$-dimensional space. Let us look at: {\bf i.} internal space of fermions
as the superposition of odd products of the Clifford objects $\gamma^a$'s,
{\bf ii.} internal space of the corresponding gauge fields as the superposition of
even products of the  Clifford objects $\gamma^a$'s. 

Choosing the ''basis vectors'' to be eigenvectors of all the members of the Cartan 
subalgebra of the Lorentz algebra and correspondingly the products of nilpotents 
and projectors (Statement 1.)  one finds  the ''basis vectors'' presented in 
Table~\ref{Table Clifffourplet.}. The table presents the eigenvalues of  the ''basis 
vectors'' for each member of the Cartan subalgebra  for the group $SO(5,1)$.



The  {\it odd I} group (is chosen to) present the ''basis vectors''  describing the 
internal space of fermions. Their Hermitian conjugated partners are then the
''basis vectors'' presented in  the group {\it odd II}.  

The {\it even I} and {\it even II} represent commuting Clifford even ''basis vectors'', 
representing bosons, the gauge fields of fermions.

We shall analyse both kinds of ''basis vectors'' through the subgroups of the $SO(5,1)$ 
group.  The choices  of $SU(2)\times SU(2)\times U(1)$  and $SU(3) \times U(1)$ 
subgroups of the $SO(5,1)$ group will also be discussed just to see the differences 
in properties from the properties of the $SO(5,1)$ group.

In Table~\ref{Table Clifffourplet.} the properties of ''basis vectors'' are  presented 
as products of nilpotents $\stackrel{ab}{(+i)}$  ($\stackrel{ab}{(+i)}^2=0$) and projectors $\stackrel{ab}{[+]}$($\stackrel{ab}{[+]}^2=$ $\stackrel{ab}{[+]}$).
''Basis vectors'' for fermions  contain an odd number of nilpotents,  ''basis vectors'' 
for bosons contain an even number of nilpotents. In both cases  nilpotents 
$\stackrel{ab}{(+i)}$  and projectors $\stackrel{ab}{[+]}$ are chosen to 
be the ''eigenvectors'' of the Cartan subalgebra. Eq.~(\ref{cartangrasscliff}), of the 
Lorentz algebra.
%
%

The ''basis  vectors'', determining the creation operators for fermions and their
Hermitian conjugated partners, $\hat{b}^{m \dagger}_{f}$ and $\hat{b}^{m}_{f}$, 
respectively, as we shall see in Subsect.~\ref{odd5+1} they are 
superposition of odd products of  $\gamma^{a}$, algebraically 
anticommute, due to the properties of the Clifford algebra 
\begin{eqnarray}
\label{gammatildeantiher0}
\{\gamma^{a}, \gamma^{b}\}_{+}&=&2 \eta^{a b}= \{\tilde{\gamma}^{a}, 
\tilde{\gamma}^{b}\}_{+}\,, \nonumber\\
\{\gamma^{a}, \tilde{\gamma}^{b}\}_{+}&=&0\,,\quad
 (a,b)=(0,1,2,3,5,\cdots,d)\,, \nonumber\\
(\gamma^{a})^{\dagger} &=& \eta^{aa}\, \gamma^{a}\, , \quad 
(\tilde{\gamma}^{a})^{\dagger} =  \eta^{a a}\, \tilde{\gamma}^{a}\,,\nonumber\\
\gamma^a \gamma^a &=& \eta^{aa}\,, \quad 
\gamma^a (\gamma^a)^{\dagger} =I\,,\quad
 \tilde{\gamma}^a  \tilde{\gamma}^a = \eta^{aa} \,,\quad
 \tilde{\gamma}^a  (\tilde{\gamma}^a)^{\dagger} =I\,,
\end{eqnarray}
where $I$ represents the unit operator.


\begin{table*}
\begin{tiny}
\caption{\label{Table Clifffourplet.}  $2^d=64$ "eigenvectors" of the Cartan subalgebra
of the Clifford  odd and even algebras in $d=(5+1)$-dimensional space are presented, 
divided into four groups. The first group, $odd \,I$, is chosen to represent "basis vectors", 
named ${\hat b}^{m \dagger}_f$, appearing in $2^{\frac{d}{2}-1}=4$ 
"families"  ($f=1,2,3,4$), each ''family'' with $2^{\frac{d}{2}-1}=4$  
''family'' members ($m=1,2,3,4$). 
The second group, $odd\,II$, contains Hermitian conjugated partners of the first 
group for each  family separately, ${\hat b}^{m}_f=$ 
$({\hat b}^{m \dagger}_f)^{\dagger}$.
The "family" quantum  numbers of ${\hat b}^{m \dagger}_f$, that is the eigenvalues of 
$(\tilde{S}^{03}, \tilde{S}^{12},\tilde{S}^{56})$,  are written   above each "family".
The properties of anticommuting ''basis vectors'' are discussed  in  
Subsects.~\ref{odd5+1},~\ref{generalbasisinternal}. 
%
%
The  two groups with the even number of $\gamma^a$'s, {\it even \,I} and {\it even \,II}, 
have their Hermitian conjugated partners within their own group each. The two groups
which are products of even number of nilpotents and even or odd number of projectors
represent the ''basis vectors'' for the corresponding boson gauge fields. Their 
properties are discussed in Subsecs.~\ref{even5+1}~and~\ref{generalbasisinternal}. 
$\Gamma^{(5+1)}$ and $\Gamma^{(3+1)}$ represent handedness in $d=(3+1)$ and
 $d=(5+1)$ space calculated as products of $\gamma^a$'s, App.~\ref{handednessGrassCliff}.}
%
%
\begin{center}
  \begin{tabular}{|c|c|c|c|c|c|r|r|r|r|r|}
\hline
$''basis $&$m$&$ f=1$&$ f=2 $&$ f=3 $&
$ f=4 $&$$&$$&$$&$$&$$\\
$vectors''$&&$(\frac{i}{2},- \frac{1}{2},-\frac{1}{2})$&$(-\frac{i}{2},-\frac{1}{2},\frac{1}{2})$&
$(-\frac{i}{2},\frac{1}{2},-\frac{1}{2})$&$(\frac{i}{2},\frac{1}{2},\frac{1}{2})$&$S^{03}$
&$S^{12}$&$S^{56}$&$\Gamma^{(5+1)}$&$\Gamma^{(3+1)}$\\
&& 
$\stackrel{03}{\;\,}\;\;\,\stackrel{12}{\;\,}\;\;\,\stackrel{56}{\;\,}$&
$\stackrel{03}{\;\,}\;\;\,\stackrel{12}{\;\,}\;\;\,\stackrel{56}{\;\,}$&
$\stackrel{03}{\;\,}\;\;\,\stackrel{12}{\;\,}\;\;\,stackrel{56}{\;\,}$&
$\stackrel{03}{\;\,}\;\;\,\stackrel{12}{\;\,}\;\;\,\stackrel{56}{\;\,}$&
&&&&\\
\hline
$odd \,I\; {\hat b}^{m \dagger}_f$&$1$& 
$\stackrel{03}{(+i)}\stackrel{12}{[+]}\stackrel{56}{[+]}$&
                        $\stackrel{03}{[+i]}\stackrel{12}{[+]}\stackrel{56}{(+)}$ & 
                        $\stackrel{03}{[+i]}\stackrel{12}{(+)}\stackrel{56}{[+]}$ &  
                        $\stackrel{03}{(+i)}\stackrel{12}{(+)}\stackrel{56}{(+)}$ &
                        $\frac{i}{2}$&$\frac{1}{2}$&$\frac{1}{2}$&$1$&$1$\\
$$&$2$&    $[-i](-)[+] $ & $(-i)(-)(+) $ & $(-i)[-][+] $ & $[-i][-](+) $ &$-\frac{i}{2}$&
$-\frac{1}{2}$&$\frac{1}{2}$&$1$&$1$\\
$$&$3$&    $[-i] [+](-)$ & $(-i)[+][-] $ & $(-i)(+)(-) $ & $[-i](+)[-] $&$-\frac{i}{2}$&
$\frac{1}{2}$&$-\frac{1}{2}$&$1$&$-1$ \\
$$&$4$&    $(+i)(-)(-)$ & $[+i](-)[-] $ & $[+i][-](-) $ & $(+i)[-][-]$&$\frac{i}{2}$&
$-\frac{1}{2}$&$-\frac{1}{2}$&$1$&$-1$ \\
\hline
$ $&$$&$ $&$ $&$ $&&$$&$$&$$&&\\
&&
$\stackrel{03}{\;\,}\;\;\,\stackrel{12}{\;\,}\;\;\,\stackrel{56}{\;\,}$&
$\stackrel{03}{\;\,}\;\;\,\stackrel{12}{\;\,}\;\;\,\stackrel{56}{\;\,}$&
$\stackrel{03}{\;\,}\;\;\,\stackrel{12}{\;\,}\;\;\,\stackrel{56}{\;\,}$&
$\stackrel{03}{\;\,}\;\;\,\stackrel{12}{\;\,}\;\;\,\stackrel{56}{\;\,}$&
&&&$\Gamma^{(5+1)}$&$$\\
\hline
$ $&$$&$ $&$ $&$ $&&$$&$$&$$&&\\
$odd\,II\; {\hat b}^{m}_f$&$1$ &$(-i)[+][+]$ & $[+i][+](-)$ & $[+i](-)[+]$ & $(-i)(-)(-)$&
$$&$$&$$&$-1$&$$ \\
$$&$2$&$[-i](+)[+]$ & $(+i)(+)(-)$ & $(+i)[-][+]$ & $[-i][-](-)$&
$$&$$&$$&$-1$&$$ \\
$$&$3$&$[-i][+](+)$ & $(+i)[+][-]$ & $(+i)(-)(+)$ & $[-i](-)[-]$&
$$&$$&$$&$-1$&$$ \\
$$&$4$&$(-i)(+)(+)$ & $[+i](+)[-]$ & $[+i][-](+)$ & $(-i)[-][-]$&
$$&$$&$$&$-1$&$$ \\
\hline
&&&&&&&&&&\\
\hline
$ $&$$&$ $&$ $&$ $&&$$&$$&$$&&\\
$ even\, I$&$m$&$ $&$ $&$ $&
$ $&$S^{03}$&$S^{12}$&$S^{56}$&$\Gamma^{(5+1)}$&
$\Gamma^{(3+1)}$\\ 
&&$(-\frac{i}{2},\frac{1}{2},\frac{1}{2})$&$(\frac{i}{2},-\frac{1}{2},\frac{1}{2})$&
$(-\frac{i}{2},-\frac{1}{2},-\frac{1}{2})$&$(\frac{i}{2},\frac{1}{2},-\frac{1}{2})$&
&&&&\\
&& 
$\stackrel{03}{\;\,}\;\;\,\stackrel{12}{\;\,}\;\;\,\stackrel{56}{\;\,}$&
$\stackrel{03}{\;\,}\;\;\,\stackrel{12}{\;\,}\;\;\,\stackrel{56}{\;\,}$&
$\stackrel{03}{\;\,}\;\;\,\stackrel{12}{\;\,}\;\;\,\stackrel{56}{\;\,}$&
$\stackrel{03}{\;\,}\;\;\,\stackrel{12}{\;\,}\;\;\,\stackrel{56}{\;\,}$&
&&&&\\
\hline
$$&$1$&$[+i](+)(+) $ & $(+i)[+](+) $ & $[+i][+][+] $ & $(+i)(+)[+] $ &$\frac{i}{2}$&
$\frac{1}{2}$&$\frac{1}{2}$&$1$&$1$ \\
$$&$2$&$(-i)[-](+) $ & $[-i](-)(+) $ & $(-i)(-)[+] $ & $[-i][-][+] $ &$-\frac{i}{2}$&
$-\frac{1}{2}$&$\frac{1}{2}$&$1$&$1$ \\
$$&$3$&$(-i)(+)[-] $ & $[-i][+][-] $ & $(-i)[+](-) $ & $[-i](+)(-) $&$-\frac{i}{2}$&
$\frac{1}{2}$&$-\frac{1}{2}$&$1$&$-1$ \\
$$&$4$&$[+i][-][-] $ & $(+i)(-)[-] $ & $[+i](-)(-) $ & $(+i)[-](-) $&$\frac{i}{2}$&
$-\frac{1}{2}$&$-\frac{1}{2}$&$1$&$-1$ \\ 
\hline
$ $&$$&$ $&$ $&$ $&&$$&$$&$$&&\\
$ even\, II$&$m$&$ $&$$&$ $&
$ $&$S^{03}$&$S^{12}$&$S^{56}$&$\Gamma^{(5+1)}$&
$\Gamma^{(3+1)}$\\
&&$(\frac{i}{2},\frac{1}{2},\frac{1}{2})$&$(-\frac{i}{2},-\frac{1}{2},\frac{1}{2})$&
$(\frac{i}{2},-\frac{1}{2},-\frac{1}{2})$&$(-\frac{i}{2},\frac{1}{2},-\frac{1}{2})$&
&&&&\\
&& 
$\stackrel{03}{\;\,}\;\;\,\stackrel{12}{\;\,}\;\;\,\stackrel{56}{\;\,}$&
$\stackrel{03}{\;\,}\;\;\,\stackrel{12}{\;\,}\;\;\,\stackrel{56}{\;\,}$&
$\stackrel{03}{\;\,}\;\;\,\stackrel{12}{\;\,}\;\;\,\stackrel{56}{\;\,}$&
$\stackrel{03}{\;\,}\;\;\,\stackrel{12}{\;\,}\;\;\,\stackrel{56}{\;\,}$&
&&&&\\
\hline
$$&$1$& $[-i](+)(+) $ & $(-i)[+](+) $ & $[-i][+][+] $ & $(-i)(+)[+] $ &$-\frac{i}{2}$&
$\frac{1}{2}$&$\frac{1}{2}$&$-1$&$-1$ \\ 
$$&$2$&    $(+i)[-](+) $ & $[+i](-)(+) $ & $(+i)(-)[+] $ & $[+i][-][+] $ &$\frac{i}{2}$&
$-\frac{1}{2}$&$\frac{1}{2}$&$-1$&$-1$ \\
$$&$3$&    $(+i)(+)[-] $ & $[+i][+][-] $ & $(+i)[+](-) $ & $[+i](+)(-) $&$\frac{i}{2}$&
$\frac{1}{2}$&$-\frac{1}{2}$&$-1$&$1$ \\
$$&$4$&    $[-i][-][-] $ & $(-i)(-)[-] $ & $[-i](-)(-) $ & $(-i)[-](-) $&$-\frac{i}{2}$&
$-\frac{1}{2}$&$-\frac{1}{2}$&$-1$&$1$ \\
\hline
 \end{tabular}
\end{center}
\end{tiny}
\end{table*}
%

%
\subsubsection{''Basis vectors'' of odd products of $\gamma^a$'s in $d=(5+1)$}
\label{odd5+1}

Let us see in more details properties of  the Clifford odd ''basis vectors'',  analysing
them also with respect to two kinds of the subgroups $SO(3,1) \times U(1)$ and
$SU(3) \times U(1)$ of the  group $SO(5,1)$, with the same number of Cartan 
subalgebra members in all three cases, $\frac{d}{2}=3$. We use the 
expressions for the commuting operators for the subgroup $SO(3,1) \times U(1)$ 
\begin{eqnarray}
\label{so1+3 5+1}
&& N^3_{\pm}(= N^3_{(L,R)}): = \,\frac{1}{2} (%
 S^{12}\pm i S^{03} )\,,\quad
\tilde{N}^3_{\pm}(=\tilde{N}^3_{(L,R)}): = \,\frac{1}{2} (%
\tilde{S}^{12}\pm i \tilde{S}^{03} )\,,
\end{eqnarray}
and for the commuting generators for the subgroup  $SU(3)$ and $U(1)$
%
 \begin{eqnarray}
 \label{so64 5+1}
 \tau^{3}: = &&\frac{1}{2} \,(%
 -S^{1\,2} - iS^{0\,3})\, , \qquad 
\tau^{8}= \frac{1}{2\sqrt{3}} (-i S^{0\,3} + S^{1\,2} -  2 S^{13\;14})\,,\nonumber\\
 \tau^{4}: = &&-\frac{1}{3}(-i S^{0\,3} + S^{1\,2} + S^{5\,6})\,.
%
 \end{eqnarray}
 The corresponding relations for $\tilde{\tau}^{3}, \tilde{\tau}^{8}$ and 
 $\tilde{\tau}^4$ can be read from Eq.~(\ref{so64 5+1}), if replacing $S^{ab}$  
 by $\tilde{S}^{ab}$. 
 Recognizing that ${\bf {\cal S}}^{ab}=S^{ab}+ \tilde{S}^{ab}$ one reproduces 
 all the relations  for the corresponding  $\vec{\bf {\cal \tau}}$ and 
 ${\bf {\cal N}}^3_{\pm}$.\\

 The rest of generators of both kinds of subgroups of the group $SO(5,1)$ can be 
 found in Eqs.~(\ref{so1+3}, \ref{so64}) of App.~\ref{grassmannandcliffordfermions}.\\

In Table~\ref{cliff basis5+1.} the properties of the odd ''basis vectors'' 
$\hat{b}^{m \dagger}_{f}$ are presented with respect to the generators of the 
group  {\bf i.} $SO(5,1)$ (with $15$ generators, $3$ of them forming the 
corresponding Commuting among subalgebra),  
{\bf ii.} $SO(4)\times U(1)$ (with $7$ generators and $3$ of the corresponding
Cartan subalgebra members) and {\bf iii.} $SU(3)\times U(1)$ (with $9$ generators 
and $3$ of the corresponding Cartan subalgebra members), together with the 
eigenvalues of the commuting generators. These ''basis vectors'' are already 
presented as a part of Table~\ref{Table Clifffourplet.}. They fulfil together
with their Hermitian conjugated partners the anticommutation relations of 
Eqs.~(\ref{almostDirac}, \ref{should}).

%
%
\begin{table}
\begin{tiny}
 \begin{center}
\begin{minipage}[t]{16.5 cm}
\caption{The basic creation operators, ''basis vectors'' --- 
 $\hat{b}^{m=(ch,s)\dagger}_{f}$ (each is a product of projectors and an odd 
product of nilpotents, and is the "eigenvector" of all the Cartan subalgebra members, 
$S^{03}$, $S^{12}$, $S^{56}$ and $\tilde{S}^{03}$, $\tilde{S}^{12}$, 
$\tilde{S}^{56}$, Eq.~(\ref{cartangrasscliff})
($ch$ (charge), the eigenvalue of $S^{56}$, and $s$ (spin), the eigenvalues of 
$S^{03}$ and $S^{12}$, explain index $m$, $f$ determines family quantum 
numbers, the eigenvalues of ($\tilde{S}^{03}$, $\tilde{S}^{12}$, 
$\tilde{S}^{56}$) ---  
are presented for $d= (5+1)$-dimensional case. This table represents also the 
eigenvalues of the three commuting operators $N^3_{L,R}$ and $S^{56 }$ of the 
subgroups $SU(2)\times SU(2)\times U(1)$ and the eigenvalues of the three 
commuting operators $\tau^3, \tau^8$ and $\tau^{4}$ of the subgroups 
 $SU(3)\times U(1)$, in these two last cases index $m$ represents the eigenvalues 
 of the corresponding commuting generators. $\Gamma^{(5+1)}=-\gamma^0 
 \gamma^1 \gamma^2 \gamma^3\gamma^5\gamma^6$, $\Gamma^{(3+1)}
 = i\gamma^0 \gamma^1 \gamma^2 \gamma^3$.
Operators $\hat{b}^{m=(ch,s) \dagger}_{f}$ 
 and $\hat{b}^{m=(ch, s)}_{f}$
 fulfil the anticommutation relations of Eqs.~(\ref{almostDirac}, \ref{should}).
\vspace{2mm}}
\label{cliff basis5+1.}
\end{minipage}
 \begin{tabular}{|r|l r|r|r|r|r|r|r|r|r|r|r|r|r|r|r|}
 \hline
$\, f $&$m $&$=(ch,s)$&$\hat{b}^{ m=(ch,s) \dagger}_f$
&$S^{03}$&$ S^{1 2}$&$S^{5 6}$&$\Gamma^{3+1}$ &$N^3_L$&$N^3_R$&
$\tau^3$&$\tau^8$&$\tau^4$&
$\tilde{S}^{03}$&$\tilde{S}^{1 2}$& $\tilde{S}^{5 6}$\\
\hline
$I$&$1$&$(\frac{1}{2},\frac{1}{2})$&$
\stackrel{03}{(+i)}\,\stackrel{12}{[+]}| \stackrel{56}{[+]}$&
$\frac{i}{2}$&
$\frac{1}{2}$&$\frac{1}{2}$&$1$&
$0$&$\frac{1}{2}\frac{1}{2}$&$0$&$0$&$-\frac{1}{2}$&$\frac{i}{2}$&$-\frac{1}{2}$&$-\frac{1}{2}$\\
$$ &$2$&$(\frac{1}{2},-\frac{1}{2})$&$
\stackrel{03}{[-i]}\,\stackrel{12}{(-)}|\stackrel{56}{[+]}$&
$-\frac{i}{2}$&$-\frac{1}{2}$&$\frac{1}{2}$&$1$
&$0$&$-\frac{1}{2}$&$0$&$-\frac{1}{\sqrt{3}}$&$\frac{1}{6}$&$\frac{i}{2}$&$-\frac{1}{2}$&$-\frac{1}{2}$\\
$$ &$3$&$(-\frac{1}{2},\frac{1}{2})$&$
\stackrel{03}{[-i]}\,\stackrel{12}{[+]}|\stackrel{56}{(-)}$&
$-\frac{i}{2}$&$ \frac{1}{2}$&$-\frac{1}{2}$&$-1$
&$\frac{1}{2}$&$0$&$-\frac{1}{2}$&$\frac{1}{2\sqrt{3}}$&$\frac{1}{6}$&$\frac{i}{2}$&$-\frac{1}{2}$&$-\frac{1}{2}$\\
$$ &$4$&$(-\frac{1}{2},-\frac{1}{2})$&$
\stackrel{03}{(+i)}\, \stackrel{12}{(-)}|\stackrel{56}{(-)}$&
$\frac{i}{2}$&$- \frac{1}{2}$&$-\frac{1}{2}$&$-1$
&$-\frac{1}{2}$&$0$&$\frac{1}{2}$&$\frac{1}{2\sqrt{3}}$&$\frac{1}{6}$&$\frac{i}{2}$&$-\frac{1}{2}$&$-\frac{1}{2}$\\ 
\hline 
$II$&$1$&$(\frac{1}{2},\frac{1}{2})$&$
\stackrel{03}{[+i]}\,\stackrel{12}{(+)}| \stackrel{56}{[+]}$&
$\frac{i}{2}$&
$\frac{1}{2}$&$\frac{1}{2}$&$1$&
$0$&$\frac{1}{2}$&$0$&$0$&$-\frac{1}{2}$&$-\frac{i}{2}$&$\frac{1}{2}$&$-\frac{1}{2}$\\
$$ &$2$&$(\frac{1}{2},-\frac{1}{2})$&$
\stackrel{03}{(-i)}\,\stackrel{12}{[-]}|\stackrel{56}{[+]}$&
$-\frac{i}{2}$&$-\frac{1}{2}$&$\frac{1}{2}$&$1$
&$0$&$-\frac{1}{2}$&$0$&$-\frac{1}{\sqrt{3}}$&$\frac{1}{6}$&$-\frac{i}{2}$&$\frac{1}{2}$&$-\frac{1}{2}$\\
$$ &$3$&$(-\frac{1}{2},\frac{1}{2})$&$
\stackrel{03}{(-i)}\,\stackrel{12}{(+)}|\stackrel{56}{(-)}$&
$-\frac{i}{2}$&$ \frac{1}{2}$&$-\frac{1}{2}$&$-1$
&$\frac{1}{2}$&$0$&$-\frac{1}{2}$&$\frac{1}{2\sqrt{3}}$&$\frac{1}{6}$&$-\frac{i}{2}$&$\frac{1}{2}$&$-\frac{1}{2}$\\
$$ &$4$&$(-\frac{1}{2},-\frac{1}{2})$&$
\stackrel{03}{[+i]} \stackrel{12}{[-]}|\stackrel{56}{(-)}$&
$\frac{i}{2}$&$- \frac{1}{2}$&$-\frac{1}{2}$&$-1$
&$-\frac{1}{2}$&$0$&$\frac{1}{2}$&$\frac{1}{2\sqrt{3}}$&$\frac{1}{6}$&$-\frac{i}{2}$&$\frac{1}{2}$&$-\frac{1}{2}$\\
\hline
$III$&$1$&$(\frac{1}{2},\frac{1}{2})$&$
\stackrel{03}{[+i]}\,\stackrel{12}{[+]}| \stackrel{56}{(+)}$&
$\frac{i}{2}$&
$\frac{1}{2}$&$\frac{1}{2}$&$1$&
$0$&$\frac{1}{2}$&$0$&$0$&$-\frac{1}{2}$&$-\frac{i}{2}$&$-\frac{1}{2}$&$\frac{1}{2}$\\
$$ &$2$&$(\frac{1}{2},-\frac{1}{2})$&$
\stackrel{03}{(-i)}\,\stackrel{12}{(-)}|\stackrel{56}{(+)}$&
$-\frac{i}{2}$&$-\frac{1}{2}$&$\frac{1}{2}$&$1$
&$0$&$-\frac{1}{2}$&$0$&$-\frac{1}{\sqrt{3}}$&$\frac{1}{6}$&$-\frac{i}{2}$&$-\frac{1}{2}$&$\frac{1}{2}$\\
$$ &$3$&$(-\frac{1}{2},\frac{1}{2})$&$
\stackrel{03}{(-i)}\,\stackrel{12}{[+]}|\stackrel{56}{[-]}$&
$-\frac{i}{2}$&$ \frac{1}{2}$&$-\frac{1}{2}$&$-1$
&$\frac{1}{2}$&$0$&$-\frac{1}{2}$&$\frac{1}{2\sqrt{3}}$&$\frac{1}{6}$&$-\frac{i}{2}$&$-\frac{1}{2}$&$\frac{1}{2}$\\
$$ &$4$&$(-\frac{1}{2},-\frac{1}{2})$&$
\stackrel{03}{[+i]}\, \stackrel{12}{(-)}|\stackrel{56}{[-]}$&
$\frac{i}{2}$&$- \frac{1}{2}$&$-\frac{1}{2}$&$-1$
&$-\frac{1}{2}$&$0$&$\frac{1}{2}$&$\frac{1}{2\sqrt{3}}$&$\frac{1}{6}$&$-\frac{i}{2}$&$-\frac{1}{2}$&$\frac{1}{2}$\\ 
\hline
$IV$&$1$&$(\frac{1}{2},\frac{1}{2})$&$
\stackrel{03}{(+i)}\,\stackrel{12}{(+)}| \stackrel{56}{(+)}$&
$\frac{i}{2}$&$\frac{1}{2}$&$\frac{1}{2}$&$1$
&$0$&$\frac{1}{2}$&$0$&$0$&$-\frac{1}{2}$&$\frac{i}{2}$&$\frac{1}{2}$&
$\frac{1}{2}$\\
$$ &$2$&$(\frac{1}{2},-\frac{1}{2})$&$
\stackrel{03}{[-i]}\,\stackrel{12}{[-]}|\stackrel{56}{(+)}$&
$-\frac{i}{2}$&$-\frac{1}{2}$&$\frac{1}{2}$&$1$
&$0$&$-\frac{1}{2}$&$0$&$-\frac{1}{\sqrt{3}}$&$\frac{1}{6}$&$\frac{i}{2}$&$\frac{1}{2}$&$\frac{1}{2}$\\
$$ &$3$&$(-\frac{1}{2},\frac{1}{2})$&$
\stackrel{03}{[-i]}\,\stackrel{12}{(+)}|\stackrel{56}{[-]}$&
$-\frac{i}{2}$&$ \frac{1}{2}$&$-\frac{1}{2}$&$-1$
&$\frac{1}{2}$&$0$&$-\frac{1}{2}$&$\frac{1}{2\sqrt{3}}$&$\frac{1}{6}$&$\frac{i}{2}$&$\frac{1}{2}$&$\frac{1}{2}$\\
$$ &$4$&$(-\frac{1}{2},-\frac{1}{2})$&$
\stackrel{03}{(+i)}\,\stackrel{12}{[-]}|\stackrel{56}{[-]}$&
$\frac{i}{2}$&$- \frac{1}{2}$&$-\frac{1}{2}$&$-1$
&$-\frac{1}{2}$&$0$&$\frac{1}{2}$&$\frac{1}{2\sqrt{3}}$&$\frac{1}{6}$&$\frac{i}{2}$&$\frac{1}{2}$&$\frac{1}{2}$\\
\hline
 \end{tabular}
 \end{center}
\end{tiny}
\end{table}

The right handed, 
 $\Gamma^{(5+1)}=1$, fourthplet of the fourth family of 
 Table~\ref{cliff basis5+1.}  can be found in the first  four  lines of 
Table~\ref{Table so13+1.} if only the $d=(5+1)$ part is taken into 
 account. The left handed fourthplet of the  fourth family of 
 Table~\ref{cliff basis5+1left.} can be found in four lines from line 33 to line
 36, again  if only the $d=(5+1)$ part is taken into  account. \\

 {\bf Statement 5.}  {\it In a chosen $d$=dimensional space there is the choice that the ''basis vectors'' are right handed. Their Hermitian conjugated  partners are correspondingly left handed. One could make the  opposite choice, like in Table~\ref{cliff basis5+1left.}. } 
 
Then the  ''basis vectors'' of Table~\ref{cliff basis5+1.} would be the Hermitian 
conjugated partners to the left handed ''basis vectors'' of Table~\ref{cliff basis5+1left.}. 
For the left handed  ''basis vectors'' the vacuum state $|\psi_{oc}>$, Eq.~(\ref{vaccliff}), chosen as  the   $\sum_{f} \hat{b}^{m}_{f} *_{A}  \hat{b}^{m \dagger}_{f} $, 
has to be changed, since the vacuum state must have the property that 
$ \hat{b}^{m}_{f} $$|\psi_{oc}>=0$  and  $\hat{b}^{m \dagger}_{f}$ $|\psi_{oc}>=
\hat{b}^{m \dagger}_{f}$.\\


 One can notice that:\\
 {\bf i.} The family members of ''basis vectors'' have the same properties in 
 all the families, independently whether one observes the group $SO(d-1,1)$
 ($SO(5,1 )$ in the case of $d=(5+1)$) or  of the subgroups with the same 
  number of commuting operators ($SU(2)\times SU(2)\times U(1)$ or 
  $\times SU(3)\times U(1)$ in $d=(5+1)$case).
 The families carry different family quantum numbers. This is true for right,
  ($\Gamma^{(5+1)}=1$), and for left  ($\Gamma^{(5+1)}=-1$), 
representations. \\
{\bf ii.} The sum of all the eigenvalues of all the commuting operators over 
the  $2^{\frac{d}{2}-1}$ family members is equal to zero for each of
 $2^{\frac{d}{2}-1}$ families, separately for left and 
separately for right handed representations, independently whether the
group $SO(d-1,1)$ ($SO(5,1)$) or  the  subgroups ($SU(2)\times SU(2)
\times U(1)$ or $\times SU(3)\times U(1)$) are considered.\\
{\bf iii.} The sum of the family quantum numbers over the four families is 
zero as well.\\
{\bf iv.} The properties of the left handed family members differ strongly 
from the right handed ones.
It is easy to recognize this  in our $d=(5+1)$ case when looking at 
$SU(3)\times U(1)$ quantum numbers since the right handed realization 
manifests the ''colour'' properties of ''quarks'' and ''leptons'' and the left 
handed the ''colour'' properties of  ''antiquarks'' and ''antileptons''.\\
{\bf v.} For a chosen even $d$ there is a choice for either right or left handed
family members. The choice of the handedness of the family members determine
also the vacuum state for the chosen ''basis vectors''. 
%

Let me add that the ''basis vectors'' and their Hermitian conjugated partners
fulfil the anticommutation relations postulated by Dirac for the second 
quantized fermion fields. When forming tensor products, $*_T$, of these
''basis vectors'' and the basis of ordinary, momentum or coordinate, space 
the single fermion creation and annihilation operators fulfil all the requirements
of the Dirac's second quantized fermion fields, explaining therefore the 
postulates of  Dirac, Sect.~\ref{fermionsbosons}.\\

 

 
%
\subsubsection{''Basis vectors'' of even products of $\gamma^a$'s in $d=(5+1)$}
\label{even5+1}

The Clifford even ''basis vectors'', they are products of an even number of nilpotents,
 $\stackrel{ab}{(k)}$, and the rest up to $\frac{d}{2}$ of projectors,  
 $\stackrel{ab}{[k]}$, commute since  even products of (anticommuting) 
 $\gamma^a$'s commute.\\
  
Let us see in more details several properties of  the Clifford even ''basis vectors'':\\

{\bf A.} $\quad$  The properties of the algebraic, $*_{A}$, application of the Clifford 
even ''basis vectors'' on the Clifford odd ''basis vectors'' $\hat{b}^{m \dagger}_{f}$, 
presented in Table~\ref{cliff basis5+1.}, teaches us that the Clifford 
even ''basis vectors''  describe the internal space of the gauge fields of 
 $\hat{b}^{m \dagger}_{f}$.\\
 
 {\bf A.i.}\\
Let $\hat{b}^{m ` \dagger}_{f `}$ represents the $m `^{th}$ Clifford $odd \,I$
''basis vector'' (the part of the creation operators which determines the  internal 
part of the fermion state) of the $f `^{th}$ family and let 
${\hat{\cal A}}^{m \dagger}_{f}$ 
denotes the $m^{th}$ Clifford $even \,II$ ''basis vector''  of the $f ^{th}$ 
irreducible representation with respect to $S^{ab}$ --- but not with respects to 
${\cal S}^{ab}= S^{ab} + \tilde{S}^{ab}$, which includes all $2^{\frac{d}{2}-1}
\times 2^{\frac{d}{2}-1}$ members. Let us evaluate the algebraic products 
${\hat{\cal A}}^{m \dagger}_{f }$ on $\hat{b}^{m `\dagger}_{f  `}$ for any 
$(m, m')$ and $(f, f ')$.

Taking into account Eq.~(\ref{graficcliff}) and Tables~(\ref{Table Clifffourplet.}, 
\ref{cliff basis5+1.}) one can easily evaluate the algebraic products 
${\hat{\cal A}}^{m \dagger}_{f}$ on $\hat{b}^{m' \dagger}_{f '}$ for any $(m, m')$
and $(f, f ')$. Starting with $\hat{b}^{1\dagger}_{1}$ one finds the non zero 
contributions only if applying ${\hat{\cal A}}^{m \dagger}_{3}$, $m=(1,2,3,4)$ on 
$\hat{b}^{1 \dagger}_{1}$
\begin{small}
\begin{eqnarray}
&& {\hat{\cal A}}^{m \dagger}_{3}*_A \hat{b}^{1 \dagger}_{1} 
(\equiv \stackrel{03}{(+i)} \stackrel{12}{[+]} \stackrel{56}{[+]}):\nonumber\\
&&{\hat{\cal A}}^{1 \dagger}_{3} (\equiv \stackrel{03}{[+i]}
\stackrel{12}{[+]} \stackrel{56}{[+]})  *_{A} \hat{b}^{1 \dagger}_{1} 
(\equiv \stackrel{03}{(+i)} \stackrel{12}{[+]} \stackrel{56}{[+]}) \rightarrow
\hat{b}^{1 \dagger}_{1}\,,\nonumber\\
&&{\hat{\cal A}}^{2 \dagger}_{3} (\equiv \stackrel{03}{(-i)}
\stackrel{12}{(-)} \stackrel{56}{[+]}) *_{A} \hat{b}^{1 \dagger}_{1}
\rightarrow \hat{b}^{2 \dagger}_{1} 
(\equiv \stackrel{03}{[-i]} \stackrel{12}{(-)} \stackrel{56}{[+]})\,, \nonumber\\ 
&& {\hat{\cal A}}^{3 \dagger}_{3} (\equiv \stackrel{03}{(-i)}
\stackrel{12}{[+]} \stackrel{56}{(-)}) *_{A} \hat{b}^{1 \dagger}_{1}
\rightarrow \hat{b}^{3 \dagger}_{1} 
(\equiv \stackrel{03}{[-i]} \stackrel{12}{[+]} \stackrel{56}{(-)})\,, \nonumber\\
&&{\hat{\cal A}}^{4 \dagger}_{3} (\equiv \stackrel{03}{[+i]}
\stackrel{12}{(-)} \stackrel{56}{(-)}) *_{A} \hat{b}^{1 \dagger}_{1}
\rightarrow \hat{b}^{4 \dagger}_{1}
(\equiv \stackrel{03}{(+i)} \stackrel{12}{(-)} \stackrel{56}{(-)})\,.
\label{calAb1}
\end{eqnarray}
\end{small}
The products of an even number of nilpotents and even or an odd number of projectors, 
represented by even products of $\gamma^a$'s, applying on family members of a 
particular family, obviously transform family members, representing fermions of one 
particular family, into the same or another family member of the same family.
                   
All the rest of ${\hat {\cal A}}^{m}_f, f\ne 3,$ applying on  $\hat{b}^{1 \dagger}_{1}$,
give zero for any family $f$. 

Let us comment the above events, concerning only the internal space of  fermions and, 
obviously, bosons: If the fermion, the internal space of which is described by Clifford odd
''basis vector'' $\hat{b}^{1 \dagger}_{1}$, absorbs the boson ${\hat{\cal A}}^{1}_{3}$
(with ${\cal{S}}^{03}=0, {\cal{S}}^{12}=0, {\cal{S}}^{56}=0$), its  ''basis 
vector'' $\hat{b}^{1 \dagger}_{1}$ remains unchanged.

 The fermion with the ''basis vector'' $\hat{b}^{1 \dagger}_{1}$, if absorbing the boson 
with ${\hat{\cal A}}^{2}_{3}$ (with ${\cal{S}}^{03}=-i, {\cal{S}}^{12}=-1, 
{\cal{S}}^{56}=0$), changes its internal ''basis vector'' $\hat{b}^{1 \dagger}_{1}$ into
the ''basis vector'' $\hat{b}^{2 \dagger}_{1}$ (which carries now  $S^{03}=-\frac{i}{2},
S^{12}=-\frac{1}{2}$, and the same $S^{56}=\frac{1}{2}$ as before).  The fermion
with ''basis vector'' $\hat{b}^{1 \dagger}_{1}$  absorbing the boson with the 
''basis vector'' ${\hat{\cal A}}^{3}_{3}$ changes its ''basis vector'' to 
$\hat{b}^{3 \dagger}_ {1}$, while  the   fermion with the ''basis vector''
$\hat{b}^{1 \dagger}_{1}$  absorbing the boson with the ''basis vector'' 
${\hat{\cal A}}^{4}_{3}$ changes its 
''basis vector'' to $\hat{b}^{4 \dagger}_ {1}$.

Let us see how do the rest of ${\hat{\cal A}}^{m}_{f}$, $m=(1,2,3,4)$, $f=(1,2,3,4)$
change the properties of $\hat{b}^{n \dagger}_{1}$, $n={2,3,4}$. 

It is easy to  evaluate if taking into account Eq	.~(\ref{graficcliff}) that
\begin{small}
\begin{eqnarray}
&& {\hat{\cal A}}^{m\dagger}_{4}*_A \hat{b}^{2 \dagger}_{1} 
(\equiv \stackrel{03}{[-i]} \stackrel{12}{(-)} \stackrel{56}{[+]}):\nonumber\\
&&{\hat{\cal A}}^{1 \dagger}_{4} (\equiv \stackrel{03}{(+i)}
\stackrel{12}{(+)} \stackrel{56}{[+]})  *_{A} \hat{b}^{2 \dagger}_{1} 
(\equiv \stackrel{03}{[-i]} \stackrel{12}{(-)} \stackrel{56}{[+]}) \rightarrow
\hat{b}^{1 \dagger}_{1}\,,\nonumber\\
&&{\hat{\cal A}}^{2 \dagger}_{4} (\equiv \stackrel{03}{[-i]}
\stackrel{12}{[-]} \stackrel{56}{[+]}) *_{A} \hat{b}^{2 \dagger}_{1}
\rightarrow \hat{b}^{2 \dagger}_{1} 
(\equiv \stackrel{03}{[-i]} \stackrel{12}{(-)} \stackrel{56}{[+]})\,, \nonumber\\ 
&& {\hat{\cal A}}^{3 \dagger}_{4} (\equiv \stackrel{03}{[-i]}
\stackrel{12}{(+)} \stackrel{56}{(-)}) *_{A} \hat{b}^{2 \dagger}_{1}
\rightarrow \hat{b}^{3 \dagger}_{1} 
(\equiv \stackrel{03}{[-i]} \stackrel{12}{[+]} \stackrel{56}{(-)})\,, \nonumber\\
&&{\hat{\cal A}}^{4 \dagger}_{4} (\equiv \stackrel{03}{(+i)}
\stackrel{12}{[-]} \stackrel{56}{(-)}) *_{A} \hat{b}^{2 \dagger}_{1}
\rightarrow \hat{b}^{4 \dagger}_{1}
(\equiv \stackrel{03}{(+i)} \stackrel{12}{(-)} \stackrel{56}{(-)})\,,\nonumber\\
&& {\hat{\cal A}}^{m \dagger}_{2}*_A \hat{b}^{3 \dagger}_{1} 
(\equiv \stackrel{03}{[-i]} \stackrel{12}{[+]} \stackrel{56}{(-)}):\nonumber\\
&&{\hat{\cal A}}^{1 \dagger}_{2} (\equiv \stackrel{03}{(+i)}
\stackrel{12}{[+]} \stackrel{56}{(+)} ) *_{A} \hat{b}^{3 \dagger}_{1} 
(\equiv \stackrel{03}{[-i]} \stackrel{12}{[+]} \stackrel{56}{(-)]}) \rightarrow
\hat{b}^{1 \dagger}_{1}\,,\nonumber\\
&&{\hat{\cal A}}^{2 \dagger}_{2} (\equiv \stackrel{03}{[-i]}
\stackrel{12}{(-)} \stackrel{56}{(+) } ) *_{A} \hat{b}^{3 \dagger}_{1}
\rightarrow \hat{b}^{2 \dagger}_{1} 
(\equiv \stackrel{03}{[-i]} \stackrel{12}{(-)} \stackrel{56}{[+]})\,, \nonumber\\ 
&& {\hat{\cal A}}^{3 \dagger}_{2} (\equiv \stackrel{03}{[-i]}
\stackrel{12}{[+] } \stackrel{56}{[-]}) *_{A} \hat{b}^{3 \dagger}_{1}
\rightarrow \hat{b}^{3 \dagger}_{1} 
(\equiv \stackrel{03}{[-i]} \stackrel{12}{[+]} \stackrel{56}{(-)})\,, \nonumber\\
&&{\hat{\cal A}}^{4 \dagger}_{2} (\equiv \stackrel{03}{(+i)}
\stackrel{12}{(-)} \stackrel{56}{[-]}) *_{A} \hat{b}^{1 \dagger}_{1}
\rightarrow \hat{b}^{4 \dagger}_{1}
(\equiv \stackrel{03}{(+i)} \stackrel{12}{(-)} \stackrel{56}{(-)})\,,'\nonumber\\
&& {\hat{\cal A}}^{m \dagger}_{1}*_A \hat{b}^{4 \dagger}_{1} 
(\equiv \stackrel{03}{(+i)} \stackrel{12}{(-)} \stackrel{56}{(-)}):\nonumber\\
&&{\hat{\cal A}}^{1\dagger}_{1} (\equiv \stackrel{03}{[+i]}
\stackrel{12}{(+)} \stackrel{56}{(+)})  *_{A} \hat{b}^{4 \dagger}_{1} 
(\equiv \stackrel{03}{(+i)} \stackrel{12}{(-)} \stackrel{56}{(-)}) \rightarrow
\hat{b}^{1 \dagger}_{1}\,,\nonumber\\
&&{\hat{\cal A}}^{2\dagger}_{1} (\equiv \stackrel{03}{(-i)}
\stackrel{12}{[-]} \stackrel{56}{(+)}) *_{A} \hat{b}^{4 \dagger}_{1}
\rightarrow \hat{b}^{2 \dagger}_{1} 
(\equiv \stackrel{03}{[-i]} \stackrel{12}{(-)} \stackrel{56}{[+]})\,, \nonumber\\ 
&& {\hat{\cal A}}^{3 \dagger}_{1} (\equiv \stackrel{03}{(-i)}
\stackrel{12}{(+)} \stackrel{56}{[-]}) *_{A} \hat{b}^{4 \dagger}_{1}
\rightarrow \hat{b}^{3 \dagger}_{1} 
(\equiv \stackrel{03}{[-i]} \stackrel{12}{[+]} \stackrel{56}{(-)})\,, \nonumber\\
&&{\hat{\cal A}}^{4 \dagger}_{1} (\equiv \stackrel{03}{[+i]}
\stackrel{12}{[-]} \stackrel{56}{[-]}) *_{A} \hat{b}^{4 \dagger}_{1}
\rightarrow \hat{b}^{4 \dagger}_{1}
(\equiv \stackrel{03}{(+i)} \stackrel{12}{(-)} \stackrel{56}{(-)})\,.
\label{calAb234}
\end{eqnarray}
\end{small}

All the rest of ${\hat {\cal A}}^{m}_f$, applying on  $\hat{b}^{n \dagger}_{1}$,
give zero for any other  $f$ except the one presented in Eqs.~(\ref{calAb1}, 
\ref{calAb234}). 

We can repeat this calculation for all four family members $\hat{b}^{m ` \dagger}_{f `}$
of any of families $f `$. concluding
\begin{eqnarray}
&&{\hat{\cal A}}^{m \dagger}_{3} *_{A} \hat{b}^{1 \dagger}_{f}
\rightarrow \hat{b}^{m \dagger}_{f}\,,\nonumber\\
&&{\hat{\cal A}}^{m \dagger}_{4} *_{A} \hat{b}^{2 \dagger}_{f}
\rightarrow \hat{b}^{m \dagger}_{f}\,,\nonumber\\
&&{\hat{\cal A}}^{m \dagger}_{2} *_{A} \hat{b}^{3 \dagger}_{f}
\rightarrow \hat{b}^{m \dagger}_{f}\,,\nonumber\\
&&{\hat{\cal A}}^{m \dagger}_{1} *_{A} \hat{b}^{4 \dagger}_{f}
\rightarrow \hat{b}^{m \dagger}_{f}\,.
\label{calAb1234}
\end{eqnarray}
The recognition of this subsection concerns so far only internal space of fermions, not yet 
its dynamics in ordinary space. Let us interpret what  is noticed:\\
\vspace{1mm}

{\bf Statement 6.} {\it A fermion with the ''basis vector'' $\hat{b}^{m ` \dagger}_{f `}$, ''absorbing'' one of the commuting Clifford even objects, ${\hat{\cal A}}^{m \dagger}_{f }$, transforms into another family member of the same family, to 
$\hat{b}^{m \dagger}_{f}$, changing correspondingly the family member quantum numbers  and keeping the same family quantum number or remains unchanged.}

The application of  the Clifford even ''basis vector'' ${\hat{\cal A}}^{m \dagger}_{f}$ 
on the Clifford odd ''basis vector'' does not cause the change of the family of the Clifford 
odd ''basis vector''.\\

{\bf A.ii.}\\ 
We need to know the quantum numbers of the  Clifford even ''basis vectors'', which 
obviously manifest properties of the boson fields since they bring to the  Clifford 
odd ''basis vectors'' --- representing the internal space of fermions  --- the  quantum 
numbers which cause transformation into another  fermion with a different  Clifford 
odd ''basis vectors'' of the same family $f$.  
The  Clifford even ''basis vectors'' do not cause the change of the  family of 
fermions.

Let us point out that the Clifford odd ''basis vectors''  appear in $2^{\frac{d}{2}-1}$ 
families with $2^{\frac{d}{2}-1}$ family members in each family, four members in four families in the $d=(5+1)$ case, while the Hermitian conjugated partners belong to 
another group of $2^{\frac{d}{2}-1}\times 2^{\frac{d}{2}-1}$ Clifford  odd ''basis 
vectors'', (to $odd II$ in Table~\ref{Table Clifffourplet.}), while the Clifford even ''basis vectors'' have their Hermitian conjugated partners within the same group of 
$2^{\frac{d}{2}-1}\times 2^{\frac{d}{2}-1}$ members (appearing in our treating 
case in $even II$ in Table~\ref{Table Clifffourplet.}). Since we found in Eqs.~(\ref{calAb1234}, \ref{calAb1}, \ref{calAb1}) that the Clifford even ''basis vector'' transforms  the Clifford odd ''basis vector'' into another member of the same family, 
changing the family members quantum numbers for an integer, they must carry the 
integer quantum numbers.  

One can see in Table~\ref{Table Clifffourplet.} that the members  of the group
$even II$, for example, are Hermitian conjugated to one another in pairs and 
four of them  are self adjoint. Correspondingly ${}^{\dagger}$  has no special 
meaning, it is only the decision that all the Clifford even
''basis vector'' are equipped with ${}^{\dagger}$: ${\cal \hat{A}}^{m \dagger}_{f}$.
 
Let us therefore calculate  the quantum numbers  of 
 ${\cal \hat{A}}^{m \dagger}_{f}$, 
where $m$ and $f$ distinguish among different Clifford even ''basis vectors'' (with 
$f$ which does not really denote the family, since  ${\cal S}^{ab}= S^{ab} + 
\tilde{\cal S}^{ab}$ defines the whole irreducible representation of 
 $2^{\frac{d}{2}-1}\times 2^{\frac{d}{2}-1}$  ''basis vectors'') 
with the Cartan subalgebra operators ${\cal S}^{ab}= S^{ab} + 
\tilde{\cal S}^{ab}$, presented in Eqs.~(\ref{cartangrasscliff}).

In Table \ref{cliff basis5+1even.}  the eigenvalues of the Cartan subalgebra
members of ${\cal S}^{ab}$ are presented, as well as the eigenvalues  of the 
commuting operators of subgroups  $SU(2)\times  SU(2) \times U(1)$, that is the eigenvalues of (${\cal N}^{3}_L, {\cal N}^{3}_R, {\cal S}^{03}$),
and of $SU(3) \times U(1)$, that is the eigenvalues of (${\cal \tau}^{3}, 
{\cal \tau}^{8}, {\cal \tau}^{4}$), expressions for which can be found in
Eqs.~(\ref{so1+3 5+1}, \ref{so64 5+1}) if one takes into account that 
${\cal S}^{ab}= S^{ab} + \tilde{\cal S}^{ab}$.
The algebraic application of any member of a group $f$ on the self adjoint operator 
(denoted in Table~\ref{cliff basis5+1even.} by $\bigcirc$) of this group $f$, gives 
the same member back. 



The vacuum state of the Clifford even ''basis vectors'' is correspondingly  the 
normalized sum of all the  self adjoint operators of these Clifford even group 
$even II$. Each of ${\hat \cal{A}}^{m \dagger}_{f}$ when applying on such a 
vacuum state gives the same ${\cal \hat{A}}^{m \dagger}_{f}$.
\begin{eqnarray}
\label{vaceven}
&&|\phi_{oc_{even}}>= \frac{1}{2}(\stackrel{03}{[+i]} \,\stackrel{12}{[-]}
\,\stackrel{56}{[-]} +\stackrel{03}{[-i]} \,\stackrel{12}{[+]}
\,\stackrel{56}{[-]} + \stackrel{03}{[+i]} \,\stackrel{12}{[+]}
\,\stackrel{56}{[+]} + \stackrel{03}{[-i]} \,\stackrel{12}{[-]}
\,\stackrel{56}{[+]})\,.
\end{eqnarray}

The pairs of ''basis vectors''  ${\hat {\cal{A}}}^{m \dagger}_{f}$, which are 
Hermitian conjugated to each other, are in Table~\ref{cliff basis5+1even.} 
pointed out  by the same symbols.  This property is independent  of the group 
or subgroups which we choose to observe properties of the  ''basis vectors''.  
If treating the subgroup $SU(3)\times U(1)$ one finds the $8$ members  of 
${\hat \cal{A}}^{m \dagger}_{f}$, which belong to
the group $SU(3)$ forming octet which has $\tau^{4}=0$, six of them appear
in three pairs Hermitian conjugated to each other, two of them are self 
adjoint members of the octet, with eigenvalues of all the Cartan subalgebra 
members equal to zero. There are also two singlets with eigenvalues of all the 
Cartan subalgebra members equal to zero. And there is the sextet,  with three 
pairs which are mutually Hermitian conjugated.
\begin{table}
\begin{tiny}
 \begin{center}
\begin{minipage}[t]{16.5 cm}
\caption{The ''basis vectors''  ${\cal \hat{A}}^{m \dagger}_{f}$,  each is the product of 
projectors and an even number of nilpotents,  and is the "eigenvector" of all the 
Cartan subalgebra members, ${\cal S}^{03}$, ${\cal S}^{12}$, ${\cal S}^{56}$, Eq.~(\ref{cartangrasscliff}), are presented for $d= (5+1)$-dimensional case. 
Indexes $m$ and $f$ determine  $2^{\frac{d}{2}-1}\times 2^{\frac{d}{2}-1}$
different members  ${\cal \hat{A}}^{m \dagger}_{f}$. 
In the third column   the  ''basis vectors'' ${\cal \hat{A}}^{m \dagger}_{f}$ which are Hermitian conjugated partners to each other, and can therefore annihilate each other, 
are pointed out with the same symbol. For example
with $\star$ are equipped the first member with $m=1$ and $f=1$ and the last 
member with $m=4$ and $f=3$.
The sign $\bigcirc$ denotes the ''basis vectors'' which are self adjoint 
$( {\cal \hat{A}}^{m \dagger}_{f})^{\dagger}$ $={\cal \hat{A}}^{m \dagger}_{f}$.
This table represents also the 
eigenvalues of the three commuting operators ${\cal N}^3_{L,R}$ and 
${\cal S}^{56 }$ of the subgroups $SU(2)\times SU(2)\times U(1)$ of the group
$SO(5,1)$ and the eigenvalues of the three 
commuting operators ${\cal \tau}^3, {\cal \tau}^8$ and ${\cal \tau}^{4}$ of the 
subgroups  $SU(3)\times U(1)$.
\vspace{2mm}
}
%
%
\label{cliff basis5+1even.}
\end{minipage}
 \begin{tabular}{|r|l r|r|r|r|r|r|r|r|r|r|r|}
 \hline
$\, f $&$m $&$*$&${\cal \hat{A}}^{m \dagger}_f$
&${\cal S}^{03}$&$ {\cal S}^{1 2}$&${\cal S} ^{5 6}$&
${\cal N}^3_L$&${\cal N}^3_R$&
${\cal \tau}^3$&${\cal \tau}^8$&${\cal \tau}^4$
\\
\hline
$I$&$1$&$\star \star$&$
\stackrel{03}{[+i]}\,\stackrel{12}{(+)} \stackrel{56}{(+)}$&
$0$&$1$&$1$
&$\frac{1}{2}$&$\frac{1}{2}$&$-\frac{1}{2}$&$-\frac{1}{2\sqrt{3}}$&$-\frac{2}{3}$
\\
$$ &$2$&$\bigtriangleup$&$
\stackrel{03}{(-i)}\,\stackrel{12}{[-]}\,\stackrel{56}{(+)}$&
$- i$&$0$&$1$
&$\frac{1}{2}$&$-\frac{1}{2}$&$-\frac{1}{2}$&$-\frac{3}{2\sqrt{3}}$&$0$
\\
$$ &$3$&$\ddagger$&$
\stackrel{03}{(-i)}\,\stackrel{12}{(+)}\,\stackrel{56}{[-]}$&
$-i$&$ 1$&$0$
&$1 $&$0$&$-1$&$0$&$0$
\\
$$ &$4$&$\bigcirc$&$
\stackrel{03}{[+i]}\,\stackrel{12}{[-]}\,\stackrel{56}{[-]}$&
$0$&$0$&$0$
&$0$&$0$&$0$&$0$&$0$
\\
\hline 
$II$&$1$&$\bullet$&$
\stackrel{03}{(+i)}\,\stackrel{12}{[+]}\, \stackrel{56}{(+)}$&
$i$&$0$&$1$&
$-\frac{1}{2}$&$\frac{1}{2}$&$\frac{1}{2}$&$-\frac{1}{2\sqrt{3}}$&$-\frac{2}{3}$\\
$$ &$2$&$\otimes$&$
\stackrel{03}{[-i]}\,\stackrel{12}{(-)}\,\stackrel{56}{(+)}$&
$0$&$-1$&$1$
&$-\frac{1}{2}$&$-\frac{1}{2}$&$\frac{1}{2}$&$-\frac{3}{2\sqrt{3}}$&$0$
\\
$$ &$3$&$\bigcirc$&$
\stackrel{03}{[-i]}\,\stackrel{12}{[+]}\,\stackrel{56}{[-]}$&
$0$&$ 0$&$0$
&$0$&$0$&$0$&$0$&$0$
\\
$$ &$4$&$\ddagger$&$
\stackrel{03}{(+i)}\, \stackrel{12}{(-)}\,\stackrel{56}{[-]}$&
$i$&$-1$&$0$
&$-1$&$0$&$1$&$0$&$0$
\\ 
%
%
 \hline
$III$&$1$&$\bigcirc$&$
\stackrel{03}{[+i]}\,\stackrel{12}{[+]}\, \stackrel{56}{[+]}$&
$0$&$0$&$0$&
$0$&$0$&$0$&$0$&$0$\\
$$ &$2$&$\odot \odot$&$
\stackrel{03}{(-i)}\,\stackrel{12}{(-)}\,\stackrel{56}{[+]}$&
$-i$&$-1$&$0$
&$0$&$-1$&$0$&$-\frac{1}{\sqrt{3}}$&$\frac{2}{3}$\\
$$ &$3$&$\bullet$&$
\stackrel{03}{(-i)}\,\stackrel{12}{[+]}\,\stackrel{56}{(-)}$&
$-i$&$ 0$&$-1$
&$\frac{1}{2}$&$-\frac{1}{2}$&$-\frac{1}{2}$&$\frac{1}{2\sqrt{3}}$&$\frac{2}{3}$
\\
$$ &$4$&$\star \star$&$
\stackrel{03}{[+i]} \stackrel{12}{(-)}\,\stackrel{56}{(-)}$&
$0$&$- 1$&$- 1$
&$-\frac{1}{2}$&$-\frac{1}{2}$&$\frac{1}{2}$&$\frac{1}{2\sqrt{3}}$&$\frac{2}{3}$
\\
\hline
$IV$&$1$&$\odot \odot $&$
\stackrel{03}{(+i)}\,\stackrel{12}{(+)}\, \stackrel{56}{[+]}$&
$i$&$1$&$0$&
$0$&$1$&$0$&$\frac{1}{\sqrt{3}}$&$-\frac{2}{3}$
\\
$$ &$2$&$\bigcirc$&$
\stackrel{03}{[-i]}\,\stackrel{12}{[-]}\,\stackrel{56}{[+]}$&
$0$&$0$&$0$
&$0$&$0$&$0$&$0$&$0$
\\
$$ &$3$&$\otimes$&$
\stackrel{03}{[-i]}\,\stackrel{12}{(+)}\,\stackrel{56}{(-)}$&
$0$&$ 1$&$-1$
&$\frac{1}{2}$&$\frac{1}{2}$&$-\frac{1}{2}$&$\frac{3}{2\sqrt{3}}$&$0$
\\
$$ &$4$&$\bigtriangleup$&$
\stackrel{03}{(+i)}\, \stackrel{12}{[-]}\,\stackrel{56}{(-)}$&
$i$&$0$&$-1$
&$-\frac{1}{2}$&$\frac{1}{2}$&$\frac{1}{2}$&$\frac{3}{2\sqrt{3}}$&$0$\\ 
\hline 
 \end{tabular}
 \end{center}
\end{tiny}
\end{table}
One can notice that the sum of all the eigenvalues of all the Cartan subalgebra
members over the $16$ members  ${\cal \hat{A}}^{m \dagger}_{f}$ is equal 
to zero, independent of whether we treat the group $SO(5,1)$, $SU(2)\times SU(2)
\times U(1)$, or  $SU(3)\times U(1)$.\\

{\bf A.iii.}\\
In {\bf A.i.} we saw that the application of  ${\cal \hat{A}}^{m \dagger}_{f}$ 
on the fermion ''basis vectors'' $\hat{b}^{m \dagger}_{f}$  transforms the particular 
member $\hat{b}^{m \dagger}_{f}$  to one of the members of the same family $f$, 
changing eigenvalues of the Cartan subalgebra members for an integer. We 
found in {\bf A.ii} the eigenvalues of  the Cartan subalgebra members for 
each of $2^{\frac{d}{2}-1}\times ^{\frac{d}{2}-1}$ (equal to $16$ in $d=(5+1)$) 
${\cal \hat{A}}^{m \dagger}_{f}$, recognizing that they do have properties of the 
boson fields. 

It remains to look for the behaviour of these Clifford even ''basis vector'' when they 
apply on each other. Let us denote the self adjoint member in each group of
''basis vectors'' of particular $f$  as ${\cal \hat{A}}^{m_{0} \dagger}_{f}$.
 We easily see that
\begin{eqnarray}
\label{evenproperties}
\{{\cal \hat{A}}^{m \dagger}_{f}\,, {\cal \hat{A}}^{m' \dagger}_{f}\}_{-}&=&0\,,
\quad {\rm if }\,  (m,m') \, \ne m_0  \,{\rm or}\, m=m_0=m'\,, \forall \, f\,,\nonumber\\
{\cal \hat{A}}^{m \dagger}_{f}*_{A} {\cal \hat{A}}^{m_0 \dagger}_{f}&=&
{\cal \hat{A}}^{m \dagger}_{f}\,, \quad \forall \, m \,, \,\forall \,f\,.
\end{eqnarray}
Two ''basis vectors'' ${\cal \hat{A}}^{m \dagger}_{f}$  and 
${\cal \hat{A}}^{m ' \dagger}_{f}$ of the same $f$ and of $(m, m')\ne m_0$ are orthogonal.

The two ''basis vectors'' ${\cal \hat{A}}^{m \dagger}_{f}$  and 
 ${\cal \hat{A}}^{m' \dagger}_{f '}$, the algebraic product, $*_{A}$, of which 
 gives nonzero contribution, like  ${\cal \hat{A}}^{1 \dagger}_{1}$ 
  $*_A \,{\cal \hat{A}}^{4 \dagger}_{2}=$   ${\cal \hat{A}}^{1 \dagger}_{2}$,   
 ''scatter'' into the third one, or annihilate into vacuum $|\phi_{oc_{even}}>$, Eq.~(\ref{vaceven}), like ${\cal \hat{A}}^{2 \dagger}_{2}$ 
  $*_A \,{\cal \hat{A}}^{3 \dagger}_{4}=$   ${\cal \hat{A}}^{2 \dagger}_{4}$.
\footnote{I use ''scatter''  in quotation marks since the  ''basis vectors'' 
${\cal \hat{A}}^{m \dagger}_{f}$ determine only the internal space of bosons,
as also the ''basis vectors'' $\hat{b}^{m \dagger}_{f}$ determine only the internal space
of fermions.}
To generate creation and annihilation operators the tensor products, $*_T,$ of the
''basis vectors'' ${\cal \hat{A}}^{m \dagger}_{f}$, as well as of the
''basis vectors'' $\hat{b}^{m \dagger}_f$,  with the basis in ordinary, 
momentum or coordinate, space is needed.\\

{\bf Statement 7.} {\it  Two ''basis vectors'' ${\cal \hat{A}}^{m \dagger}_{f}$  
and ${\cal \hat{A}}^{m'_{0} \dagger}_{f}$ of the same $f$ and of 
$(m, m')\ne m_0$ are orthogonal.  The two ''basis vectors'' with nonzero
algebraic product, $*_{A}$,  ''scatter'' into the third one, or annihilate into 
vacuum.}\\

{\bf B.} $\quad$
Let us point out that the choice of the Clifford odd ''basis vectors'', $odd\, I$, describing  
the internal space of fermions, and consequently the choice of the Clifford even ''basis vectors'',  $even \, II$, describing  the internal space of their gauge fields, is ours. If  we 
choose in Table~\ref{Table Clifffourplet.} $odd \,II $ to represent  the ''basis vectors''
describing the internal space of fermions, then the corresponding ''basis vectors''
representing the internal space of bosonic partners are those of $even\, I$. \\

For a different choice of handedness of the Clifford odd ''basis vectors'' for describing 
fermions --- making a choice of  the left handedness instead 
of  the right handedeness --- Table~\ref{cliff basis5+1.} should be replaced by 
Table~\ref{cliff basis5+1left.} and correspondingly also 
{\bf A.i.}, {\bf A.ii.}, {\bf A.iii.} should be rewritten.

{\it For an even $d$ there is a choice for either right or left handed
family members.} The choice of the handedness of the family members determine
also the vacuum state for the chosen ''basis vectors'' for either --- Clifford odd
''basis vectors'' of fermions or for the corresponding  Clifford even ''basis vectors'' of
the corresponding gauge boson fields. \\

{\bf C.} $\quad$
The Clifford even ''basis vectors'' ${\cal \hat{A}}^{m \dagger}_{f}$, representing 
the boson gauge fields to the corresponding Clifford odd ''basis vectors'' 
$\hat{b}^{m \dagger}_{f}$, have the properties that they transform Clifford
odd ''basis vectors'' $ \hat{b}^{m \dagger}_{f}$ of each family within the family 
members. There are the additional Clifford even ''basis vectors'' 
${\cal \hat{\tilde{A}}}^{m \dagger}_{f}$ which transform each family member of particular family into the same family member of some of the rest families.

These Clifford even ''basis vectors'' ${\cal \hat{\tilde{A}}}^{m \dagger}_{f}$ 
are products of an even number of nilpotents and of projectors, which are 
eigenvectors of the Cartan subalgebra operators $\tilde{S}^{03}, \tilde{S}^{12},
\tilde{S}^{56},\dots,\tilde{S}^{d-1\,d}$. The table like  
Table~\ref{cliff basis5+1even.} should be prepared and their properties described 
as in the case of {\bf A.i.}, {\bf A.ii.}, {\bf A.iii.}. A short illustration is to help 
understanding the role of these  Clifford even ''basis vectors'' 
${\cal \hat{\tilde{A}}}^{m \dagger}_{f}$.

Let us use for the Clifford even ''basis vectors'' ${\cal \hat{\tilde{A}}}^{m \dagger}_{f}$
the same arrangement with products of nilpotents and projectors as the one, chosen
for the Clifford even ''basis vectors'' ${\cal \hat{A}}^{m \dagger}_{f}$ in the case of 
$d=(5+1)$ in Table~\ref{cliff basis5+1even.}, except that now nilpotents and 
projectors are eigenvectors of the Cartan subalgebra  operators 
$\tilde{S}^{03}, \tilde{S}^{12}, \tilde{S}^{56}$, and are correspondingly written 
in terms of nilpotents $\stackrel{ab}{\tilde{(k)}}$ and projectors $\stackrel{ab}
{\tilde{[k]}}$.
The application of these nilpotents and projectors on nilpotents and projectors 
appearing in $\hat{b }^{m \dagger}_f$ are presented in
Eq.~(\ref{graphbinomsfamilies}). 
Making a choice of 
${\cal \hat{\tilde{A}}}^{1 \dagger}_{1} (\equiv \stackrel{03}{\tilde{[+i]}}
\,\stackrel{12}{\tilde{(+)}}\,\stackrel{56}{\tilde{(+)}}) $, with quantum numbers
$({\cal S}^{03}=0, {\cal S}^{12}= 1, {\cal S}^{56}=1)$,  on $\hat{b}^{4 \dagger}_{1}
(\equiv  \stackrel{03}{(+i)}\,\stackrel{12}{(-)}\,\stackrel{56}{(-)}) $ with the 
family members quantum numbers $ (S^{03}=\frac{i}{2}, S^{12}= - \frac{1}{2},
S^{56}=  - \frac{1}{2})$ and the family quantum numbers $ (\tilde{S}^{03}=
\frac{i}{2}, \tilde{S}^{12}= - \frac{1}{2}, \tilde{S}^{56}=  - \frac{1}{2})$ it follows
%
\begin{eqnarray}
\label{bosontilde}
{\cal \hat{\tilde{A}}}^{1 \dagger}_{1} (\equiv \stackrel{03}{\tilde{[+i]}}
\,\stackrel{12}{\tilde{(+)}}\,\stackrel{56}{\tilde{(+)}}) \,*_{A}\,
\hat{b}^{4 \dagger}_{1} (\equiv  \stackrel{03}{(+i)}\,
\stackrel{12}{(-)}\,\stackrel{56}{(-)}) \rightarrow
\hat{b}^{4 \dagger}_{4} (\equiv  \stackrel{03}{(+i)}\,
\stackrel{12}{[-]}\,\stackrel{56}{[-]})\,.
\end{eqnarray}
%
$\hat{b}^{4 \dagger}_{4} (\equiv  \stackrel{03}{(+i)}\,
\stackrel{12}{[-]}\,\stackrel{56}{[-]})$ carry the same family members 
quantum numbers as $ \hat{b}^{4 \dagger}_{1}$ $(S^{03}=\frac{i}{2}, S^{12}=
 -\frac{1}{2}, S^{56}= - \frac{1}{2})$) but belongs to the different family with 
the family quantum numbers $ (\tilde{S}^{03}=
\frac{i}{2}, \tilde{S}^{12}= \frac{1}{2}, \tilde{S}^{56}= \frac{1}{2})$. 



The detailed analyse of these last two cases {\bf B.} and {\bf C.} will be studied after
this Bled proceedings.\\

We can conclude that the  Clifford even ''basis vectors'' 
${\hat{\cal A}}^{m \dagger}_{f} $: 

{\bf a.} Have the quantum numbers determined 
by  the Cartan subalgebra members of the Lorentz group of ${\cal S}^{ab}=S^{ab} +
\tilde{S}^{ab}$.  Applying algebraically, $*_{A}$, ${\hat{\cal A}}^{m \dagger}_{f} $
on the Clifford odd ''basis vectors'' $\hat{b}^{m \dagger}_{f}$, 
${\hat{\cal A}}^{m \dagger}_{3} $ transform these
''basis vectors'' to another ones with the same family quantum numbers, 
$\hat{b}^{m ` \dagger}_{f }$. 

{\bf b.} In any irreducibly representation of ${\cal S}^{ab}$
${\hat{\cal A}}^{m \dagger}_{f} $ appear in pairs, which are Hermitian conjugated 
to each other or they are self adjoint. 

{\bf c.} The self adjoint members ${\hat{\cal A}}^{m \dagger}_{f} $ define the 
vacuum state of the second quantized boson fields. 

{\bf d.} Applying ${\hat{\cal A}}^{m \dagger}_{f} $ algebraically to each other  
these commuting Clifford even ''basis vector'' forming another Clifford even 
''basis vector'' or annihilate into the vacuum.

{\bf e.} The choice of the left or the right handedness of the  ''basis vectors'' of an odd Clifford character, describing the internal space of fermions, is ours. The left and the right handed ''basis vectors'' of an odd Clifford character are namely Hermitian conjugated to 
each other. With the choice of the handedness of the fermion ''basis 
vectors'' also the choice of boson Clifford even ''basis vectors'' --- which are their corresponding gauge fields  --- are chosen.

{\bf f.} There exist the Clifford even ''basis vectors'' 
${\cal \hat{\tilde{A}}}^{m \dagger}_{f} $ (like
${\cal \hat{\tilde{A}}}^{1 \dagger}_{1} (\equiv \stackrel{03}{\tilde{[+i]}}
\,\stackrel{12}{\tilde{(+)}}\,\stackrel{56}{\tilde{(1)}})$) which transform the 
Clifford odd ''basis vectors'' $\hat{b}^{m \dagger}_{f}$, representing the internal 
space of fermions, into the Clifford odd ''basis vectors'' $\hat{b}^{m \dagger}_{f '}$ 
with the same family member $m$ belonging to another family $f `$.

%
\subsection{''Basis vectors'' describing internal space of fermions and bosons in any
even dimensional space}
\label{generalbasisinternal}


In  Subsect.~\ref{cliffordoddevenbasis5+1} the properties of  the ''basis vectors'', 
describing internal space of fermions and bosons in a toy model  with $d=(5+1)$
are presented in order to simplify (to make more illustrative) the discussions 
on the properties of the Clifford odd ''basis vectors'' describing the internal space 
of fermions and the Clifford even ''basis vectors'' describing the internal space of corresponding bosons, the gauge fields of fermions.

The generalization to any even $d$ is straightforward. For the description of the 
internal space of fermions I follow here Ref.~\cite{nh2021RPPNP}. \\
 
 {\bf a.} The  ''basis vectors'' offering the description of the internal space of 
fermions, $\hat{b}^{m \dagger}_f$, must contain an odd  product of nilpotents  
$\stackrel{ab}{(k)}$, $2n'+1$, in $d=2(2n +1)$,  
$n'=(0,1,2,\dots, \frac{1}{2}(\frac{d}{2}-1)$, and the rest is the product of $n''$ 
projectors $\stackrel{ab}{[k]}$, $n''=\frac{d}{2}-(2n'+1)$. 
Nilpotents and projectors are chosen to be ''eigenvectors'' of the $\frac{d}{2}$
 members of the Cartan subalgebra. 
 
After the reduction of the two kinds of the Clifford algebras to only one, 
$\gamma^a$'s, the generators   $S^{ab}$ of the Lorentz transformations in 
the internal space of fermions described by $\gamma^a$'s, determine the 
$2^{\frac{d}{2}-1}$ family members for each  of  $2^{\frac{d}{2}-1}$ families,  
while $\tilde{S}^{ab}$'s determine the $\frac{d}{2}$ numbers (the eigenvalues 
of the Cartan subalgebra members of the $2^{\frac{d}{2}-1}$ families). 
 
The Cliford odd ''basis vectors'' $\hat{b}^{m \dagger}_{f}$ obey the postulates of Dirac 
for the second quantized fermion fields
\begin{eqnarray}
\{ \hat{b}^{m}_{f}, \hat{b}^{m' \dagger}_{f'} \}_{*_{A}+}\, |\psi_{oc}> 
&=& \delta^{m m'} \, \delta_{ff'} \,  |\psi_{oc}>\,,\nonumber\\
\{ \hat{b}^{m}_{f}, \hat{b}^{m'}_{f'} \}_{*_{A}+}  \,  |\psi_{oc}>
&=& 0 \,\cdot\,  |\psi_{oc}>\,,\nonumber\\
\{\hat{b}^{m  \dagger}_{f},\hat{b}^{m' \dagger}_{f'}\}_{*_{A}+} \, |\psi_{oc}>
&=& 0 \, \cdot\, |\psi_{oc}>\,,\nonumber\\
 \hat{b}^{m \dagger}_{f} \,{}_{*_{A}} |\psi_{oc}>&=& |\psi^{m}_{f}>\,, \nonumber\\
 \hat{b}^{m}_{f}   \,{*_{A}}  |\psi_{oc}>&=& 0 \,\cdot\,  |\psi_{oc}>\,,
\label{alphagammatildeprod}
\end{eqnarray}
with ($m,m'$) denoting the "family" members and ($f,f '$) denoting "families",
${*_{A}}$ represents the algebraic multiplication of $ \hat{b}^{m \dagger}_{f} $
with  their Hermitian conjugated objects $ \hat{b}^{m}_{f } $, with the vacuum 
state $|\psi_{oc}>$,  Eq.~(\ref{vaccliff}), and $ \hat{b}^{m \dagger}_{f} $ or
$ \hat{b}^{m `}_{f `} $ among themselves. It is not difficult 
to prove the above relations if taking into account 
Eq.~(\ref{gammatildeantiher0}). 

The Clifford odd ''basis vectors'' $ \hat{b}^{m \dagger}_{f} $'s and their Hermitian conjugated partners  $ \hat{b}^{m}_{f} $'s appear in two independent groups, 
each with $2^{\frac{d}{2}-1}\times$  $2^{\frac{d}{2}-1}$ members, Hermitian 
conjugated to each other.

It is our choice which one of these two groups with $2^{\frac{d}{2}-1}\times$  
$2^{\frac{d}{2}-1}$ members to take  as  ''basis vectors'' 
$ \hat{b}^{\dagger m}_{f} $'s. Making the opposite choice the  ''basis vectors'' 
change handedness.\\

{\bf b.} The  ''basis vectors'' for bosons, $\hat{\cal A}^{m \dagger}_f$, must 
contain an  even  number of nilpotents  $\stackrel{ab}{(k)}$, $2n'$. In 
$d=2(2n+1)$, $n'=(0,1,2,\dots, \frac{1}{2} (\frac{d}{2}-1$)), the rest,  $n''$, 
are projectors   $\stackrel{ab}{[k]}$,  $n''=(\frac{d}{2}-(2n'))$. 
 
The  ''basis vectors'' are either self adjoint or have the Hermitian conjugated 
partners within the same group of $2^{\frac{d}{2}-1}\times$  
$2^{\frac{d}{2}-1}$ members.
 
They do not form families, $m$ and $f$ only note a particular ''basis vector''. One 
of the members of particular $f $ is self adjoint and participates to the vacuum 
state which has $2^{\frac{d}{2}-1}$ summands, Eq.~(\ref{vaceven}).
 
The  Clifford even ''basis vectors'' ${\cal \hat{A}}^{m \dagger}_{f}$ commute, 
$\{{\cal \hat{A}}^{m \dagger}_{f}\,,  {\cal \hat{A}}^{m' \dagger}_{f}\}_{-}=0$, if both have the same index $f$  and none of them or both of them are self adjoint operators.
\begin{eqnarray}
\label{evenproperties1}
\{{\cal \hat{A}}^{m \dagger}_{f}\,, {\cal \hat{A}}^{m' \dagger}_{f}\}_{-}&=&0\,,
\quad {\rm if }\,  (m,m') \, \ne m_0  \,{\rm or}\, m=m_0=m'\,, \forall \, f\,,\nonumber\\
{\cal \hat{A}}^{m \dagger}_{f}*_{A} {\cal \hat{A}}^{m_0 \dagger}_{f}&=&
{\cal \hat{A}}^{m \dagger}_{f}\,, \quad \forall \, m \,, \,\forall \,f\,.
\end{eqnarray}
The two ''basis vectors'', ${\cal \hat{A}}^{m \dagger}_{f}$  and 
 ${\cal \hat{A}}^{m' \dagger}_{f '}$, the algebraic product, $*_{A}$, of which 
 gives nonzero contribution, ''scatter'' into the third one, or annihilate into the
 vacuum  $|\phi_{oc_{even}}>$.
 
 Quantum numbers of ${\cal \hat{A}}^{m \dagger}_{f}$ are determined by 
 the Cartan subalgebra members of the Lorentz group ${\cal S}^{ab}=S^{ab} +
 \tilde{S}^{ab}$.

 If a fermion with the ''basis vector'' $\,\hat{b}^{m \dagger}_{f}$ ''absorbs'' 
one of the commuting Clifford even objects, ${\hat{\cal A}}^{m ` \dagger}_{f '}$, it 
transforms into another family member of the same family, to 
$\hat{b}^{m ' \dagger}_{f}$,
changing correspondingly the family member quantum numbers, keeping the
family quantum number the same,  or remains unchanged.
 
 The remaining group of $2^{\frac{d}{2}-1}\times 2^{\frac{d}{2}-1}$ Clifford
 even ''basis vectors'', presented in Table~\ref{Table Clifffourplet.} do not  influence the chosen  Clifford odd ''basic vectors'', but rather their Hermitian conjugated partners 
 $\hat{b}^{m}_{f}$. 
 
 There are the even ''basis vectors'' ${\cal \hat{\tilde{A}}}^{m \dagger}_{f}$, 
 the nilpotents and projectors of which are  $\stackrel{ab}{\tilde{(k)}}$, 
 $\stackrel{ab}{\tilde{[k]}}$, respectively. These ''basis vectors'' 
 ${\cal \hat{\tilde{A}}}^{m \dagger}_{f}$, if applying on the Clifford odd ''basis 
 vectors'' $\hat{b}^{m \dagger}_{f}$, transform  these ''basis  vectors'' 
 into ''basis  vectors'' $\hat{b}^{m \dagger}_{f '}$ belonging to different family $f $, 
 while the family member  quantum number $m$ remains unchanged.

 Exchanging  the role of the Clifford odd ''basis vector'' $\hat{b}^{m \dagger}_{f}$
and their Hermitian conjugated partners $\hat{b}^{m}_{f}$ (what means in the case 
of $d=(5+1)$ the exchange of $odd \,I$, which is right handed,  with $odd \,II$, which is lefthanded, in Table~\ref{Table Clifffourplet.}), not only causes the change of the handedness  of the new $\hat{b}^{n \dagger}_{f}$, but also the change of 
the role of  the Clifford even ''basis vectors'' (what means in the case 
of $d=(5+1)$ the exchange of $even \,II$ with $even \,I$).

 

%
\section{Second quantized fermion  and boson fields with internal space described 
by Clifford algebra}
\label{fermionsbosons}

After the {\bf reduction} of the Clifford space to only the part determined by 
$\gamma^a$'s, the ''basis vectors'', which are superposition of odd products
of  $\gamma^a$'s, determine the internal space of fermions. The ''basis vectors''
are orthogonal and appear in even dimensional spaces in $2^{\frac{d}{2}-1}$ 
families, each with $2^{\frac{d}{2}-1}$ family members. Quantum numbers 
of family members are determined by $S^{ab}$, quantum numbers of families 
are determined by $\tilde{\gamma}^a$'s, or better by $\tilde{S}^{ab}$'s. 
$\tilde{\gamma}^a$'s anticommute among themselves and with $\gamma^a$'s, 
as they did before the reduction of the Clifford space, 
Eq.~(\ref{alphagammatildeprod}).


''Basis vectors'' $\hat{b}^{m \dagger}_{f}$, determining internal space of 
fermions, are in even dimensional spaces products of an odd number of 
nilpotents and an even number of 
projectors, chosen to be eigenvectors of the  $\frac{d}{2}$ Cartan  subalgebra 
members of the Lorentz algebra $S^{ab}$, Table~\ref{cliff basis5+1.}.  
There are $ 2^{\frac{d}{2}-1}\times 2^{\frac{d}{2}-1}$  Hermitian 
conjugated partners  of ''basis vectors'', denoted by $\hat{b}^{m}_{f} (=$ 
$(\hat{b}^{m \dagger}_{f})^{\dagger}$.  It is our choice which one of these
two groups of $2^{\frac{d}{2}-1} \times2^{\frac{d}{2}-1}$ members 
are ''basis vectors'' and which one are their Hermitian conjugated partners.
These two groups differ in handedness as  can be seen in 
Table~\ref{Table Clifffourplet.}, if observing $odd \,I$ and $odd \,II$, 
as well as if we compare Table~\ref{cliff basis5+1.} and 
Table~\ref{cliff basis5+1left.}.\\

{\it The Clifford odd  anticommuting ''basis vectors'', describing the internal 
space of fermions, obey together with their Hermitian conjugated partners 
the postulates of Dirac for the second quantized fermion fields,} Eq.~(\ref{alphagammatildeprod}).

{\it The Clifford even products of   $\gamma^a$'s (with the even number of 
nilpotents) form twice $2^{\frac{d}{2}-1} \times 2^{\frac{d}{2}-1}$ ''basis 
vectors'', ${\cal \hat{A}}^{m \dagger}_{f}$, describing properties of bosons, }
Table~\ref{cliff basis5+1even.}. {\it Each of the two groups are commuting 
objects due to the fact that even number of  $\gamma^a$'s commute. } \\

Also the Clifford even ''basis vectors'' are chosen to be the eigenvectors of the 
Cartan subalgebra of the Lorentz group, this time determined by ${\cal S}^{ab}
=S^{ab} +\tilde{S}^{ab}$, Eqs.~(\ref{calAb1234}, \ref{evenproperties}). 
While the Clifford odd ''basis vectors'' and their Hermitian conjugated partners 
form two independent groups, the Clifford even ''basis vectors''  have their Hermitian conjugated partners within each of the two groups. 

The choice of the ''basis vectors'' among  the two groups of the Clifford odd products of nilpotents and projectors for the description of the internal space of fermions, 
distinguishing also in handedness and other properties (Table~\ref{cliff basis5+1.} and 
Table~\ref{cliff basis5+1left.})
made as well the choice foe the Clifford even ''basis vectors'' describing the 
corresponding boson fields. We notice in Table~\ref{Table Clifffourplet.} that the 
choice of $odd \,I$ for the description of the internal space of fermions makes $even \,II$
to be  the corresponding boson field. 

The remaining group of  the $2^{\frac{d}{2}-1}\times 2^{\frac{d}{2}-1}$ Clifford
 even ''basis vectors'', presented in Table~\ref{Table Clifffourplet.} as $even\, I$ are not  
the  boson partners to the chosen  Clifford odd $odd \,I$ ''basic vectors'', but rather to their  Hermitian conjugated partners  $\hat{b}^{m}_{f}$, presented  as $odd \,II$ in the same 
 Table~\ref{Table Clifffourplet.}.\\

{\it The creation operators, either for creating fermions or for creating bosons, must 
have besides the ''basis vectors'' defining the internal space of fermions and bosons 
also the basis in ordinary space in momentum or coordinate representation}. I follow here 
shortly Ref.~\cite{nh2021RPPNP}. \\

Let us briefly present the relations  concerning the momentum or coordinate part of 
the  single particle states. The longer version is presented in Ref.~(\cite{nh2021RPPNP} in 
Subsect.~3.3 and in App. J)
\begin{eqnarray}
\label{creatorp}
|\vec{p}>&=& \hat{b}^{\dagger}_{\vec{p}} \,|\,0_{p}\,>\,,\quad 
<\vec{p}\,| = <\,0_{p}\,|\,\hat{b}_{\vec{p}}\,, \nonumber\\
<\vec{p}\,|\,\vec{p}'>&=&\delta(\vec{p}-\vec{p}')=
<\,0_{p}\,|\hat{b}_{\vec{p}}\; \hat{b}^{\dagger}_{\vec{p}'} |\,0_{p}\,>\,, 
\nonumber\\
&&{\rm leading \;to\;} \nonumber\\
\hat{b}_{\vec{p'}}\, \hat{b}^{\dagger}_{\vec{p}} &=&\delta(\vec{p'}-\vec{p})\,,
\end{eqnarray}
where the normalization $<\,0_{p}\, |\,0_{p}\,>=1$ to identity is assumed. 
While the quantized operators $\hat{\vec{p}}$ and  $\hat{\vec{x}}$ commute
 $\{\hat{p}^i\,, \hat{p}^j \}_{-}=0$ and  $\{\hat{x}^k\,, \hat{x}^l \}_{-}=0$, 
 this is not the case for  $\{\hat{p}^i\,, \hat{x}^j \}_{-}=i \eta^{ij}$. It
 therefore follows
 \begin{eqnarray}
 \label{eigenvalue10}
 <\vec{p}\,| \,\vec{x}>&=&<0_{\vec{p}}\,|\,\hat{b}_{\vec{p}}\;
\hat{b}^{\dagger}_{\vec{x}} 
 |0_{\vec{x}}\,>=(<0_{\vec{x}}\,|\,\hat{b}_{\vec{x}}\;
\hat{b}^{\dagger}_{\vec{p}} \,
 |0_{\vec{p}}\,>)^{\dagger}\, \nonumber\\
 \{\hat{b}^{\dagger}_{\vec{p}}\,,  \,
\hat{b}^{\dagger}_{\vec{p}\,'}\}_{-}&=&0\,,\qquad 
\{\hat{b}_{\vec{p}},  \,\hat{b}_{\vec{p}\,'}\}_{-}=0\,,\qquad
\{\hat{b}_{\vec{p}},  \,\hat{b}^{\dagger}_{\vec{p}\,'}\}_{-}=0\,,
\nonumber\\
\{\hat{b}^{\dagger}_{\vec{x}},  \,\hat{b}^{\dagger}_{\vec{x}\,'}\}_{-}&=&0\,,
\qquad 
\{\hat{b}_{\vec{x}},  \,\hat{b}_{\vec{x}\,'}\}_{-}=0\,,\qquad
\{\hat{b}_{\vec{x}},  \,\hat{b}^{\dagger}_{\vec{x}\,'}\}_{-}=0\,,
\nonumber\\
{\rm while}&&\nonumber\\
\{\hat{b}_{\vec{p}},  \,\hat{b}^{\dagger}_{\vec{x}}\}_{-}&=&
 e^{i \vec{p} \cdot \vec{x}} \frac{1}{\sqrt{(2 \pi)^{d-1}}}\,,\qquad,
\{\hat{b}_{\vec{x}},  \,\hat{b}^{\dagger}_{\vec{p}}\}_{-}=
 e^{-i \vec{p} \cdot \vec{x}} \frac{1}{\sqrt{(2 \pi)^{d-1}}}\,,
\end{eqnarray}

{\bf Statement  8.} {\it While the internal space of either fermions or bosons has 
 the finite degrees of freedom --- $2^{\frac{d}{2}-1}\times 2^{\frac{d}{2}-1}$ --- 
 the momentum basis has obviously continuously infinite degrees of freedom.}\\

Let us use the common symbol $\hat{a}^{m}_{f}$ for both ''basis vectors'' 
 $\hat{b}^{ m\dagger}_{f}$ and ${\cal \hat{A}}^{m \dagger}_{f}$. And let be
 taken into account that either fermion or boson second quantized states are
 solving equations of motion, which relate $p^0$ and $\vec{p}$: $p^0 =|\vec{p}|$.
 Then the solution of the equations of motion can be written as the superposition
 of the tensor products, $*_{T}$, of a finite number of  ''basis vectors'' describing 
 the internal space of a second quantized single particle state, $\hat{a}^{m}_{f}$, 
 and  the continuously infinite momentum basis
%
%
 \begin{eqnarray}
\label{wholespacegeneral}
\{{\bf \hat{a}}^{s \dagger}_{f} (\vec{p}) \,&=& \sum_m c^{s m}{}_{f}\,(\vec{p})\,
\hat{b}^{\dagger}_{\vec{p}}\,*_{T}\,\hat{a}^{m \dagger}_{f}\} \,
|vac_c>\,*_{T}\, |0_{\vec{p}}> \,,                                                                                                 
 \end{eqnarray}
where $\vec{p}$ determines the momentum in ordinary space and $s$ determines 
all the rest of quantum numbers. 
The state written here as $|vac_{o} >\,*_{T}\, |0_{\vec{p}} >$
is considered as the vacuum for a starting  single particle state from which one 
obtains the other single particle states by the operators, like  
$\hat{b}^{\dagger}_{\vec{p}}$, 
which pushes the momentum by an amount $\vec{p}$ and the vacuum for either
fermions $|\psi_{oc} >$, Eq.~(\ref{vaccliff}), or bosons $|\phi_{oc_{even}}>$, Eq.~(\ref{vaceven}). 

The creation operators for fermions can be therefore written as
 \begin{eqnarray}
\label{wholespacefermions}
\{{\bf \hat{b}}^{s \dagger}_{f} (\vec{p}) \,&=& \sum_m c^{s m}{}_{f}\,(\vec{p})\,
\hat{b}^{\dagger}_{\vec{p}}\,*_{T}\,\hat{b}^{m\dagger}_{f}\} \,
|\psi_{oc}>\,*_{T}\, |0_{\vec{p}}> \,,                                                                                                 
 \end{eqnarray}
while for the corresponding gauge bosons it follows
 \begin{eqnarray}
\label{wholespacebosons}
\{{\bf {\cal \hat{A}}}^{s \dagger}_{f} (\vec{p}) \,&=& 
\sum_m {\cal C}^{s m}{}_{f}\,(\vec{p})\,
\hat{b}^{\dagger}_{\vec{p}}\,*_{T}\,{\cal \hat{A}}^{m \dagger}_{f}\} \,
|\phi_{oc_{even}}>\,*_{T}\, |0_{\vec{p}}> \,.                                                                                               
 \end{eqnarray}

Since the ''basis vectors'' $\hat{b}^{m \dagger}_{f}$, describing the internal space of fermion, and their Hermitian conjugated 
partners do fulfil the anticommuting properties of Eq.~(\ref{alphagammatildeprod}), 
then also ${\bf \hat{b}}^{s \dagger}_{f} (\vec{p})$ and 
$({\bf \hat{b}}^{s \dagger}_{f} (\vec{p}))^{\dagger}$, 
Eq.~(\ref{wholespacegeneral}),  fulfil the 
anticommutation relations of Eq.~(\ref{alphagammatildeprod}) due the 
commutativity of operators $\hat{b}^{\dagger}_{\vec{p}} 
= (\hat{b}^{\dagger}_{-\vec{p}})^{\dagger}=\hat{b}_{-\vec{p}}$ and 
anticommutativity of ''basis vectors''.

The ''basis vectors'' for fermions bring to the second quantized fermions, 
that is to the creation and correspondingly to the annihilation operators 
operating on the vacuum state, the {\it anticomutativity} and  
$2^{\frac{d}{2}-1}\times 2^{\frac{d}{2}-1}$ quantum numbers of family 
members and of families for each of continuously $\infty$ many $\vec{p}$. 
The fermion single particle states therefore already anticommute.\\

The $2^{\frac{d}{2}-1}\times 2^{\frac{d}{2}-1}$ Clifford even ''basis vectors'' 
${\cal \hat{A}}^{m \dagger}_{f}$,  appearing in pairs which are Hermitian 
conjugated to each other, fulfil the commuting properties of 
Eq.~(\ref{evenproperties}), transfering these commuting properties also to 
$2^{\frac{d}{2}-1}\times 2^{\frac{d}{2}-1}$ members of 
${\cal \hat{A}}^{s \dagger}_{f} (\vec{p})$, Eq.~(\ref{wholespacegeneral}),  
for any of continuously $\infty$ $\vec{p}$,
so that ${\cal \hat{A}}^{s \dagger}_{f} (\vec{p})$ fulfil the 
commutation relations of Eq.~(\ref{evenproperties}) according to
commutativity properties of operators ${\cal \hat{A}}^{m \dagger}_{\vec{p}}$.


{\bf Statement 9.} {\it The odd products of the Clifford objects $\gamma^a$'s offer 
the ''basis vectors'' to describe the internal space of  the second quantized fermion 
fields. The even products of the Clifford objects $\gamma^a$'s offer the ''basis 
vectors'' to describe the internal space of  the second quantized boson fields. They 
are the gauge fields of the  fermion fields described by the odd Clifford objects. }\\

{\bf Statement 9.a} {\it The description of the internal space of fermions with
the odd Clifford algebra explains the second quantization postulates of Dirac.
The quantized single fermion states anticommute.}\\

The ${\cal \hat{A}}^{s \dagger}_{f} (\vec{p})$ ''basis vectors'' bring to the second quantized bosons, that is to the creation operators and annihilation operators, appearing in pairs or as self adjoint operators, operating on the vacuum state,  
the {\it commutativity} properties and  $2^{\frac{d}{2}-1}\times 2^{\frac{d}{2}-1}$ 
quantum numbers, explaining properties of boson particles. The ordinary basis,
$\hat{b}^{\dagger}_{\vec{p}}$, brings to the creation operators the continuously 
infinite degrees of freedom.\\

{\bf Statement 9.b} {\it The description of the internal space of bosons with
the even Clifford algebra explains the second quantization postulates  for gauge fields.
The quantized single boson states commute.}\\

Let us represent here the anticommutation relations for the creation and annihilation operators of the second quantized fermion  fields 
$\hat{\bf b}^{s \dagger }_{f}(\vec{p})$ and $\hat{\bf b}^{s }_{f }(\vec{p})$
by taking into account Eq.~(\ref{alphagammatildeprod})
\begin{eqnarray}
\{  \hat{\bf b}^{s' }_{f `}(\vec{p'})\,,\, 
\hat{\bf b}^{s \dagger}_{f }(\vec{p}) \}_{+} \,|\psi_{oc}> |0_{\vec{p}}>&=&
\delta^{s s'} \delta_{f f'}\,\delta(\vec{p}' - \vec{p})\, |\psi_{oc}> |0_{\vec{p}}>
\,,\nonumber\\
\{  \hat{\bf b}^{s' }_{f `}(\vec{p'})\,,\, 
\hat{\bf b}^{s}_{f }(\vec{p}) \}_{+} \,|\psi_{oc}> |0_{\vec{p}}>&=&0\,
 |\psi_{oc}> |0_{\vec{p}}>
\,,\nonumber\\
\{  \hat{\bf b}^{s' \dagger}_{f '}(\vec{p'})\,,\, 
\hat{\bf b}^{s \dagger}_{f }(\vec{p}) \}_{+}\, |\psi_{oc}> |0_{\vec{p}}>&=&0
\,|\psi_{oc}> |0_{\vec{p}}>
\,,\nonumber\\
 \hat{\bf b}^{s \dagger}_{f }(\vec{p}) \,|\psi_{oc}> |0_{\vec{p}}>&=&
|\psi^{s}_{f}(\vec{p})>\,\nonumber\\
 \hat{\bf b}^{s}_{f }(\vec{p}) \, |\psi_{oc}> |0_{\vec{p}}>&=&0
 \,|\psi_{oc}> |0_{\vec{p}}>\nonumber\\
 |p^0| &=&|\vec{p}|\,.
\label{Weylpp'comrel}
\end{eqnarray}
The creation operators $  \hat{\bf b}^{s \dagger}_{f }(\vec{p}, p^0) )$  and 
their Hermitian conjugated partners annihilation operators  
$\hat{\bf b}^{s}_{f }(\vec{p}, p^0) )$, creating and annihilating the single fermion 
state, respectively, fulfil when applying on the vacuum state,  
$|\psi_{oc}>|0_{\vec{p}}>$, the anticommutation relations for the second quantized fermions, postulated by Dirac (Ref.~\cite{nh2021RPPNP}, Subsect.~3.3.1, Sect.~5).
 
 The anticommutation relations of Eq.~(\ref{Weylpp'comrel}) are valid also if we 
 replace  the vacuum state,  $|\psi_{oc}>|0_{\vec{p}}>$, by the Hilbert space of Clifford fermions generated by the tensor product multiplication, $*_{T_{H}}$, of 
any number of the Clifford odd fermion states of all possible internal quantum 
numbers and all possible momenta (that is of any number of 
$ \hat {\bf b}^{s \, \dagger}_{f} (\vec{p})$ of any
 $(s,f, \vec{p})$), Ref.~(\cite{nh2021RPPNP}, Sect. 5.).\\



The commutation relations among boson creation operators 
$ {\bf {\cal \hat{A}}}^{s \dagger}_{f} (\vec{p})$ can be written as
\begin{eqnarray}
\label{evenpropertieswhole}
\{{\bf {\cal \hat{A}}}^{s \dagger}_{f} (\vec{p})\,, 
{\bf {\cal \hat{A}}}^{s' \dagger}_{f } (\vec{p}')\}_{-}&=&f^{s s' s''f f `f''}
{\bf {\cal \hat{A}}}^{s'' \dagger}_{f''} \,\delta(\vec{p}- \vec{p}')\,.
\end{eqnarray}
Let us present an example with $\vec{p}=(0,0,p^3,0,0)$ and the choice
${\bf {\cal \hat{A}}}^{3 \dagger}_{1} (\vec{p})$ and
${\bf {\cal \hat{A}}}^{2 \dagger}_{2} (\vec{p}')$, taken from
Table~\ref{cliff basis5+1even.}, one finds
\begin{eqnarray}
\label{evenpropertieswholeE}
\{{\bf {\cal \hat{A}}}^{3 \dagger}_{1} (\vec{p})\,,
{\bf {\cal \hat{A}}}^{1 \dagger}_{2} (\vec{p}')\}_{-}&=& -
\delta(\vec{p}- \vec{p}') \,{\bf {\cal \hat{A}}}^{2 \dagger}_{1}  (\vec{p})\,.
\end{eqnarray}
One can notice that the sums over each of the quantum numbers 
(${\cal S}^{03}, {\cal S}^{12}, {\cal S}^{56}, {\cal N}^3_{L}, {\cal N}^3_{R}, 
{\cal \tau}^3, {\cal \tau}^8, {\cal \tau}^4$) of the left hand side  are equal to the corresponding quantum numbers on the right hand side. \\


 The study of properties  of the second quantized bosons with the internal space of
which is described by the Clifford even algebra has just started and needs further
consideration. 

Let us point out that when breaking symmetries, like in the case of $d=(5+1)$
into $SU(2)\times SU(2) \times U(1)$, one easily sees that the same, either the 
right or the left representations appear within the same, only the right, 
Table~\ref{cliff basis5+1.}, or only the left, Table~\ref{cliff basis5+1left.},
representation, manifesting the right (left) hand fermions and the left (right) 
handed antifermions~\cite{nhds}. The same observation demonstrates also Table~\ref{Table so13+1.}, in which in each octet of $u$-quarks and $d$-quarks 
of any colour and in the octet of colourless leptons the left and the right members 
of fermions and antifermions appear.




%

%
%
%


%
\subsection{Simple action for fermion and boson  fields}
\label{fermionbosonaction}\
%

Let the space be $d=2(2n+1)$-dimensional. The {\it spin-charge-family} theory 
proposes $d=(13+1)$-dimensional space, or larger, so that the ''basis vectors''. 
describing the internal space of fermions and bosons, offers the properties of the 
observed quarks and leptons and their antiquarks and antileptons, as well as the
corresponding boson fields, as we learn in thic contribution. 

The action for the second quantized massless fermion and antifermion fields, and 
the corresponding massless boson fields in  $d=2(2n+1)$-dimensional space is 
therefore 
\begin{eqnarray}
{\cal A}\,  &=& \int \; d^dx \; E\;\frac{1}{2}\, (\bar{\psi} \, \gamma^a p_{0a} \psi) 
+ h.c. +
\nonumber\\  
               & & \int \; d^dx \; E\; (\alpha \,R + \tilde{\alpha} \, \tilde{R})\,,
\nonumber\\
               p_{0a } &=& f^{\alpha}{}_a p_{0\alpha} + \frac{1}{2E}\, \{ p_{\alpha},
E f^{\alpha}{}_a\}_- \,,\nonumber\\
          p_{0\alpha} &=&  p_{\alpha}  - \frac{1}{2}  S^{ab} \omega_{ab \alpha} - 
                    \frac{1}{2}  \tilde{S}^{ab}   \tilde{\omega}_{ab \alpha} \,,
                    \nonumber\\                    
R &=&  \frac{1}{2} \, \{ f^{\alpha [ a} f^{\beta b ]} \;(\omega_{a b \alpha, \beta} 
- \omega_{c a \alpha}\,\omega^{c}{}_{b \beta}) \} + h.c. \,, \nonumber \\
\tilde{R}  &=&  \frac{1}{2} \, \{ f^{\alpha [ a} f^{\beta b ]} 
\;(\tilde{\omega}_{a b \alpha,\beta} - \tilde{\omega}_{c a \alpha} \,
\tilde{\omega}^{c}{}_{b \beta})\} + h.c.\,.               
\label{wholeaction}
\end{eqnarray}
Here~\footnote{$f^{\alpha}{}_{a}$ are inverted vielbeins to 
$e^{a}{}_{\alpha}$ with the properties $e^a{}_{\alpha} f^{\alpha}{\!}_b = 
\delta^a{\!}_b,\; e^a{\!}_{\alpha} f^{\beta}{\!}_a = \delta^{\beta}_{\alpha} $, 
$ E = \det(e^a{\!}_{\alpha}) $.
Latin indices  
$a,b,..,m,n,..,s,t,..$ denote a tangent space (a flat index),
while Greek indices $\alpha, \beta,..,\mu, \nu,.. \sigma,\tau, ..$ denote an Einstein 
index (a curved index). Letters  from the beginning of both the alphabets
indicate a general index ($a,b,c,..$   and $\alpha, \beta, \gamma,.. $ ), 
from the middle of both the alphabets   
the observed dimensions $0,1,2,3$ ($m,n,..$ and $\mu,\nu,..$), indexes from 
the bottom of the alphabets
indicate the compactified dimensions ($s,t,..$ and $\sigma,\tau,..$). 
We assume the signature $\eta^{ab} =
diag\{1,-1,-1,\cdots,-1\}$.} 
$f^{\alpha [a} f^{\beta b]}= f^{\alpha a} f^{\beta b} - f^{\alpha b} f^{\beta a}$.


It is proven in Refs.~\cite{nd2017,IARD2020} that the spin connection gauge fields 
manifest in $d=(3+1)$ as the ordinary gravity, the known vector gauge fields and 
the scalar gauge fields, offering the (simple) explanation for the origin of higgs  
assumed by the {\it standard model}, explaining as well the Yukawa 
couplings.

\section{Conclusions}
\label{conclusions}
%


In the {\it spin-charge-family} theory the Clifford algebra is used to describe the 
internal space of fermion fields, what brings new insights, new recognitions 
about properties of fermion and boson 
fields~(\cite{nh2021RPPNP} and references therein): \\
The use of the odd Clifford algebra elements $\gamma^{a}$'s to describe 
the internal space of fermions offers not only the explanation for all the 
assumptions of the {\it standard model}, with the appearance of  the families 
of quarks and leptons and antiquarks and antileptons included,  but also for the 
appearance of the dark matter in the universe, for the explanation of the second 
quantized postulates for fermions of Dirac, for the matter/antimatter 
asymmetry in the universe,  and for several other observed phenomena, making
several predictions.

This article is the first trial to describe  the internal space of bosons  while using 
the even products of Clifford algebra  objects $\gamma^a$'s.
 
Although this study of the internal space of boson fields with the even Clifford 
algebra objects needs further considerations, yet the properties demonstrated in 
this paper are at least very promising.\\

Let me repeat briefly what I hope that we have learned.\\ 
{\bf i.} There are two kinds of the anticommuting algebras, the Grassmann algebra, 
offering in $d$-dimensional space $2\cdot 2^d$ operators, and the two Clifford 
algebras, each with $ 2^d$ operators. The Grassmann algebra operators are 
expressible with the operators of the two Clifford algebras and opposite,  Eq.~(\ref{clifftheta1}), and opposite. 
The two Clifford algebras are independent of each other, Eq.~(\ref{gammatildeantiher}),
forming two independent spaces. \\
{\bf ii.} Either the Grassmann algebra or the two Clifford 
algebras can be used to describe the internal space of anticommuting objects, if the
odd products of operators are used to describe the internal space of these objects, and of commuting objects, if the even products of operators are used to describe the internal 
space of these objects. \\
{\bf iii.} The ''basis vectors'' can be found in each of these algebras, which are 
eigenvectors of
the Cartan subalgebras,  Eq.~(\ref{cartangrasscliff}), of the corresponding Lorentz 
algebras ${\cal S}^{ab}$, $S^{ab}$ and $\tilde{S}^{ab}$, 
Eq.~(\ref{eigencliffcartan}).\\
{\bf iv.} After the reduction of the two Clifford algebras to only one --- 
$\gamma^{ab}$'s --- assuming how does $ \tilde{\gamma}^a$  apply  on
 $ \gamma^a$: $\{ \tilde{\gamma}^a B =(-)^B\, i \, B \gamma^a\}\, |\psi_{oc}>$, 
 with $(-)^B = -1$, if $B$ is (a function of) an odd products of $\gamma^a$'s,
 otherwise $(-)^B = 1$, there remain  twice $2^{\frac{d}{2}-1}$ iredducible representations  of $S^{ab}$, each with the  $2^{\frac{d}{2}-1}$ members.
 $\tilde{\gamma}^a$'s operate on superposition of products of $\gamma^{a}$'s.\\
 {\bf v.}  The ''basis vectors'', which are superposition of odd products of  
 $\gamma^{a}$'s, can be arranged to  fulfil the anticommutation relations, 
 postulated by Dirac, explaining correspondingly the anticommutation postulates 
 of Dirac, Eqs.~(\ref{almostDirac}, \ref{should}).\\
{\bf v.a.} The Clifford odd $2^{\frac{d}{2}-1}$ members of each 
of the  $2^{\frac{d}{2}-1}$ irreducible representations of ''basis vectors'' have 
their Hermitian conjugated partners in another set of $2^{\frac{d}{2}-1}$ 
$\cdot 2^{\frac{d}{2}-1}$ ''basis vectors'', Tables~(\ref{Table Clifffourplet.},
\ref{cliff basis5+1.}). The two sets of ''basis vectors'' differ in handedness, 
Tables~(\ref{cliff basis5+1.}, \ref{cliff basis5+1left.}). \\
{\bf v.b.} It is our choice which set we use to describe the creation operators and 
which one to describe the annihilation operators. Correspondingly we have either 
left or right handed creation operators. \\
{\bf v.c.} The family members of ''basis vectors'' have the same properties in 
 all the families. The sum of all the eigenvalues of all the commuting operators over 
the  $2^{\frac{d}{2}-1}$ family members is equal to zero for each of
 $2^{\frac{d}{2}-1}$ families, separately for left and 
separately for right handed representations. The sum of the family quantum numbers 
over the four families is zero.\\
{\bf vi.} The Clifford even ''basis vectors'', which are superposition of even
 products of $\gamma^a$'s, commute. \\ 
{\bf vi.a.} The Clifford even  ''basis vectors'' have their Hermitian conjugated partners
within the same group of $2^{\frac{d}{2}-1}$$\times2^{\frac{d}{2}-1}$ members,
Table~\ref{cliff basis5+1even.}, or are self adjoint.\\
{\bf vi.b. } Each of the two groups of the Clifford even $2^{\frac{d}{2}-1}\times$
$2^{\frac{d}{2}-1}$ ''basis vectors'' applies algebraically on only one of the two 
Clifford odd ''basis vectors'', (in Table~\ref{Table Clifffourplet.} Clifford $even \,II$ 
''basis vectors'' apply on Clifford $odd \,I$ ''basis vectors''), conserving the quantum 
numbers of the internal space.\\
{\bf vi.c.} The Clifford even ''basis vectors'', applying algebraically on the Clifford
odd ''basis vectors'', transform the Clifford odd ''basis vector'' into another member 
of the same family, Eqs.~(\ref{calAb1}, \ref{calAb234}, \ref{calAb1234}).\\
{\bf vi.d.} The Clifford even ''basis vectors'' have obviously the quantum numbers
of the adjoint representations with respect to the fundamental representation of the 
Clifford odd partners of the Clifford even ''basis vectors'', 
Table~\ref{cliff basis5+1even.}.\\
{\bf vi.e.} The sum of all the eigenvalues of all the Cartan subalgebra members 
over the members  of Clifford even ''basis vectors'' 
is equal to zero, independent of the choice of the subgroups (with the same number 
of the Cartan subalgeba), Table~\ref{cliff basis5+1even.}. \\
{\bf vi.f.} Two Clifford even ''basis vectors'' (${\cal \hat{A}}^{m \dagger}_{f}$  and 
 ${\cal \hat{A}}^{m ' \dagger}_{f}$) of the same $f$ and of 
$(m, m')\ne m_0$ are orthogonal.  The two ''basis vectors'' with non zero
algebraic product, $*_{A}$,  ''scatter'' into the third one, or annihilate into 
the vacuum,.\\
{\bf vi.g.} The superposition of products of  even number of $\tilde{\gamma}^a$'s
transform the member of the Clifford odd ''basis vector'' of particular family into the 
same family member of another family.\\
{\bf vii.} The creation and annihilation operators for either the Clifford odd or 
the Clifford even fields,  contain besides the corresponding ''basis vectors''
also the basis in ordinary, coordinate or momentum, space, 
Eqs.~(\ref{wholespacegeneral}, \ref{wholespacefermions}, \ref{wholespacebosons}). \\
{\bf vii.a.} The tensor products, $*_{T}$, of the ''basis vectors'' describing the internal 
space of fermions or bosons and the basis in ordinary space have the properties of 
creation and annihilation operators for either fermion or boson fields, defining the states when applying on the corresponding vacuum states, Eqs.~(\ref{vaccliff}, 
\ref{vaceven}).\\
{\bf vii.b.} While the internal  space of either fermions or bosons has  the finite 
degrees of freedom --- $2^{\frac{d}{2}-1}\times 2^{\frac{d}{2}-1}$ --- 
 the momentum basis has obviously continuously infinite degrees of freedom.
 Correspondingly the single particle states have  continuously infinite degrees of 
 freedom.\\
{\bf vii.c.} There are the  ''basis vectors'' describing the internal spaces of either
fermions or bosons, which bring commutativity or anticommutativity to creation and 
annihilation operators.\\ 
{\bf vii.d.} The single particle states described by applying the Clifford odd creation 
operators on the vacuum state, anticommute, while the single particle states  
described by  applying the Clifford even creation operators on the vacuum state
commute. The same rules are valid also when applying creation operators on 
the corresponding Hilbert spaces, Ref.~(nh2021RPPNP), Sect.~5.\\
{\bf vii.e.} Fermion fields described by using the Clifford odd creation operators 
interact with exchange of the corresponding boson fields described  by
the Clifford even creation operators, Eq.~(\ref{calAb1234}).   Bosons fields
interacts on both ways, with boson fields (if  the corresponding two ''basis vectors'' 
have non zero algebraic product, $*_{A}$), as well as with fermions.\\
{\bf vii.f.} The application of the creation operators with the Clifford even 
''basis vectors'', in which all the $\gamma^a$'s are replaced by 
$\tilde{\gamma}^a$'s, on the fermion creation operators, transform the fermion 
creation operator to another one, belonging to different family with the unchanged
family members of the ''basis vectors'', Subsect.~(\ref{generalbasisinternal},  
part {\bf b.}).\\ 

Let me conclude this contribution by saying that so far the description of  the internal 
space of the second quantized fermions with  the Clifford odd ''basis vectors'' offers
a new insight into the Hilbert space of the second quantized fermions (although 
there are still open questions waiting to be discussed, like it is the appearance 
of the Dirac sea in the usual approaches), the equivalent description of  the 
internal space of the second quantized boson fields with  the Clifford 
even ''basis vectors'' needs, although to my opinion very promising, a lot of further 
study.

\appendix

\section{Eigenstates of Cartan subalgebra of Lorentz algebra}
\label{eigencartan}

The eigenvectors of $S^{ab}$ 
and  $\tilde{S}^{ab}$ in the space  determined by $\gamma^a$'s
is as follows
\begin{eqnarray}
S^{ab} \frac{1}{2} (\gamma^a + \frac{\eta^{aa}}{ik} \gamma^b) &=& \frac{k}{2}  \,
\frac{1}{2} (\gamma^a + \frac{\eta^{aa}}{ik} \gamma^b)\,,\nonumber\\
S^{ab} \frac{1}{2} (1 +  \frac{i}{k}  \gamma^a \gamma^b) &=&  \frac{k}{2}  \,
 \frac{1}{2} (1 +  \frac{i}{k}  \gamma^a \gamma^b)\,,\nonumber\\
\tilde{S}^{ab} \frac{1}{2} (\tilde{\gamma}^a + \frac{\eta^{aa}}{ik} \tilde{\gamma}^b) &=& 
\frac{k}{2}  \,\frac{1}{2} (\tilde{\gamma}^a + \frac{\eta^{aa}}{ik} \tilde{\gamma}^b)\,,
\nonumber\\
\tilde{S}^{ab} \frac{1}{2} (1 +  \frac{i}{k}  \tilde{\gamma}^a \tilde{\gamma}^b) &=& 
- \frac{k}{2}  \, \frac{1}{2} (1 +  \frac{i}{k} \tilde{\gamma}^a \tilde{\gamma}^b)\,.
\label{eigencliffcartanapp}
\end{eqnarray}
with $k^2 = \eta^{aa} \eta^{bb}$.

The proof of the first two equations of Eq.(\ref{eigencliffcartanapp}) goes as follows, 
$a\ne b$ is assumed:

$\frac{i}{2}\gamma^a \gamma^b \frac{1}{2}  (\gamma^a + 
\frac{\eta^{aa}}{ik} \gamma^b)=\frac{i}{2} \frac{1}{2}  (-\eta^{aa}\gamma^b + 
\frac{\eta^{aa}\eta^{bb}}{ik} \gamma^a) = \frac{k}{2}\frac{1}{2}(\gamma^a-
\eta^{aa}\frac{i}{k} \gamma^b)$.

$\frac{i}{2}\gamma^a \gamma^b \frac{1}{2}  ( 1+\frac{i}{k}\gamma^a \gamma^b)
=\frac{i}{2} \frac{1}{2}(\gamma^a \gamma^b -\frac{i}{k}\eta^{aa} \eta^{bb})=
\frac{k}{2} \frac{1}{2}(1 +  \frac{i}{k} \gamma^a \gamma^b)$.\\

 For proving the second two equations it must be recognized that after the 
reduction of the Clifford space to only  the part spent by $\gamma^a$'s, that is 
after  requiring 

$\{ \tilde{\gamma}^{a} B= (-)^B\, i \, B \gamma^a \}\,|\psi_{oc}>$,\\
 with $(-)^B = -1$, if $B$ is (a function of) an odd product of $\gamma^a$'s,
 otherwise $(-)^B = 1$~\cite{nh03}, the relations of  Eq.~(\ref{gammatildeantiher}) remain unchanged. 
 
 One can see this as follows (I follow here Ref.~\cite{nh2021RPPNP}, Statement 3a. of App.I)\\
  $\{ \tilde{\gamma}^{a}, \tilde{\gamma}^{b}\}_{+}= 2\eta^{ab}=$
$\tilde{\gamma}^{a} \tilde{\gamma}^{b}+\tilde{\gamma}^{b}
\tilde{\gamma}^{a}=$ $ \tilde{\gamma}^{a} i\gamma^b +\tilde{\gamma}^{b} i \gamma^a=$
 $ i \gamma^b (-i)\gamma^a + i\gamma^a(-i)\gamma^b= 2\eta^{ab} $.\\
 $\{ \tilde{\gamma}^{a}, \gamma^b\}_{+}= 0=$
 $\tilde{\gamma}^{a} \gamma^b+\gamma^b \tilde{\gamma}^{a}=$
 $ \gamma^b (-i)\gamma^a+  \gamma^b i \gamma^a=0$.
  
Taking this into account it follows \\
$\tilde{S}^{ab}\frac{1}{2}(\gamma^a + \frac{\eta^{aa}}{ik}\gamma^b)= 
\frac{i}{2} \tilde{\gamma}^a \tilde{\gamma}^b \frac{1}{2}(\gamma^a +
 \frac{\eta^{aa}}{ik}\gamma^b) = \frac{i}{2} \frac{1}{2}(\gamma^a + 
\frac{\eta^{aa}}{ik}\gamma^b) \gamma^b \gamma^a= $
$\frac{i}{2} \frac{1}{2}(-\eta^{aa} \gamma^b +\frac{\eta^{aa} \eta^{bb}}{ik}
\gamma^a)$ $=\frac{k}{2} \frac{1}{2}(\gamma^a + 
\frac{\eta^{aa}}{ik}\gamma^b)$,\\
$\tilde{S}^{ab}\frac{1}{2}( 1+  \frac{i}{k} \gamma^a\gamma^b)= 
 \frac{i}{2} \frac{1}{2}(1 + \frac{i}{k}\gamma^a\gamma^b)
 \gamma^b \gamma^a= $ $\frac{i}{2} \frac{1}{2}(- \gamma^a\gamma^b + 
\frac{i}{k} \eta^{aa} \eta^{bb}) =- \frac{k}{2}\frac{1}{2} ( 1+  \frac{i}{k}
 \gamma^a\gamma^b)$,\\
 where it is taken into account that $k^2=\eta^{aa} \eta^{bb}$.

\section{Clifford odd and even ''basis vectors'' continue}
\label{cliffordoddlefthanded}
%



In Table~\ref{cliff basis5+1.} the  Clifford odd ''basis vectors'' of the right handedness
were chosen for the description of the internal space of fermions in 
$d=(5+1)$-dimensional space, noted  in Table~\ref{Table Clifffourplet.} as $odd \,I$.

If we make a choice of $odd \,II$ for the Clifford odd ''basis vectors'' in 
Table~\ref{Table Clifffourplet.}, and take the $odd\, I$ as their Hermitian conjugated partners, then these ''basis vectors'' are left (not right) handed  and have properties presented in Table~\ref{cliff basis5+1left.}. We can compare their properties by the 
properties of the right handed ''basis vectors'' appearing in Table~\ref{cliff basis5+1.}.
 The two groups $odd \,I$ and $odd \,II$ are Hermitian conjugated to each other.
\begin{table}
\begin{tiny}
 \begin{center}
\begin{minipage}[t]{16.5 cm}
\caption{The ''basis vectors'',  this time left handed --- 
 $\hat{b}^{m=(ch,s)\dagger}_{f}$ (each is a product of projectors and an odd 
number of nilpotents, and is the "eigenstate" of all the Cartan subalgebra members, 
$S^{03}$, $S^{12}$, $S^{56}$ and $\tilde{S}^{03}$, $\tilde{S}^{12}$, 
$\tilde{S}^{56}$, Eq.~(\ref{cartangrasscliff})
($ch$ (charge), the eigenvalue of $S^{56}$, and $s$ (spin), the eigenvalues of 
$S^{03}$ and $S^{12}$, explain index $m$, $f$ determines family quantum 
numbers, the eigenvalues of ($\tilde{S}^{03}$, $\tilde{S}^{12}$, 
$\tilde{S}^{56}$) ---  
are presented for $d= (5+1)$-dimensional case.  Their Hermitian conjugated  partners
--- $\hat{b}^{m=(ch,s)}_{f}$ --- can be found in Table~\ref{cliff basis5+1.} 
as ''basis vectors''. 
This table represents also the 
eigenvalues of the three commuting operators $N^3_{L,R}$ and $S^{56 }$ of the 
subgroups $SU(2)\times SU(2)\times U(1)$ and the eigenvalues of the three 
commuting operators $\tau^3, \tau^8$ and $\tau^{4}$ of the subgroups 
 $SU(3)\times U(1)$, in these two last cases index $m$ represents the eigenvalues 
 of the corresponding commuting generators. $\Gamma^{(5+1)}=-\gamma^0 
 \gamma^1 \gamma^2 \gamma^3\gamma^5\gamma^6= -1$, $\Gamma^{(3+1)}
 = i\gamma^0 \gamma^1 \gamma^2 \gamma^3$.
Operators $\hat{b}^{m=(ch,s) \dagger}_{f}$ 
 and $\hat{b}^{m=(ch, s)}_{f}$
 fulfil the anticommutation relations of Eqs.~(\ref{almostDirac}, \ref{should}).
\vspace{2mm}}
\label{cliff basis5+1left.}
\end{minipage}
 \begin{tabular}{|r|l r|r|r|r|r|r|r|r|r|r|r|r|r|r|r|}
 \hline
$\, f $&$m $&$=(ch,s)$&$\hat{b}^{ m=(ch,s) \dagger}_f$
&$S^{03}$&$ S^{1 2}$&$S^{5 6}$&$\Gamma^{3+1}$ &$N^3_L$&$N^3_R$&
$\tau^3$&$\tau^8$&$\tau^4$&
$\tilde{S}^{03}$&$\tilde{S}^{1 2}$& $\tilde{S}^{5 6}$\\
\hline
$I$&$1$&$(\frac{1}{2},\frac{1}{2})$&$
\stackrel{03}{(-i)}\,\stackrel{12}{(+)}| \stackrel{56}{(+)}$&
$- \frac{i}{2}$&$\frac{1}{2}$&$\frac{1}{2}$&$-1$
&$\frac{1}{2}$&$0$&$- \frac{1}{2}$&$-\frac{1}{2\sqrt{3}}$&$-\frac{1}{6}$&
$-\frac{i}{2}$&$\frac{1}{2}$&$\frac{1}{2}$\\
$I$ &$2$&$(\frac{1}{2},-\frac{1}{2})$&$
\stackrel{03}{[+i]}\,\stackrel{12}{[-]}|\stackrel{56}{(+)}$&
$\frac{i}{2}$&$-\frac{1}{2}$&$\frac{1}{2}$&$-1$
&$-\frac{1}{2}$&$0$&$\frac{1}{2}$&$-\frac{1}{2\sqrt{3}}$&$-\frac{1}{6}$&
$-\frac{i}{2}$&$\frac{1}{2}$&$\frac{1}{2}$\\
$I$ &$3$&$(-\frac{1}{2},\frac{1}{2})$&$
\stackrel{03}{[+i]}\,\stackrel{12}{(+)}|\stackrel{56}{[-]}$&
$\frac{i}{2}$&$ \frac{1}{2}$&$-\frac{1}{2}$&$ 1$
&$0$&$\frac{1}{2}$&$0$&$\frac{1}{\sqrt{3}}$&$-\frac{1}{6}$&$-\frac{i}{2}$&$\frac{1}{2}$&$\frac{1}{2}$\\
$I$ &$1$&$(-\frac{1}{2},-\frac{1}{2})$&$
\stackrel{03}{(-i)}\,\stackrel{12}{[-]}|\stackrel{56}{[-]}$&
$-\frac{i}{2}$&$- \frac{1}{2}$&$-\frac{1}{2}$&$ 1$
&$0$&$-\frac{1}{2}$&$0$&$0$&$\frac{1}{2}$&$-\frac{i}{2}$&$\frac{1}{2}$&$\frac{1}{2}$\\
\hline 
$II$&$1$&$(\frac{1}{2},\frac{1}{2})$&$
\stackrel{03}{[-i]}\,\stackrel{12}{[+]}| \stackrel{56}{(+)}$&
$-\frac{i}{2}$& $\frac{1}{2}$&$\frac{1}{2}$&$-1$&
$\frac{1}{2}$&$0$&$-\frac{1}{2}$&$-\frac{1}{2\sqrt{3}}$&$-\frac{1}{6}$&$\frac{i}{2}$&$-\frac{1}{2}$&$\frac{1}{2}$\\
$II$ &$2$&$(\frac{1}{2},-\frac{1}{2})$&$
\stackrel{03}{(+i)}\,\stackrel{12}{(-)}|\stackrel{56}{(+)}$&
$\frac{i}{2}$&$-\frac{1}{2}$&$\frac{1}{2}$&$-1$
&$-\frac{1}{2}$&$0$&$\frac{1}{2}$&$-\frac{1}{2\sqrt{3}}$&$-\frac{1}{6}$&$\frac{i}{2}$&$-\frac{1}{2}$&$\frac{1}{2}$\\
$II$ &$3$&$(-\frac{1}{2},\frac{1}{2})$&$
\stackrel{03}{(+i)}\,\stackrel{12}{[+]}|\stackrel{56}{[-]}$&
$\frac{i}{2}$&$ \frac{1}{2}$&$-\frac{1}{2}$&$1$
&$0$&$\frac{1}{2}$&$0$&$\frac{1}{\sqrt{3}}$&$-\frac{1}{6}$&$\frac{i}{2}$&$-\frac{1}{2}$&$\frac{1}{2}$\\
$II$ &$4$&$(-\frac{1}{2},-\frac{1}{2})$&$
\stackrel{03}{[-i]}\, \stackrel{12}{(-)}|\stackrel{56}{[-]}$&
$-\frac{i}{2}$&$- \frac{1}{2}$&$-\frac{1}{2}$&$1$
&$0$&$-\frac{1}{2}$&$0$&$0$&$\frac{1}{2}$&$\frac{i}{2}$&$-\frac{1}{2}$&$\frac{1}{2}$\\ 
%
%
 \hline
$III$&$1$&$(\frac{1}{2},\frac{1}{2})$&$
\stackrel{03}{[-i]}\,\stackrel{12}{(+)}| \stackrel{56}{[+]}$&
$-\frac{i}{2}$& $\frac{1}{2}$&$\frac{1}{2}$&$-1$&
$\frac{1}{2}$&$0$&$-\frac{1}{2}$&$-\frac{1}{2\sqrt{3}}$&$-\frac{1}{6}$&$\frac{i}{2}$&$\frac{1}{2}$&$-\frac{1}{2}$\\
$III$ &$2$&$(\frac{1}{2},-\frac{1}{2})$&$
\stackrel{03}{(+i)}\,\stackrel{12}{[-]}|\stackrel{56}{[+]}$&
$\frac{i}{2}$&$-\frac{1}{2}$&$\frac{1}{2}$&$-1$
&$-\frac{1}{2}$&$0$&$\frac{1}{2}$&$-\frac{1}{2\sqrt{3}}$&
$-\frac{1}{6}$&$\frac{i}{2}$&$\frac{1}{2}$&$-\frac{1}{2}$\\
$III$ &$3$&$(-\frac{1}{2},\frac{1}{2})$&$
\stackrel{03}{(+i)}\,\stackrel{12}{(+)}|\stackrel{56}{(-)}$&
$-\frac{i}{2}$&$ \frac{1}{2}$&$-\frac{1}{2}$&$1$
&$0$&$\frac{1}{2}$&$0$&$\frac{1}{\sqrt{3}}$&$-\frac{1}{6}$&$\frac{i}{2}$&$\frac{1}{2}$&$-\frac{1}{2}$\\
$III$ &$4$&$(-\frac{1}{2},-\frac{1}{2})$&$
\stackrel{03}{[-i]} \stackrel{12}{[-]}|\stackrel{56}{(-)}$&
$-\frac{i}{2}$&$- \frac{1}{2}$&$-\frac{1}{2}$&$1$
&$0$&$-\frac{1}{2}$&$0$&$0$&$\frac{1}{2}$&$\frac{i}{2}$&$\frac{1}{2}$&$-\frac{1}{2}$\\
\hline
$IV$&$1$&$(\frac{1}{2},\frac{1}{2})$&$
\stackrel{03}{(-i)}\,\stackrel{12}{[+]}| \stackrel{56}{[+]}$&
$-\frac{i}{2}$&$\frac{1}{2}$&$\frac{1}{2}$&$-1$&
$\frac{1}{2}$&$0$&$-\frac{1}{2}$&$-\frac{1}{2\sqrt{3}}$&$-\frac{1}{6}$&
$-\frac{i}{2}$&$-\frac{1}{2}$&$-\frac{1}{2}$\\
$IV$ &$2$&$(\frac{1}{2},-\frac{1}{2})$&$
\stackrel{03}{[+i]}\,\stackrel{12}{(-)}|\stackrel{56}{[+]}$&
$\frac{i}{2}$&$-\frac{1}{2}$&$\frac{1}{2}$&$-1$
&$-\frac{1}{2}$&$0$&$\frac{1}{2}$&$-\frac{1}{2\sqrt{3}}$&$-\frac{1}{6}$&
$-\frac{i}{2}$&$-\frac{1}{2}$&$-\frac{1}{2}$\\
$IV$ &$3$&$(-\frac{1}{2},\frac{1}{2})$&$
\stackrel{03}{[+i]}\,\stackrel{12}{[+]}|\stackrel{56}{(-)}$&
$\frac{i}{2}$&$ \frac{1}{2}$&$-\frac{1}{2}$&$1$
&$0$&$\frac{1}{2}$&$0$&$\frac{1}{\sqrt{3}}$&$-\frac{1}{6}$&$-\frac{i}{2}$&$-\frac{1}{2}$&$-\frac{1}{2}$\\
$IV$ &$4$&$(-\frac{1}{2},-\frac{1}{2})$&$
\stackrel{03}{(-i)}\, \stackrel{12}{(-)}|\stackrel{56}{(-)}$&
$-\frac{i}{2}$&$- \frac{1}{2}$&$-\frac{1}{2}$&$1$
&$0$&$-\frac{1}{2}$&$0$&$0$&$\frac{1}{2}$&$-\frac{i}{2}$&$-\frac{1}{2}$&$-\frac{1}{2}$\\ 
\hline 
 \end{tabular}
 \end{center}
\end{tiny}
\end{table}
We clearly see if comparing both tables, Table~\ref{cliff basis5+1.} and
Table~\ref{cliff basis5+1left.}, that they do differ in properties. In particular 
the difference among these two kinds of ''basis vectors'' is easily seen in the 
$SU(3)\times U(1) $ subgroup, that is in $(\tau^3, \tau^8, \tau^4)$ values.

In Table~\ref{Table so13+1.} one finds the left and the right handed content of 
one of the families, the fourth ones, presented in Ref.~\cite{nh2021RPPNP}, Table~5,
if $d=(5+1)$ is taken as the subspace of the space $d=(13+1)$.

\section{Some useful relations 
in Grassmann and Clifford space, needed also in App.~\ref{13+1representation} }
\label{grassmannandcliffordfermions}

The generator of the Lorentz transformation in Grassmann space is defined as follows~\cite{norma93}
\begin{eqnarray}
\label{Lorentztheta}
{\cal {\bf S}}^{ab} &=&  (\theta^a p^{\theta b} - \theta^b p^{\theta a})\,\nonumber\\
&=& S^{ab} +\tilde{S}^{ab} \,, \quad  \{S^{ab}, \tilde{S}^{cd}\}_{-} =0\,,
\end{eqnarray}
where $S^{ab}$ and $\tilde{S}^{ab}$ are the corresponding two generators of the Lorentz
 transformations in the Clifford space, forming orthogonal representations with respect to each other.

We make a choice of the Cartan subalgebra of the Lorentz algebra as follows 
\begin{eqnarray}
&& {\cal {\bf S}}^{03}, {\cal {\bf S}}^{12}, {\cal {\bf S}}^{56}, \cdots, 
{\cal {\bf S}}^{d-1\; d}\,, \nonumber\\
&& S^{03}, S^{12}, S^{56}, \cdots, S^{d-1\; d}\,,\nonumber\\
&& \tilde{S}^{03}, \tilde{S}^{12}, \tilde{S}^{56}, \cdots, \tilde{S}^{d-1\; d}\,,\nonumber\\
 &&{\rm if } \quad d = 2n\,.
\label{choicecartan}
\end{eqnarray}
We find the infinitesimal generators of the Lorentz transformations in 
 Clifford space
\begin{eqnarray}
\label{Lorentzgammatilde}
S^{ab} &=& \frac{i}{4} (\gamma^a \gamma^b - \gamma^b \gamma^a)\,, \quad
S^{ab \dagger} = \eta^{aa} \eta^{bb} S^{ab}\,,\nonumber\\
\tilde{S}^{ab} &=& \frac{i}{4} (\tilde{\gamma}^a \tilde{\gamma}^b -
\tilde{\gamma}^b\tilde{\gamma}^a) \,,  \quad \tilde{S}^{ab \dagger} =
\eta^{aa} \eta^{bb} \tilde{S}^{ab}\,,
\end{eqnarray}
where $\gamma^a$ and $\tilde{\gamma}^a$ are defined in Eqs.~(\ref{clifftheta1},
\ref{gammatildeantiher}). 
The commutation relations for either ${\cal {\bf S}}^{ab}$ or $S^{ab}$ or $\tilde{S}^{ab}$,
${\cal {\bf S}}^{ab} = S^{ab} + \tilde{S}^{ab}$, are 
%
%
%
\begin{eqnarray}
\label{LorentzthetaCliffcom}
\{S^{ab}, \tilde{S}^{cd}\}_{-}&=& 0\,, \nonumber\\
\{S^{ab},S^{cd}\}_{-} &=& i (\eta^{ad} S^{bc} + \eta^{bc} S^{ad} -
 \eta^{ac} S^{bd} - \eta^{bd} S^{ac})\,,\nonumber\\
\{\tilde{S}^{ab},\tilde{S}^{cd}\}_{-} &=& i(\eta^{ad} \tilde{S}^{bc} + 
\eta^{bc} \tilde{S}^{ad} 
- \eta^{ac} \tilde{S}^{bd} - \eta^{bd} \tilde{S}^{ac})\,.
\end{eqnarray}
The infinitesimal generators of the two invariant subgroups of the group $SO(3,1)$ can be expressed as follows
\begin{eqnarray}
\label{so1+3}
\vec{N}_{\pm}(= \vec{N}_{(L,R)}): &=& \,\frac{1}{2} (S^{23}\pm i S^{01},S^{31}\pm i S^{02}, 
S^{12}\pm i S^{03} )\,.
\end{eqnarray}
The infinitesimal generators of the two invariant subgroups of the group $SO(4)$ are expressible with
$S^{ab}, (a,b) = (5,6,7,8)$ as follows 
 \begin{eqnarray}
 \label{so42}
 \vec{\tau}^{1}:&=&\frac{1}{2} (S^{58}-  S^{67}, \,S^{57} + S^{68}, \,S^{56}-  S^{78} )\,,
\nonumber\\
 \vec{\tau}^{2}:&=& \frac{1}{2} (S^{58}+  S^{67}, \,S^{57} - S^{68}, \,S^{56}+  S^{78} )\,,
 \end{eqnarray}
while the generators of the $SU(3)$ and  $U(1)$ subgroups of the group $SO(6)$ can be expressed by
$S^{ab}, (a,b) = (9,10,11,12,13,14)$
 \begin{eqnarray}
 \label{so64}
 \vec{\tau}^{3}: = &&\frac{1}{2} \,\{  S^{9\;12} - S^{10\;11} \,,
  S^{9\;11} + S^{10\;12} ,\, S^{9\;10} - S^{11\;12} ,\nonumber\\
 && S^{9\;14} -  S^{10\;13} ,\,  S^{9\;13} + S^{10\;14} \,,
  S^{11\;14} -  S^{12\;13}\,,\nonumber\\
 && S^{11\;13} +  S^{12\;14} ,\, 
 \frac{1}{\sqrt{3}} ( S^{9\;10} + S^{11\;12} - 
 2 S^{13\;14})\}\,,\nonumber\\
 \tau^{4}: = &&-\frac{1}{3}(S^{9\;10} + S^{11\;12} + S^{13\;14})\,.
 \end{eqnarray}
 The group $SO(6)$ has $\frac{d (d-1)}{2}=15$  generators and $\frac{d}{2}=3$ 
 commuting operators.  The subgroups $SU(3)$  $\times U(1)$ have the same number
 of commuting operators, expressed with $\tau^{33}$, $\tau^{38}$ and  $\tau^4$, 
 and $9$ generators, $8$ of $SU(3)$ and one of $U(1)$. The rest of $6$ generators, 
 not included in $SU(3)$  $\times U(1)$, can be expressed as $\frac{1}{2} \,\{  S^{9\;12}   +S^{10\;11},   S^{9\;11} - S^{10\;12}$, $S^{9\;14} + S^{10\;13}, S^{9\;13} - 
 S^{10\;14},   S^{11\;14} +  S^{12\;13},  S^{11\;13} -  S^{12\;14} $.\\

The hyper charge $Y$ can be defined as $Y=\tau^{23} + \tau^{4}$. 

The equivalent expressions for the "family" charges, expressed by $\tilde{S}^{ab},$ follow if in 
Eqs.~(\ref{so1+3} - \ref{so64}) $S^{ab}$ are replaced by $\tilde{S}^{ab}$.

Let us present some useful relations from Ref.~\cite{IARD2016}.
\begin{eqnarray}
\stackrel{ab}{(k)}\stackrel{ab}{(k)}& =& 0\,, \quad \quad \stackrel{ab}{(k)}\stackrel{ab}{(-k)}
= \eta^{aa}  \stackrel{ab}{[k]}\,, \quad \stackrel{ab}{(-k)}\stackrel{ab}{(k)}=
\eta^{aa}   \stackrel{ab}{[-k]}\,,\quad
\stackrel{ab}{(-k)} \stackrel{ab}{(-k)} = 0\,, \nonumber\\
\stackrel{ab}{[k]}\stackrel{ab}{[k]}& =& \stackrel{ab}{[k]}\,, \quad \quad
\stackrel{ab}{[k]}\stackrel{ab}{[-k]}= 0\,, \;\;
\quad \quad  \quad \stackrel{ab}{[-k]}\stackrel{ab}{[k]}=0\,,
 \;\;\quad \quad \quad \quad \stackrel{ab}{[-k]}\stackrel{ab}{[-k]} = \stackrel{ab}{[-k]}\,,
 \nonumber\\
\stackrel{ab}{(k)}\stackrel{ab}{[k]}& =& 0\,,\quad \quad \quad \stackrel{ab}{[k]}\stackrel{ab}{(k)}
=  \stackrel{ab}{(k)}\,, \quad \quad \quad \stackrel{ab}{(-k)}\stackrel{ab}{[k]}=
 \stackrel{ab}{(-k)}\,,\quad \quad \quad 
\stackrel{ab}{(-k)}\stackrel{ab}{[-k]} = 0\,,
\nonumber\\
\stackrel{ab}{(k)}\stackrel{ab}{[-k]}& =&  \stackrel{ab}{(k)}\,,
\quad \quad \stackrel{ab}{[k]}\stackrel{ab}{(-k)} =0\,,  \quad \quad 
\quad \stackrel{ab}{[-k]}\stackrel{ab}{(k)}= 0\,, \quad \quad \quad \quad
\stackrel{ab}{[-k]}\stackrel{ab}{(-k)} = \stackrel{ab}{(-k)}\,.
\label{graphbinoms}
\end{eqnarray}
%
%
\section{One family representation in  $d=(13+1)$-dimensional space with 
$2^{\frac{d}{2}-1}$ members representing quarks and leptons and antiquarks and antileptons 
in the {\it spin-charge-family} theory } 
\label{13+1representation}
%

In Table~{Table so13+1.} the ''basis vectors'' of one irreducible representation, one family,  
of the Clifford odd basis vectors of left handedness, $\Gamma^{(13+1)}$, is presented, 
including all the quarks and the leptons and the antiquarks and the antileptons of the {\it
standard model}. The needed definitions of the quantum numbers are presented in 
App.~\ref{grassmannandcliffordfermions}.

In Tables~\ref{Table Clifffourplet.},~\ref{cliff basis5+1.},~\ref{cliff basis5+1even.} a simple
toy model for $d=(5+1)$-dimensional space is discussed, and the properties of fermions 
(appearing in families) and their gauge boson fields (the vielbeins and the two kinds of the spin connection fields) analysed. The manifold $(5+1)$ was suggested to break either into $SU(2)\times SU(2)\times U(1)$ or to $SU(3) \times U(1)$ to study properties of the fermion and 
boson second quantized fields, with second quantization origining in the anticommutativity or 
commutativity of ''basis vectors''. 

Here only one family of ''basis vectors'' is presented to see that while the starting 
''basis vectors'' can be either left or right handed, the subgroups,of the starting group
contain left and right handed members, as it is  $SU(2)\times SU(2)\times U(1)$ of
$SO(5+1)$ in the toy model. 

The breaks of the symmetries, manifesting in Eqs.~(\ref{so1+3}, \ref{so42}, \ref{so64}), are in 
the {\it spin-charge-family} theory caused by the condensate and the nonzero vacuum expectation 
values (constant values) of the scalar fields carrying the space index $(7,8)$ 
(Refs.~\cite{normaJMP2015,IARD2016} and the references therein), all
originating in the vielbeins and the two kinds of the spin connection fields. The space 
breaks first to $SO(7,1)$
$\times SU(3) \times U(1)_{II}$ and then further to $SO(3,1)\times SU(2)_{I} \times U(1)_{I}$
$\times SU(3) \times U(1)_{II}$, what explains the connections between the weak and the hyper 
charges and the handedness of spinors.
\bottomcaption{\label{Table so13+1.}%
\tiny{
The left handed ($\Gamma^{(13,1)} = -1$~\cite{IARD2016})
multiplet of spinors --- the members of the fundamental representation of the $SO(13,1)$ group,
manifesting the subgroup $SO(7,1)$
 of the colour charged quarks and antiquarks and the colourless
leptons and antileptons --- is presented in the massless basis using the technique presented in
Refs.~\cite{nh02,nh03,IARD2016,normaJMP2015}.
It contains the left handed  ($\Gamma^{(3,1)}=-1$) 
 weak ($SU(2)_{I}$) charged  ($\tau^{13}=\pm \frac{1}{2}$, Eq.~(\ref{so42})),
and $SU(2)_{II}$ chargeless ($\tau^{23}=0$, Eq.~(\ref{so42}))
quarks and leptons and the right handed  ($\Gamma^{(3,1)}=1$) 
 weak  ($SU(2)_{I}$) chargeless and $SU(2)_{II}$
charged ($\tau^{23}=\pm \frac{1}{2}$) quarks and leptons, both with the spin $ S^{12}$  up and down ($\pm \frac{1}{2}$, respectively). 
Quarks distinguish from leptons only in the $SU(3) \times U(1)$ part: Quarks are triplets
of three colours  ($c^i$ $= (\tau^{33}, \tau^{38})$ $ = [(\frac{1}{2},\frac{1}{2\sqrt{3}}),
(-\frac{1}{2},\frac{1}{2\sqrt{3}}), (0,-\frac{1}{\sqrt{3}}) $], Eq.~(\ref{so64}))
carrying  the "fermion charge" ($\tau^{4}=\frac{1}{6}$, Eq.~(\ref{so64})).
The colourless leptons carry the "fermion charge" ($\tau^{4}=-\frac{1}{2}$).
The same multiplet contains also the left handed weak ($SU(2)_{I}$) chargeless and 
$SU(2)_{II}$
charged antiquarks and antileptons and the right handed weak ($SU(2)_{I}$) charged and
$SU(2)_{II}$ chargeless antiquarks and antileptons.
Antiquarks distinguish from antileptons again only in the $SU(3) \times U(1)$ part: Antiquarks are
antitriplets, 
 carrying  the "fermion charge" ($\tau^{4}=-\frac{1}{6}$).
The anticolourless antileptons carry the "fermion charge" ($\tau^{4}=\frac{1}{2}$).
 $Y=(\tau^{23} + \tau^{4})$ is the hyper charge, the electromagnetic charge
is $Q=(\tau^{13} + Y$).
%
The vacuum state,
on which the nilpotents and projectors operate, is presented in Eq.~(\ref{vaccliff}).
The reader can find this  Weyl representation also in
Refs.~\cite{n2014matterantimatter,normaJMP2015} and the references
therein. }
}
\tablehead{\hline
i&$$&$|^a\psi_i>$&$\Gamma^{(3,1)}$&$ S^{12}$&
$\tau^{13}$&$\tau^{23}$&$\tau^{33}$&$\tau^{38}$&$\tau^{4}$&$Y$&$Q$\\
\hline
&& ${\rm (Anti)octet},\,\Gamma^{(7,1)} = (-1)\,1\,, \,\Gamma^{(6)} = (1)\,-1$&&&&&&&&& \\
&& ${\rm of \;(anti) quarks \;and \;(anti)leptons}$&&&&&&&&&\\
\hline\hline}
\tabletail{\hline \multicolumn{12}{r}{\emph{Continued on next page}}\\}
\tablelasttail{\hline}
\begin{center}
\tiny{
\begin{supertabular}{|r|c||c||c|c||c|c||c|c|c||r|r|}
1&$ u_{R}^{c1}$&$ \stackrel{03}{(+i)}\,\stackrel{12}{[+]}|
\stackrel{56}{[+]}\,\stackrel{78}{(+)}
||\stackrel{9 \;10}{(+)}\;\;\stackrel{11\;12}{[-]}\;\;\stackrel{13\;14}{[-]} $ &1&$\frac{1}{2}$&0&
$\frac{1}{2}$&$\frac{1}{2}$&$\frac{1}{2\,\sqrt{3}}$&$\frac{1}{6}$&$\frac{2}{3}$&$\frac{2}{3}$\\
\hline
2&$u_{R}^{c1}$&$\stackrel{03}{[-i]}\,\stackrel{12}{(-)}|\stackrel{56}{[+]}\,\stackrel{78}{(+)}
||\stackrel{9 \;10}{(+)}\;\;\stackrel{11\;12}{[-]}\;\;\stackrel{13\;14}{[-]}$&1&$-\frac{1}{2}$&0&
$\frac{1}{2}$&$\frac{1}{2}$&$\frac{1}{2\,\sqrt{3}}$&$\frac{1}{6}$&$\frac{2}{3}$&$\frac{2}{3}$\\
\hline
3&$d_{R}^{c1}$&$\stackrel{03}{(+i)}\,\stackrel{12}{[+]}|\stackrel{56}{(-)}\,\stackrel{78}{[-]}
||\stackrel{9 \;10}{(+)}\;\;\stackrel{11\;12}{[-]}\;\;\stackrel{13\;14}{[-]}$&1&$\frac{1}{2}$&0&
$-\frac{1}{2}$&$\frac{1}{2}$&$\frac{1}{2\,\sqrt{3}}$&$\frac{1}{6}$&$-\frac{1}{3}$&$-\frac{1}{3}$\\
\hline
4&$ d_{R}^{c1} $&$\stackrel{03}{[-i]}\,\stackrel{12}{(-)}|
\stackrel{56}{(-)}\,\stackrel{78}{[-]}
||\stackrel{9 \;10}{(+)}\;\;\stackrel{11\;12}{[-]}\;\;\stackrel{13\;14}{[-]} $&1&$-\frac{1}{2}$&0&
$-\frac{1}{2}$&$\frac{1}{2}$&$\frac{1}{2\,\sqrt{3}}$&$\frac{1}{6}$&$-\frac{1}{3}$&$-\frac{1}{3}$\\
\hline
5&$d_{L}^{c1}$&$\stackrel{03}{[-i]}\,\stackrel{12}{[+]}|\stackrel{56}{(-)}\,\stackrel{78}{(+)}
||\stackrel{9 \;10}{(+)}\;\;\stackrel{11\;12}{[-]}\;\;\stackrel{13\;14}{[-]}$&-1&$\frac{1}{2}$&
$-\frac{1}{2}$&0&$\frac{1}{2}$&$\frac{1}{2\,\sqrt{3}}$&$\frac{1}{6}$&$\frac{1}{6}$&$-\frac{1}{3}$\\
\hline
6&$d_{L}^{c1} $&$ - \stackrel{03}{(+i)}\,\stackrel{12}{(-)}|\stackrel{56}{(-)}\,\stackrel{78}{(+)}
||\stackrel{9 \;10}{(+)}\;\;\stackrel{11\;12}{[-]}\;\;\stackrel{13\;14}{[-]} $&-1&$-\frac{1}{2}$&
$-\frac{1}{2}$&0&$\frac{1}{2}$&$\frac{1}{2\,\sqrt{3}}$&$\frac{1}{6}$&$\frac{1}{6}$&$-\frac{1}{3}$\\
\hline
7&$ u_{L}^{c1}$&$ - \stackrel{03}{[-i]}\,\stackrel{12}{[+]}|\stackrel{56}{[+]}\,\stackrel{78}{[-]}
||\stackrel{9 \;10}{(+)}\;\;\stackrel{11\;12}{[-]}\;\;\stackrel{13\;14}{[-]}$ &-1&$\frac{1}{2}$&
$\frac{1}{2}$&0 &$\frac{1}{2}$&$\frac{1}{2\,\sqrt{3}}$&$\frac{1}{6}$&$\frac{1}{6}$&$\frac{2}{3}$\\
\hline
8&$u_{L}^{c1}$&$\stackrel{03}{(+i)}\,\stackrel{12}{(-)}|\stackrel{56}{[+]}\,\stackrel{78}{[-]}
||\stackrel{9 \;10}{(+)}\;\;\stackrel{11\;12}{[-]}\;\;\stackrel{13\;14}{[-]}$&-1&$-\frac{1}{2}$&
$\frac{1}{2}$&0&$\frac{1}{2}$&$\frac{1}{2\,\sqrt{3}}$&$\frac{1}{6}$&$\frac{1}{6}$&$\frac{2}{3}$\\
\hline\hline
\shrinkheight{0.25\textheight}
9&$ u_{R}^{c2}$&$ \stackrel{03}{(+i)}\,\stackrel{12}{[+]}|
\stackrel{56}{[+]}\,\stackrel{78}{(+)}
||\stackrel{9 \;10}{[-]}\;\;\stackrel{11\;12}{(+)}\;\;\stackrel{13\;14}{[-]} $ &1&$\frac{1}{2}$&0&
$\frac{1}{2}$&$-\frac{1}{2}$&$\frac{1}{2\,\sqrt{3}}$&$\frac{1}{6}$&$\frac{2}{3}$&$\frac{2}{3}$\\
\hline
10&$u_{R}^{c2}$&$\stackrel{03}{[-i]}\,\stackrel{12}{(-)}|\stackrel{56}{[+]}\,\stackrel{78}{(+)}
||\stackrel{9 \;10}{[-]}\;\;\stackrel{11\;12}{(+)}\;\;\stackrel{13\;14}{[-]}$&1&$-\frac{1}{2}$&0&
$\frac{1}{2}$&$-\frac{1}{2}$&$\frac{1}{2\,\sqrt{3}}$&$\frac{1}{6}$&$\frac{2}{3}$&$\frac{2}{3}$\\
\hline
11&$d_{R}^{c2}$&$\stackrel{03}{(+i)}\,\stackrel{12}{[+]}|\stackrel{56}{(-)}\,\stackrel{78}{[-]}
||\stackrel{9 \;10}{[-]}\;\;\stackrel{11\;12}{(+)}\;\;\stackrel{13\;14}{[-]}$
&1&$\frac{1}{2}$&0&
$-\frac{1}{2}$&$ - \frac{1}{2}$&$\frac{1}{2\,\sqrt{3}}$&$\frac{1}{6}$&$-\frac{1}{3}$&$-\frac{1}{3}$\\
\hline
12&$ d_{R}^{c2} $&$\stackrel{03}{[-i]}\,\stackrel{12}{(-)}|
\stackrel{56}{(-)}\,\stackrel{78}{[-]}
||\stackrel{9 \;10}{[-]}\;\;\stackrel{11\;12}{(+)}\;\;\stackrel{13\;14}{[-]} $
&1&$-\frac{1}{2}$&0&
$-\frac{1}{2}$&$-\frac{1}{2}$&$\frac{1}{2\,\sqrt{3}}$&$\frac{1}{6}$&$-\frac{1}{3}$&$-\frac{1}{3}$\\
\hline
13&$d_{L}^{c2}$&$\stackrel{03}{[-i]}\,\stackrel{12}{[+]}|\stackrel{56}{(-)}\,\stackrel{78}{(+)}
||\stackrel{9 \;10}{[-]}\;\;\stackrel{11\;12}{(+)}\;\;\stackrel{13\;14}{[-]}$
&-1&$\frac{1}{2}$&
$-\frac{1}{2}$&0&$-\frac{1}{2}$&$\frac{1}{2\,\sqrt{3}}$&$\frac{1}{6}$&$\frac{1}{6}$&$-\frac{1}{3}$\\
\hline
14&$d_{L}^{c2} $&$ - \stackrel{03}{(+i)}\,\stackrel{12}{(-)}|\stackrel{56}{(-)}\,\stackrel{78}{(+)}
||\stackrel{9 \;10}{[-]}\;\;\stackrel{11\;12}{(+)}\;\;\stackrel{13\;14}{[-]} $&-1&$-\frac{1}{2}$&
$-\frac{1}{2}$&0&$-\frac{1}{2}$&$\frac{1}{2\,\sqrt{3}}$&$\frac{1}{6}$&$\frac{1}{6}$&$-\frac{1}{3}$\\
\hline
15&$ u_{L}^{c2}$&$ - \stackrel{03}{[-i]}\,\stackrel{12}{[+]}|\stackrel{56}{[+]}\,\stackrel{78}{[-]}
||\stackrel{9 \;10}{[-]}\;\;\stackrel{11\;12}{(+)}\;\;\stackrel{13\;14}{[-]}$ &-1&$\frac{1}{2}$&
$\frac{1}{2}$&0 &$-\frac{1}{2}$&$\frac{1}{2\,\sqrt{3}}$&$\frac{1}{6}$&$\frac{1}{6}$&$\frac{2}{3}$\\
\hline
16&$u_{L}^{c2}$&$\stackrel{03}{(+i)}\,\stackrel{12}{(-)}|\stackrel{56}{[+]}\,\stackrel{78}{[-]}
||\stackrel{9 \;10}{[-]}\;\;\stackrel{11\;12}{(+)}\;\;\stackrel{13\;14}{[-]}$&-1&$-\frac{1}{2}$&
$\frac{1}{2}$&0&$-\frac{1}{2}$&$\frac{1}{2\,\sqrt{3}}$&$\frac{1}{6}$&$\frac{1}{6}$&$\frac{2}{3}$\\
\hline\hline
17&$ u_{R}^{c3}$&$ \stackrel{03}{(+i)}\,\stackrel{12}{[+]}|
\stackrel{56}{[+]}\,\stackrel{78}{(+)}
||\stackrel{9 \;10}{[-]}\;\;\stackrel{11\;12}{[-]}\;\;\stackrel{13\;14}{(+)} $ &1&$\frac{1}{2}$&0&
$\frac{1}{2}$&$0$&$-\frac{1}{\sqrt{3}}$&$\frac{1}{6}$&$\frac{2}{3}$&$\frac{2}{3}$\\
\hline
18&$u_{R}^{c3}$&$\stackrel{03}{[-i]}\,\stackrel{12}{(-)}|\stackrel{56}{[+]}\,\stackrel{78}{(+)}
||\stackrel{9 \;10}{[-]}\;\;\stackrel{11\;12}{[-]}\;\;\stackrel{13\;14}{(+)}$&1&$-\frac{1}{2}$&0&
$\frac{1}{2}$&$0$&$-\frac{1}{\sqrt{3}}$&$\frac{1}{6}$&$\frac{2}{3}$&$\frac{2}{3}$\\
\hline
19&$d_{R}^{c3}$&$\stackrel{03}{(+i)}\,\stackrel{12}{[+]}|\stackrel{56}{(-)}\,\stackrel{78}{[-]}
||\stackrel{9 \;10}{[-]}\;\;\stackrel{11\;12}{[-]}\;\;\stackrel{13\;14}{(+)}$&1&$\frac{1}{2}$&0&
$-\frac{1}{2}$&$0$&$-\frac{1}{\sqrt{3}}$&$\frac{1}{6}$&$-\frac{1}{3}$&$-\frac{1}{3}$\\
\hline
20&$ d_{R}^{c3} $&$\stackrel{03}{[-i]}\,\stackrel{12}{(-)}|
\stackrel{56}{(-)}\,\stackrel{78}{[-]}
||\stackrel{9 \;10}{[-]}\;\;\stackrel{11\;12}{[-]}\;\;\stackrel{13\;14}{(+)} $&1&$-\frac{1}{2}$&0&
$-\frac{1}{2}$&$0$&$-\frac{1}{\sqrt{3}}$&$\frac{1}{6}$&$-\frac{1}{3}$&$-\frac{1}{3}$\\
\hline
21&$d_{L}^{c3}$&$\stackrel{03}{[-i]}\,\stackrel{12}{[+]}|\stackrel{56}{(-)}\,\stackrel{78}{(+)}
||\stackrel{9 \;10}{[-]}\;\;\stackrel{11\;12}{[-]}\;\;\stackrel{13\;14}{(+)}$&-1&$\frac{1}{2}$&
$-\frac{1}{2}$&0&$0$&$-\frac{1}{\sqrt{3}}$&$\frac{1}{6}$&$\frac{1}{6}$&$-\frac{1}{3}$\\
\hline
22&$d_{L}^{c3} $&$ - \stackrel{03}{(+i)}\,\stackrel{12}{(-)}|\stackrel{56}{(-)}\,\stackrel{78}{(+)}
||\stackrel{9 \;10}{[-]}\;\;\stackrel{11\;12}{[-]}\;\;\stackrel{13\;14}{(+)} $&-1&$-\frac{1}{2}$&
$-\frac{1}{2}$&0&$0$&$-\frac{1}{\sqrt{3}}$&$\frac{1}{6}$&$\frac{1}{6}$&$-\frac{1}{3}$\\
\hline
23&$ u_{L}^{c3}$&$ - \stackrel{03}{[-i]}\,\stackrel{12}{[+]}|\stackrel{56}{[+]}\,\stackrel{78}{[-]}
||\stackrel{9 \;10}{[-]}\;\;\stackrel{11\;12}{[-]}\;\;\stackrel{13\;14}{(+)}$ &-1&$\frac{1}{2}$&
$\frac{1}{2}$&0 &$0$&$-\frac{1}{\sqrt{3}}$&$\frac{1}{6}$&$\frac{1}{6}$&$\frac{2}{3}$\\
\hline
24&$u_{L}^{c3}$&$\stackrel{03}{(+i)}\,\stackrel{12}{(-)}|\stackrel{56}{[+]}\,\stackrel{78}{[-]}
||\stackrel{9 \;10}{[-]}\;\;\stackrel{11\;12}{[-]}\;\;\stackrel{13\;14}{(+)}$&-1&$-\frac{1}{2}$&
$\frac{1}{2}$&0&$0$&$-\frac{1}{\sqrt{3}}$&$\frac{1}{6}$&$\frac{1}{6}$&$\frac{2}{3}$\\
\hline\hline
25&$ \nu_{R}$&$ \stackrel{03}{(+i)}\,\stackrel{12}{[+]}|
\stackrel{56}{[+]}\,\stackrel{78}{(+)}
||\stackrel{9 \;10}{(+)}\;\;\stackrel{11\;12}{(+)}\;\;\stackrel{13\;14}{(+)} $ &1&$\frac{1}{2}$&0&
$\frac{1}{2}$&$0$&$0$&$-\frac{1}{2}$&$0$&$0$\\
\hline
26&$\nu_{R}$&$\stackrel{03}{[-i]}\,\stackrel{12}{(-)}|\stackrel{56}{[+]}\,\stackrel{78}{(+)}
||\stackrel{9 \;10}{(+)}\;\;\stackrel{11\;12}{(+)}\;\;\stackrel{13\;14}{(+)}$&1&$-\frac{1}{2}$&0&
$\frac{1}{2}$ &$0$&$0$&$-\frac{1}{2}$&$0$&$0$\\
\hline
27&$e_{R}$&$\stackrel{03}{(+i)}\,\stackrel{12}{[+]}|\stackrel{56}{(-)}\,\stackrel{78}{[-]}
||\stackrel{9 \;10}{(+)}\;\;\stackrel{11\;12}{(+)}\;\;\stackrel{13\;14}{(+)}$&1&$\frac{1}{2}$&0&
$-\frac{1}{2}$&$0$&$0$&$-\frac{1}{2}$&$-1$&$-1$\\
\hline
28&$ e_{R} $&$\stackrel{03}{[-i]}\,\stackrel{12}{(-)}|
\stackrel{56}{(-)}\,\stackrel{78}{[-]}
||\stackrel{9 \;10}{(+)}\;\;\stackrel{11\;12}{(+)}\;\;\stackrel{13\;14}{(+)} $&1&$-\frac{1}{2}$&0&
$-\frac{1}{2}$&$0$&$0$&$-\frac{1}{2}$&$-1$&$-1$\\
\hline
29&$e_{L}$&$\stackrel{03}{[-i]}\,\stackrel{12}{[+]}|\stackrel{56}{(-)}\,\stackrel{78}{(+)}
||\stackrel{9 \;10}{(+)}\;\;\stackrel{11\;12}{(+)}\;\;\stackrel{13\;14}{(+)}$&-1&$\frac{1}{2}$&
$-\frac{1}{2}$&0&$0$&$0$&$-\frac{1}{2}$&$-\frac{1}{2}$&$-1$\\
\hline
30&$e_{L} $&$ - \stackrel{03}{(+i)}\,\stackrel{12}{(-)}|\stackrel{56}{(-)}\,\stackrel{78}{(+)}
||\stackrel{9 \;10}{(+)}\;\;\stackrel{11\;12}{(+)}\;\;\stackrel{13\;14}{(+)} $&-1&$-\frac{1}{2}$&
$-\frac{1}{2}$&0&$0$&$0$&$-\frac{1}{2}$&$-\frac{1}{2}$&$-1$\\
\hline
31&$ \nu_{L}$&$ - \stackrel{03}{[-i]}\,\stackrel{12}{[+]}|\stackrel{56}{[+]}\,\stackrel{78}{[-]}
||\stackrel{9 \;10}{(+)}\;\;\stackrel{11\;12}{(+)}\;\;\stackrel{13\;14}{(+)}$ &-1&$\frac{1}{2}$&
$\frac{1}{2}$&0 &$0$&$0$&$-\frac{1}{2}$&$-\frac{1}{2}$&$0$\\
\hline
32&$\nu_{L}$&$\stackrel{03}{(+i)}\,\stackrel{12}{(-)}|\stackrel{56}{[+]}\,\stackrel{78}{[-]}
||\stackrel{9 \;10}{(+)}\;\;\stackrel{11\;12}{(+)}\;\;\stackrel{13\;14}{(+)}$&-1&$-\frac{1}{2}$&
$\frac{1}{2}$&0&$0$&$0$&$-\frac{1}{2}$&$-\frac{1}{2}$&$0$\\
\hline\hline
33&$ \bar{d}_{L}^{\bar{c1}}$&$ \stackrel{03}{[-i]}\,\stackrel{12}{[+]}|
\stackrel{56}{[+]}\,\stackrel{78}{(+)}
||\stackrel{9 \;10}{[-]}\;\;\stackrel{11\;12}{(+)}\;\;\stackrel{13\;14}{(+)} $ &-1&$\frac{1}{2}$&0&
$\frac{1}{2}$&$-\frac{1}{2}$&$-\frac{1}{2\,\sqrt{3}}$&$-\frac{1}{6}$&$\frac{1}{3}$&$\frac{1}{3}$\\
\hline
34&$\bar{d}_{L}^{\bar{c1}}$&$\stackrel{03}{(+i)}\,\stackrel{12}{(-)}|\stackrel{56}{[+]}\,\stackrel{78}{(+)}
||\stackrel{9 \;10}{[-]}\;\;\stackrel{11\;12}{(+)}\;\;\stackrel{13\;14}{(+)}$&-1&$-\frac{1}{2}$&0&
$\frac{1}{2}$&$-\frac{1}{2}$&$-\frac{1}{2\,\sqrt{3}}$&$-\frac{1}{6}$&$\frac{1}{3}$&$\frac{1}{3}$\\
\hline
35&$\bar{u}_{L}^{\bar{c1}}$&$ - \stackrel{03}{[-i]}\,\stackrel{12}{[+]}|\stackrel{56}{(-)}\,\stackrel{78}{[-]}
||\stackrel{9 \;10}{[-]}\;\;\stackrel{11\;12}{(+)}\;\;\stackrel{13\;14}{(+)}$&-1&$\frac{1}{2}$&0&
$-\frac{1}{2}$&$-\frac{1}{2}$&$-\frac{1}{2\,\sqrt{3}}$&$-\frac{1}{6}$&$-\frac{2}{3}$&$-\frac{2}{3}$\\
\hline
36&$ \bar{u}_{L}^{\bar{c1}} $&$ - \stackrel{03}{(+i)}\,\stackrel{12}{(-)}|
\stackrel{56}{(-)}\,\stackrel{78}{[-]}
||\stackrel{9 \;10}{[-]}\;\;\stackrel{11\;12}{(+)}\;\;\stackrel{13\;14}{(+)} $&-1&$-\frac{1}{2}$&0&
$-\frac{1}{2}$&$-\frac{1}{2}$&$-\frac{1}{2\,\sqrt{3}}$&$-\frac{1}{6}$&$-\frac{2}{3}$&$-\frac{2}{3}$\\
\hline
37&$\bar{d}_{R}^{\bar{c1}}$&$\stackrel{03}{(+i)}\,\stackrel{12}{[+]}|\stackrel{56}{[+]}\,\stackrel{78}{[-]}
||\stackrel{9 \;10}{[-]}\;\;\stackrel{11\;12}{(+)}\;\;\stackrel{13\;14}{(+)}$&1&$\frac{1}{2}$&
$\frac{1}{2}$&0&$-\frac{1}{2}$&$-\frac{1}{2\,\sqrt{3}}$&$-\frac{1}{6}$&$-\frac{1}{6}$&$\frac{1}{3}$\\
\hline
38&$\bar{d}_{R}^{\bar{c1}} $&$ - \stackrel{03}{[-i]}\,\stackrel{12}{(-)}|\stackrel{56}{[+]}\,\stackrel{78}{[-]}
||\stackrel{9 \;10}{[-]}\;\;\stackrel{11\;12}{(+)}\;\;\stackrel{13\;14}{(+)} $&1&$-\frac{1}{2}$&
$\frac{1}{2}$&0&$-\frac{1}{2}$&$-\frac{1}{2\,\sqrt{3}}$&$-\frac{1}{6}$&$-\frac{1}{6}$&$\frac{1}{3}$\\
\hline
39&$ \bar{u}_{R}^{\bar{c1}}$&$\stackrel{03}{(+i)}\,\stackrel{12}{[+]}|\stackrel{56}{(-)}\,\stackrel{78}{(+)}
||\stackrel{9 \;10}{[-]}\;\;\stackrel{11\;12}{(+)}\;\;\stackrel{13\;14}{(+)}$ &1&$\frac{1}{2}$&
$-\frac{1}{2}$&0 &$-\frac{1}{2}$&$-\frac{1}{2\,\sqrt{3}}$&$-\frac{1}{6}$&$-\frac{1}{6}$&$-\frac{2}{3}$\\
\hline
40&$\bar{u}_{R}^{\bar{c1}}$&$\stackrel{03}{[-i]}\,\stackrel{12}{(-)}|\stackrel{56}{(-)}\,\stackrel{78}{(+)}
||\stackrel{9 \;10}{[-]}\;\;\stackrel{11\;12}{(+)}\;\;\stackrel{13\;14}{(+)}$
&1&$-\frac{1}{2}$&
$-\frac{1}{2}$&0&$-\frac{1}{2}$&$-\frac{1}{2\,\sqrt{3}}$&$-\frac{1}{6}$&$-\frac{1}{6}$&$-\frac{2}{3}$\\
\hline\hline
41&$ \bar{d}_{L}^{\bar{c2}}$&$ \stackrel{03}{[-i]}\,\stackrel{12}{[+]}|
\stackrel{56}{[+]}\,\stackrel{78}{(+)}
||\stackrel{9 \;10}{(+)}\;\;\stackrel{11\;12}{[-]}\;\;\stackrel{13\;14}{(+)} $
&-1&$\frac{1}{2}$&0&
$\frac{1}{2}$&$\frac{1}{2}$&$-\frac{1}{2\,\sqrt{3}}$&$-\frac{1}{6}$&$\frac{1}{3}$&$\frac{1}{3}$\\
\hline
42&$\bar{d}_{L}^{\bar{c2}}$&$\stackrel{03}{(+i)}\,\stackrel{12}{(-)}|\stackrel{56}{[+]}\,\stackrel{78}{(+)}
||\stackrel{9 \;10}{(+)}\;\;\stackrel{11\;12}{[-]}\;\;\stackrel{13\;14}{(+)}$
&-1&$-\frac{1}{2}$&0&
$\frac{1}{2}$&$\frac{1}{2}$&$-\frac{1}{2\,\sqrt{3}}$&$-\frac{1}{6}$&$\frac{1}{3}$&$\frac{1}{3}$\\
\hline
43&$\bar{u}_{L}^{\bar{c2}}$&$ - \stackrel{03}{[-i]}\,\stackrel{12}{[+]}|\stackrel{56}{(-)}\,\stackrel{78}{[-]}
||\stackrel{9 \;10}{(+)}\;\;\stackrel{11\;12}{[-]}\;\;\stackrel{13\;14}{(+)}$
&-1&$\frac{1}{2}$&0&
$-\frac{1}{2}$&$\frac{1}{2}$&$-\frac{1}{2\,\sqrt{3}}$&$-\frac{1}{6}$&$-\frac{2}{3}$&$-\frac{2}{3}$\\
\hline
44&$ \bar{u}_{L}^{\bar{c2}} $&$ - \stackrel{03}{(+i)}\,\stackrel{12}{(-)}|
\stackrel{56}{(-)}\,\stackrel{78}{[-]}
||\stackrel{9 \;10}{(+)}\;\;\stackrel{11\;12}{[-]}\;\;\stackrel{13\;14}{(+)} $
&-1&$-\frac{1}{2}$&0&
$-\frac{1}{2}$&$\frac{1}{2}$&$-\frac{1}{2\,\sqrt{3}}$&$-\frac{1}{6}$&$-\frac{2}{3}$&$-\frac{2}{3}$\\
\hline
45&$\bar{d}_{R}^{\bar{c2}}$&$\stackrel{03}{(+i)}\,\stackrel{12}{[+]}|\stackrel{56}{[+]}\,\stackrel{78}{[-]}
||\stackrel{9 \;10}{(+)}\;\;\stackrel{11\;12}{[-]}\;\;\stackrel{13\;14}{(+)}$
&1&$\frac{1}{2}$&
$\frac{1}{2}$&0&$\frac{1}{2}$&$-\frac{1}{2\,\sqrt{3}}$&$-\frac{1}{6}$&$-\frac{1}{6}$&$\frac{1}{3}$\\
\hline
46&$\bar{d}_{R}^{\bar{c2}} $&$ - \stackrel{03}{[-i]}\,\stackrel{12}{(-)}|\stackrel{56}{[+]}\,\stackrel{78}{[-]}
||\stackrel{9 \;10}{(+)}\;\;\stackrel{11\;12}{[-]}\;\;\stackrel{13\;14}{(+)} $
&1&$-\frac{1}{2}$&
$\frac{1}{2}$&0&$\frac{1}{2}$&$-\frac{1}{2\,\sqrt{3}}$&$-\frac{1}{6}$&$-\frac{1}{6}$&$\frac{1}{3}$\\
\hline
47&$ \bar{u}_{R}^{\bar{c2}}$&$\stackrel{03}{(+i)}\,\stackrel{12}{[+]}|\stackrel{56}{(-)}\,\stackrel{78}{(+)}
||\stackrel{9 \;10}{(+)}\;\;\stackrel{11\;12}{[-]}\;\;\stackrel{13\;14}{(+)}$
 &1&$\frac{1}{2}$&
$-\frac{1}{2}$&0 &$\frac{1}{2}$&$-\frac{1}{2\,\sqrt{3}}$&$-\frac{1}{6}$&$-\frac{1}{6}$&$-\frac{2}{3}$\\
\hline
48&$\bar{u}_{R}^{\bar{c2}}$&$\stackrel{03}{[-i]}\,\stackrel{12}{(-)}|\stackrel{56}{(-)}\,\stackrel{78}{(+)}
||\stackrel{9 \;10}{(+)}\;\;\stackrel{11\;12}{[-]}\;\;\stackrel{13\;14}{(+)}$
&1&$-\frac{1}{2}$&
$-\frac{1}{2}$&0&$\frac{1}{2}$&$-\frac{1}{2\,\sqrt{3}}$&$-\frac{1}{6}$&$-\frac{1}{6}$&$-\frac{2}{3}$\\
\hline\hline
49&$ \bar{d}_{L}^{\bar{c3}}$&$ \stackrel{03}{[-i]}\,\stackrel{12}{[+]}|
\stackrel{56}{[+]}\,\stackrel{78}{(+)}
||\stackrel{9 \;10}{(+)}\;\;\stackrel{11\;12}{(+)}\;\;\stackrel{13\;14}{[-]} $ &-1&$\frac{1}{2}$&0&
$\frac{1}{2}$&$0$&$\frac{1}{\sqrt{3}}$&$-\frac{1}{6}$&$\frac{1}{3}$&$\frac{1}{3}$\\
\hline
50&$\bar{d}_{L}^{\bar{c3}}$&$\stackrel{03}{(+i)}\,\stackrel{12}{(-)}|\stackrel{56}{[+]}\,\stackrel{78}{(+)}
||\stackrel{9 \;10}{(+)}\;\;\stackrel{11\;12}{(+)}\;\;\stackrel{13\;14}{[-]} $&-1&$-\frac{1}{2}$&0&
$\frac{1}{2}$&$0$&$\frac{1}{\sqrt{3}}$&$-\frac{1}{6}$&$\frac{1}{3}$&$\frac{1}{3}$\\
\hline
51&$\bar{u}_{L}^{\bar{c3}}$&$ - \stackrel{03}{[-i]}\,\stackrel{12}{[+]}|\stackrel{56}{(-)}\,\stackrel{78}{[-]}
||\stackrel{9 \;10}{(+)}\;\;\stackrel{11\;12}{(+)}\;\;\stackrel{13\;14}{[-]} $&-1&$\frac{1}{2}$&0&
$-\frac{1}{2}$&$0$&$\frac{1}{\sqrt{3}}$&$-\frac{1}{6}$&$-\frac{2}{3}$&$-\frac{2}{3}$\\
\hline
52&$ \bar{u}_{L}^{\bar{c3}} $&$ - \stackrel{03}{(+i)}\,\stackrel{12}{(-)}|
\stackrel{56}{(-)}\,\stackrel{78}{[-]}
||\stackrel{9 \;10}{(+)}\;\;\stackrel{11\;12}{(+)}\;\;\stackrel{13\;14}{[-]}  $&-1&$-\frac{1}{2}$&0&
$-\frac{1}{2}$&$0$&$\frac{1}{\sqrt{3}}$&$-\frac{1}{6}$&$-\frac{2}{3}$&$-\frac{2}{3}$\\
\hline
53&$\bar{d}_{R}^{\bar{c3}}$&$\stackrel{03}{(+i)}\,\stackrel{12}{[+]}|\stackrel{56}{[+]}\,\stackrel{78}{[-]}
||\stackrel{9 \;10}{(+)}\;\;\stackrel{11\;12}{(+)}\;\;\stackrel{13\;14}{[-]} $&1&$\frac{1}{2}$&
$\frac{1}{2}$&0&$0$&$\frac{1}{\sqrt{3}}$&$-\frac{1}{6}$&$-\frac{1}{6}$&$\frac{1}{3}$\\
\hline
54&$\bar{d}_{R}^{\bar{c3}} $&$ - \stackrel{03}{[-i]}\,\stackrel{12}{(-)}|\stackrel{56}{[+]}\,\stackrel{78}{[-]}
||\stackrel{9 \;10}{(+)}\;\;\stackrel{11\;12}{(+)}\;\;\stackrel{13\;14}{[-]} $&1&$-\frac{1}{2}$&
$\frac{1}{2}$&0&$0$&$\frac{1}{\sqrt{3}}$&$-\frac{1}{6}$&$-\frac{1}{6}$&$\frac{1}{3}$\\
\hline
55&$ \bar{u}_{R}^{\bar{c3}}$&$\stackrel{03}{(+i)}\,\stackrel{12}{[+]}|\stackrel{56}{(-)}\,\stackrel{78}{(+)}
||\stackrel{9 \;10}{(+)}\;\;\stackrel{11\;12}{(+)}\;\;\stackrel{13\;14}{[-]} $ &1&$\frac{1}{2}$&
$-\frac{1}{2}$&0 &$0$&$\frac{1}{\sqrt{3}}$&$-\frac{1}{6}$&$-\frac{1}{6}$&$-\frac{2}{3}$\\
\hline
56&$\bar{u}_{R}^{\bar{c3}}$&$\stackrel{03}{[-i]}\,\stackrel{12}{(-)}|\stackrel{56}{(-)}\,\stackrel{78}{(+)}
||\stackrel{9 \;10}{(+)}\;\;\stackrel{11\;12}{(+)}\;\;\stackrel{13\;14}{[-]} $&1&$-\frac{1}{2}$&
$-\frac{1}{2}$&0&$0$&$\frac{1}{\sqrt{3}}$&$-\frac{1}{6}$&$-\frac{1}{6}$&$-\frac{2}{3}$\\
\hline\hline
57&$ \bar{e}_{L}$&$ \stackrel{03}{[-i]}\,\stackrel{12}{[+]}|
\stackrel{56}{[+]}\,\stackrel{78}{(+)}
||\stackrel{9 \;10}{[-]}\;\;\stackrel{11\;12}{[-]}\;\;\stackrel{13\;14}{[-]} $ &-1&$\frac{1}{2}$&0&
$\frac{1}{2}$&$0$&$0$&$\frac{1}{2}$&$1$&$1$\\
\hline
58&$\bar{e}_{L}$&$\stackrel{03}{(+i)}\,\stackrel{12}{(-)}|\stackrel{56}{[+]}\,\stackrel{78}{(+)}
||\stackrel{9 \;10}{[-]}\;\;\stackrel{11\;12}{[-]}\;\;\stackrel{13\;14}{[-]}$&-1&$-\frac{1}{2}$&0&
$\frac{1}{2}$ &$0$&$0$&$\frac{1}{2}$&$1$&$1$\\
\hline
59&$\bar{\nu}_{L}$&$ - \stackrel{03}{[-i]}\,\stackrel{12}{[+]}|\stackrel{56}{(-)}\,\stackrel{78}{[-]}
||\stackrel{9 \;10}{[-]}\;\;\stackrel{11\;12}{[-]}\;\;\stackrel{13\;14}{[-]}$&-1&$\frac{1}{2}$&0&
$-\frac{1}{2}$&$0$&$0$&$\frac{1}{2}$&$0$&$0$\\
\hline
60&$ \bar{\nu}_{L} $&$ - \stackrel{03}{(+i)}\,\stackrel{12}{(-)}|
\stackrel{56}{(-)}\,\stackrel{78}{[-]}
||\stackrel{9 \;10}{[-]}\;\;\stackrel{11\;12}{[-]}\;\;\stackrel{13\;14}{[-]} $&-1&$-\frac{1}{2}$&0&
$-\frac{1}{2}$&$0$&$0$&$\frac{1}{2}$&$0$&$0$\\
\hline
61&$\bar{\nu}_{R}$&$\stackrel{03}{(+i)}\,\stackrel{12}{[+]}|\stackrel{56}{(-)}\,\stackrel{78}{(+)}
||\stackrel{9 \;10}{[-]}\;\;\stackrel{11\;12}{[-]}\;\;\stackrel{13\;14}{[-]}$&1&$\frac{1}{2}$&
$-\frac{1}{2}$&0&$0$&$0$&$\frac{1}{2}$&$\frac{1}{2}$&$0$\\
\hline
62&$\bar{\nu}_{R} $&$ - \stackrel{03}{[-i]}\,\stackrel{12}{(-)}|\stackrel{56}{(-)}\,\stackrel{78}{(+)}
||\stackrel{9 \;10}{[-]}\;\;\stackrel{11\;12}{[-]}\;\;\stackrel{13\;14}{[-]} $&1&$-\frac{1}{2}$&
$-\frac{1}{2}$&0&$0$&$0$&$\frac{1}{2}$&$\frac{1}{2}$&$0$\\
\hline
63&$ \bar{e}_{R}$&$\stackrel{03}{(+i)}\,\stackrel{12}{[+]}|\stackrel{56}{[+]}\,\stackrel{78}{[-]}
||\stackrel{9 \;10}{[-]}\;\;\stackrel{11\;12}{[-]}\;\;\stackrel{13\;14}{[-]}$ &1&$\frac{1}{2}$&
$\frac{1}{2}$&0 &$0$&$0$&$\frac{1}{2}$&$\frac{1}{2}$&$1$\\
\hline
64&$\bar{e}_{R}$&$\stackrel{03}{[-i]}\,\stackrel{12}{(-)}|\stackrel{56}{[+]}\,\stackrel{78}{[-]}
||\stackrel{9 \;10}{[-]}\;\;\stackrel{11\;12}{[-]}\;\;\stackrel{13\;14}{[-]}$&1&$-\frac{1}{2}$&
$\frac{1}{2}$&0&$0$&$0$&$\frac{1}{2}$&$\frac{1}{2}$&$1$\\
\hline
\end{supertabular}
}
\end{center}

%
\section{Handedness in Grassmann and Clifford space}
\label{handednessGrassCliff}

The handedness $\Gamma^{(d)}$ is one of the invariants of the group $SO(d)$, 
with the infinitesimal generators of the Lorentz group $S^{ab}$,
defined as 
\begin{eqnarray}
\label{handedness}
\Gamma^{(d)}&=&\alpha\, \varepsilon_{a_1 a_2\dots a_{d-1} a_d}\, S^{a_1 a_2} 
\cdot S^{a_3 a_4} \cdots S^{a_{d-1} a_d}\,,
\end{eqnarray}
with $\alpha$, which is chosen so that $\Gamma^{(d)}=\pm 1$.

In the Grassmann case  $S^{ab}$  is defined in Eq.~(\ref{cartangrasscliff}), while in the Clifford case
Eq.~(\ref{handedness}) simplifies, if we take into account that $S^{ab}|_{a\ne b}= 
\frac{i}{2}\gamma^a \gamma^b$  and $\tilde{S}^{ab}|_{a\ne b}= 
\frac{i}{2}\tilde{\gamma}^a \tilde{\gamma}^b$, as follows
\begin{eqnarray}
\Gamma^{(d)} :&=&(i)^{d/2}\; \;\;\;\;\;\prod_a \quad (\sqrt{\eta^{aa}} \gamma^a), 
\quad {\rm if } \quad d = 2n\,. 
\nonumber\\
\label{hand}
\end{eqnarray}
%

\section*{Acknowledgment}
The author N.S.M.B. thanks Department of Physics, FMF, University of Ljubljana, Society of 
Mathematicians, Physicists and Astronomers of Slovenia,  for supporting the research on the 
{\it spin-charge-family} theory by offering the room and computer facilities and Matja\v z 
Breskvar of Beyond Semiconductor for donations, in particular for the annual workshops entitled "What comes beyond the standard models". 


%
\end{document}